\documentclass[a4paper,12pt]{report}

\usepackage[left=3cm,right=2cm,top=2.5cm,bottom=2.5cm]{geometry}
\usepackage{afterpage}

\usepackage[margin=\the\parindent,small,bf,rm]{caption}
\usepackage{graphicx}
\usepackage{pdfpages}
\usepackage[english]{babel}
\usepackage{microtype}
\usepackage{setspace}

\usepackage[inline]{enumitem}
\usepackage{listings,xfp}
\usepackage{anyfontsize}
\usepackage{bold-extra}
\usepackage{quotchap}

\makeatletter
{\large 
	\xdef\f@size@large{\f@size}
	\xdef\f@baselineskip@large{\f@baselineskip}
	\Large 
	\xdef\f@size@Large{\f@size}
	\xdef\f@baselineskip@Large{\f@baselineskip}
}
\newcommand{\largetoLarge}{%
	\fontsize
	{\fpeval{(\f@size@large+\f@size@Large)/2}}
	{\fpeval{(\f@baselineskip@large+\f@baselineskip@Large)/2}}%
	\selectfont
}
\makeatother
\makeatletter
{\tiny 
	\xdef\f@size@tiny{\f@size}
	\xdef\f@baselineskip@tiny{\f@baselineskip}
	\normalsize 
	\xdef\f@size@normalsize{\f@size}
	\xdef\f@baselineskip@normalsize{\f@baselineskip}
}
\newcommand{\tinytonormal}{%
	\fontsize
	{\fpeval{(\f@size@tiny+\f@size@normalsize)/2}}
	{\fpeval{(\f@baselineskip@tiny+\f@baselineskip@normalsize)/2}}%
	\selectfont
}
\makeatother
\makeatletter
{\tiny 
	\xdef\f@size@tiny{\f@size}
	\xdef\f@baselineskip@tiny{\f@baselineskip}
	\normalsize 
	\xdef\f@size@normalsize{\f@size}
	\xdef\f@baselineskip@normalsize{\f@baselineskip}
}
\newcommand{\othertinytonormal}{%
	\fontsize
	{\fpeval{(\f@size@tiny+\f@size@normalsize)*0.4}}
	{\fpeval{(\f@baselineskip@tiny+\f@baselineskip@normalsize)*0.4}}%
	\selectfont
}
\makeatother

\usepackage{textcase}
\usepackage{mfirstuc}
\usepackage{titlecaps}
\usepackage{stringstrings}
\usepackage[raggedright,rm,bf]{titlesec}
\usepackage{titlecaps}
\titlelabel{\thetitle.\quad}
\titleformat{\chapter}[display]{\LARGE\rmfamily\raggedleft}{\scshape \chaptertitlename\ \thechapter}{3pt}{\titlerule\vspace{1.5ex} \scshape \Huge \centering}[\vspace{1.5ex}\titlerule]
\titlespacing*{\chapter}{0pt}{30pt}{40pt}  

\titleformat{\section}{\rmfamily\scshape\Large}{\thesection}{1em}{}
\titleformat{\subsection}{\rmfamily\scshape\large}{\thesubsection}{1em}{}
\titleformat{\subsubsection}{\rmfamily\scshape\normalsize}{\thesubsubsection}{1em}{}

\newcommand{\ShortInTextTitle}[1]{\textbf{
		\xcapitalisewords{#1}}}

\usepackage{xcolor,titletoc}
\makeatletter
\let\originall@chapter\l@chapter
\def\l@chapter#1#2{\originall@chapter{{\rmfamily #1}}{#2}}
\makeatother
\let \savenumberline \numberline
\def \numberline#1{\savenumberline{#1.}}

\usepackage{acro}
\newlength\myitemwidth
\newlist{myacronymlist}{description}{1}
\setlist[myacronymlist]{
	labelindent = 0pt ,
	labelsep    = 0pt ,
	leftmargin  = \myitemwidth ,
	labelwidth  = \myitemwidth ,
	itemindent  = 0pt ,
	format      = \normalfont \bfseries 
}
\usepackage[intoc, english]{nomencl}

\renewcommand{\nomgroup}[1]{%
	\ifthenelse{\equal{#1}{V}}{\item[{\rmfamily\scshape\Large Variables}]}{%
		\ifthenelse{\equal{#1}{C}}{\item[{\rmfamily\scshape\Large Constants}]}{
			\ifthenelse{\equal{#1}{F}}{\item[{\rmfamily\scshape\Large Functions}]}{}}}%
}

\makenomenclature

\makeatletter
\renewcommand\tableofcontents{%
	\chapter*{\contentsname}%
	\@mkboth{\scshape \contentsname}%
	{\scshape \contentsname}%
	\@starttoc{toc}%
}
\renewcommand\listoftables{%
	\chapter*{\listtablename}%
	\@mkboth{\scshape\listtablename}%
	{\scshape\listtablename}%
	\@starttoc{lot}%
}
\renewcommand\listoffigures{%
	\chapter*{\listfigurename}%
	\@mkboth{\scshape\listfigurename}%
	{\scshape\listfigurename}%
	\@starttoc{lof}%
}
\makeatother

\usepackage[cmex10]{amsmath}
\usepackage{amssymb}
\usepackage{cancel}

\usepackage{placeins}
\usepackage{array}
\usepackage{booktabs,colortbl}
\usepackage{tabularx}
\usepackage{multirow}
\newcommand{\mytable}{
	\centering
	\small
	\renewcommand{\arraystretch}{1.2}
}

\newcommand{\PreserveBackslash}[1]{\let\temp=\\#1\let\\=\temp}
\newcolumntype{C}{>{\centering\arraybackslash}X}
\newcolumntype{L}{>{\raggedright\arraybackslash}X}
\newcolumntype{A}[1]{>{\PreserveBackslash\raggedright}p{#1}}

\setlist[description]{ 
	labelwidth=0.6\textwidth,%
	leftmargin=\labelwidth, 
	align=right,
	font=\normalfont\scshape\bfseries
}

\usepackage[linktocpage=true]{hyperref}
\hypersetup{colorlinks=true,linktoc=all,linkcolor=.,filecolor=.,urlcolor=.,citecolor=.}
\usepackage[nottoc,notlot,notlof]{tocbibind}
\usepackage{blindtext}

\usepackage{etoolbox}
\makeatletter
\patchcmd{\ttlh@hang}{\parindent\z@}{\parindent\z@\leavevmode}{}{}
\patchcmd{\ttlh@hang}{\noindent}{}{}{}
\makeatother

\usepackage[square,numbers,compress,sort]{natbib}
\usepackage{fancyhdr}
\usepackage{extramarks}
\renewcommand{\chaptermark}[1]{\markboth{\normalsize \rmfamily \thechapter.\ \scshape #1}{\normalsize \rmfamily \thechapter.\ \scshape #1}}
\renewcommand{\sectionmark}[1]{\markright{\normalsize \rmfamily \thesection.\ \scshape #1}}
\fancypagestyle{plain}{\fancyhead{}
	
}

\newcommand{\mysection}[2]{
	\let\orisectionmark\sectionmark
	\renewcommand\sectionmark[1]{}%
	\section[#1]{#2 \orisectionmark{#2}}
	\orisectionmark{#2}
	\let\sectionmark\orisectionmark
}
\newcommand{\mychapter}[2]{
	\chapter[#1]{#2 \chaptermark{#2}}
}

\usepackage[prependcaption,textsize=tiny]{todonotes}
\reversemarginpar
\definecolor{othercolor}{HTML}{CC0000}
\definecolor{oraclecolor}{HTML}{8B8589}
\definecolor{rulecolor}{HTML}{DCDCDC}
\definecolor{ashgrey}{rgb}{0.7, 0.75, 0.71}

\newcommand{\system}[1]{{\small \textsc{#1}}}
\newcommand{\tablesystem}[1]{\textsc{#1}}

\usepackage{amsthm}
\newtheorem*{definition*}{Definition}
\usepackage{changepage}

\usepackage{subcaption}
\usepackage{tcolorbox}
\usepackage{framed}
\renewenvironment{framed}[1][\hsize]
{\MakeFramed{\hsize#1\advance\hsize-\width \FrameRestore}}%
{\endMakeFramed}

\newcommand{\captionsep}{\vspace*{-5pt}}

\newcommand{\ThesisTitle}{Multilingual acoustic word embeddings for zero-resource languages}
\newcommand{\Author}{Christiaan Jacobs}

\newcommand{\Supervisor}{Prof. Herman Kamper}
\newcommand{\Department}{Department of Electrical and Electronic Engineering}
\newcommand{\Date}{December 2023}

\usepackage{textcomp}
\usepackage{graphicx}

\begin{document}
\graphicspath{{frontmatter/fig/}}

\begin{titlepage}
	\centering
%
%
	~\vspace{1.5em}
%
	\rule{\textwidth}{1.5pt} 
	\vspace{1.0em} 

	{\Huge \scshape \titlecap{\ThesisTitle} \par}
	
	\vspace{1.0em} 
	\rule{\textwidth}{1.5pt}
	
	\vspace{3em}
	{\Large by \par}
	\vspace{2em}
	
	{\LARGE \textsc{\Author} \par}
%
%
	\vspace{8em}
%
	{\large Dissertation presented for the degree of Doctor of Philosophy (Electronic Engineering) in the Faculty of Engineering at Stellenbosch University. \par}
	\vspace{6em}
	{\large \textsc{Supervisor}: \Supervisor \\
		\Department \par}
	\vspace{3em}	
	{\Large \Date}	
\end{titlepage}

%
%
%
%

\pagenumbering{roman}
\chapter*{Abstract}
\addcontentsline{toc}{chapter}{Abstract}
\makeatletter\@mkboth{}{\scshape Abstract}\makeatother

Developing speech applications with neural networks require large amounts of transcribed speech data. 
The scarcity of labelled speech data therefore restricts the development of speech applications to only a few well-resourced languages.
To address this problem, researchers are taking steps towards developing speech models for languages where no labelled data is available.
In this \textit{zero-resource} setting, researchers are developing methods that aim to learn meaningful linguistic structures from unlabelled speech alone.

Many zero-resource speech applications require speech segments of different durations to be compared.
Acoustic word embeddings (AWEs) are fixed-dimensional representations of variable-duration speech segments.
Proximity in vector space should indicate similarity between the original acoustic segments.
This allows fast and easy comparison between spoken words.

To produce AWEs for a zero-resource language, one approach is to use unlabelled data from the target language.
Another approach is to exploit the benefits of supervised learning by training a single multilingual AWE model on data from multiple well-resourced languages, and then applying the resulting model to an unseen target language.
Previous studies have shown that the supervised multilingual transfer approach outperforms the unsupervised monolingual approach.
However, the multilingual approach is still far from reaching the performance of supervised AWE approaches that are trained on the target language itself.

In this thesis, we make five specific contributions to the development of AWE models and their downstream application.
First, we introduce a novel AWE model called the \system{ContrastiveRNN}.
We compare this model against state-of-the-art AWE models.
On a word discrimination task, we show that the \system{ContrastiveRNN} outperforms all existing models in the unsupervised monolingual setting with an absolute improvement in average precision ranging from 3.3\% to 17.8\% across six evaluation languages.
In the multilingual transfer setting, the \system{ContrastiveRNN} performs on par with existing models.

As our second contribution, we propose a new adaptation strategy.
After a multilingual model is trained, instead of directly applying it to a target language, we first fine-tune the model using unlabelled data from the target language.
The \system{ContrastiveRNN}, although performing on par with multilingual variants, showed the highest increase after adaptation, giving an improvement of roughly 5\% in average precision on five of the six evaluation languages.

As our third contribution, we take a step back and question the effect a particular set of training languages have on a target language.
We specifically investigate the impact of training a multilingual model on languages that belong to the same language family as the target language.
We perform multiple experiments on African languages which show the benefit of using related languages over unrelated languages. 
For example, a multilingual model trained on one-tenth of the data from a related language outperforms a model trained on all the available training data from unrelated languages. 

As our fourth contribution, we showcase the applicability of AWEs by applying them to a real downstream task: 
we develop an AWE-based keyword spotting system (KWS) for hate speech detection in radio broadcasts.
We validate performance using actual Swahili radio audio extracted from radio stations in Kenya, a country in Sub-Saharan Africa.
In developmental experiments, our system falls short of
a speech recognition-based KWS system using five minutes of annotated target data.
However, when applying the system to real in-the-wild radio broadcasts, our AWE-based system (requiring less than a minute of template audio) proves to be more robust, nearly matching  the performance of a 30-hour speech recognition model.

In the fifth and final contribution, we introduce three novel semantic AWE models. 
The goal here is that the resulting embeddings should not only be similar for words from the same type but also for words sharing contextual meaning, similar to how textual word embeddings are grouped together based on semantic relatedness.
For instance, spoken instances of ``football'' and ``soccer'', although acoustically different, should have similar acoustic embeddings.
We specifically propose leveraging a pre-trained multilingual AWE model to assist semantic modelling.
Our best approach involves clustering word segments using a multilingual AWE model, deriving soft pseudo-word labels from the cluster centroids, and then training a classifier model on the soft vectors.
In an intrinsic word similarity task measuring semantics, this multilingual transfer approach outperforms all previous semantic AWE methods.
We also show---for the first time---that AWEs can be used for downstream semantic query-by-example search.

\acuseall
\clearpage
\renewcommand{\contentsname}{Table of Contents}
\addcontentsline{toc}{chapter}{Table of contents}
\tableofcontents
\clearpage


\markboth{}{\scshape \nomname}
\setlength{\nomlabelwidth}{3cm}
\printnomenclature


\newpage


\newpage

\pagenumbering{arabic}
\acresetall

\mychapter{Introduction}{Introduction}
\label{chap:introduction}

Over the last few years, we have seen great strides in the advancement of automatic speech recognition systems.
Most state-of-the-art speech applications rely on neural networks with millions or even billions of parameters.
However, with the increase in network sizes, the amount of required training data also increases.

Developing speech systems requires large amounts of transcribed speech data.
For most languages, a sufficient amount of labelled speech data does not exist; some languages do not even have a written form \cite{besacier_automatic_2014}.
To put this into perspective, Google Assistant only supports twelve languages\footnote{\footnotesize{\url{https://support.google.com/googlenest/answer/7550584}}}; yet there are roughly 7\,000 languages spoken throughout the world \cite{eberhard_ethnologue_2021}.
Clearly, we are in need of methods that rely on less transcribed speech data in order to accommodate the majority of under-resourced languages. 

Most existing technologies rely on supervised training techniques.
Collecting labelled data for under-resourced languages is time-consuming and expensive to the extent that it is unfeasible in many cases.
Consequently, much work has been done on developing training strategies using speech data with no labels, known as the \emph{zero-resource} setting \cite{versteegh_zero_2016}.
In this setting, the goal is to develop methods that can discover linguistic structures and representations directly from unlabelled speech data \cite{jansen_summary_2013, versteegh_zero_2016}.
This research direction has strong links with studies on language acquisition in infants since infants also acquire their native language without explicit supervision \cite{rasanen_computational_2012}.

\mysection{Motivation}{Motivation}
\label{sec:methodology_and_goals}
Although full speech recognition is not possible in most zero-resource settings, researchers have proposed methods for applications such as speech search \cite{levin_segmental_2015, huang_improved_2018, yuan_learning_2018}, word discovery \cite{park_unsupervised_2008, jansen_efficient_2011, ondel_bayesian_2019, rasanen_unsupervised_2020}, and segmentation and clustering \cite{kamper_embedded_2017, seshadri_sylnet_2019, kreuk_self-supervised_2020}.
Many of these applications require speech segments of different durations to be compared. 
This is conventionally done using alignment, for example with dynamic time warping (DTW), but this is computationally expensive and can be inaccurate \cite{rabiner_considerations_1978}.

Acoustic word embeddings (AWEs) emerged as an alignment-free alternative for measuring the similarity between two speech segments \cite{levin_fixed-dimensional_2013}.
AWEs are fixed-dimensional vector representations of variable-length speech segments where instances of the same word type should have similar embeddings. 
Given an accurate AWE model, the similarity between two speech segments can easily be determined by simply calculating the distance between their embeddings.

Currently, the most successful AWE approaches employ deep neural networks.
For the zero-resource setting, many unsupervised AWE approaches have been explored, mainly relying on autoencoder-based neural models trained on unlabelled data in
the target language \cite{holzenberger_learning_2018, chung_unsupervised_2016, kamper_deep_2016, kamper_truly_2019}.
However, there still exists a large performance gap between these unsupervised models and their supervised counterparts \cite{levin_fixed-dimensional_2013, kamper_truly_2019}, where word labels and word boundaries are available.
A recent alternative for obtaining AWEs in a zero-resource language is to use multilingual transfer learning \cite{kamper_multilingual_2020, kamper_improved_2021, hu_acoustic_2021, hu_multilingual_2020}.
The goal is to have the benefits of supervised learning by training a model on labelled data from multiple well-resourced languages, but to then apply the model to an unseen target zero-resource language without fine-tuning it---a form of transductive transfer learning \cite{ruder_neural_2019}. This multilingual transfer approach has been shown to outperform unsupervised monolingual AWE models \cite{kamper_improved_2021}.

The universal embeddings resulting from multilingual transfer are of great interest, as
it allows fast and easy development of speech applications in a new language.
In addition to the direct use of these embeddings, for example in search applications, these representations could also potentially be used as an additional signal in other upstream tasks.

\mysection{Goals and methodology}{Goals and methodology}
\label{sec:g_and_m}

This thesis mainly focuses on recent advancements in AWE modelling following a multilingual transfer approach.
In the sections that follow we highlight our new contributions in AWE-base modelling and their application in practical speech systems.
But before we delve into the contributions, we first familiarise the reader with the fundamentals of AWE representations.

\subsection[Acoustic word embeddings]{Acoustic word embeddings}
\label{sec:intro_awe}

To start with, we want to make a distinction between textual word embeddings (TWEs) and AWEs.
TWEs are fixed-dimensional vector representations of {written} words.
The use of TWEs are ubiquitous in modern natural language processing applications and research, where the first use of TWEs dates back to 1986~\cite{rumelhart_learning_1986}.
State-of-the-art word embedding methods~\cite{mikolov_distributed_2013, pennington_glove_2014} exploit the co-occurrence statistics of words in a text corpora to learn word representations such that words with the same meaning have similar representations.
Consequently, for TWEs, words that appear in a similar context will appear close to each other in vector space.
Furthermore, every instance of the same word will be represented by only one embedding.

On the other hand, AWEs are fixed-dimensional vector representations of \emph{spoken} words.
AWEs are learned speech representations that capture the acoustic properties of spoken words.
In contrast to TWEs, since every instance of a spoken word will be different due to the continuous nature of speech, every unique realisation of a word also has a unique vector representation.
Ideally, instances of the same word type should appear close to each other in vector space and further away from instances of different word types.

\begin{figure}[!b]
	\begin{subfigure}[c]{.49\linewidth}
		\centering
		\includegraphics[width=0.99\linewidth]{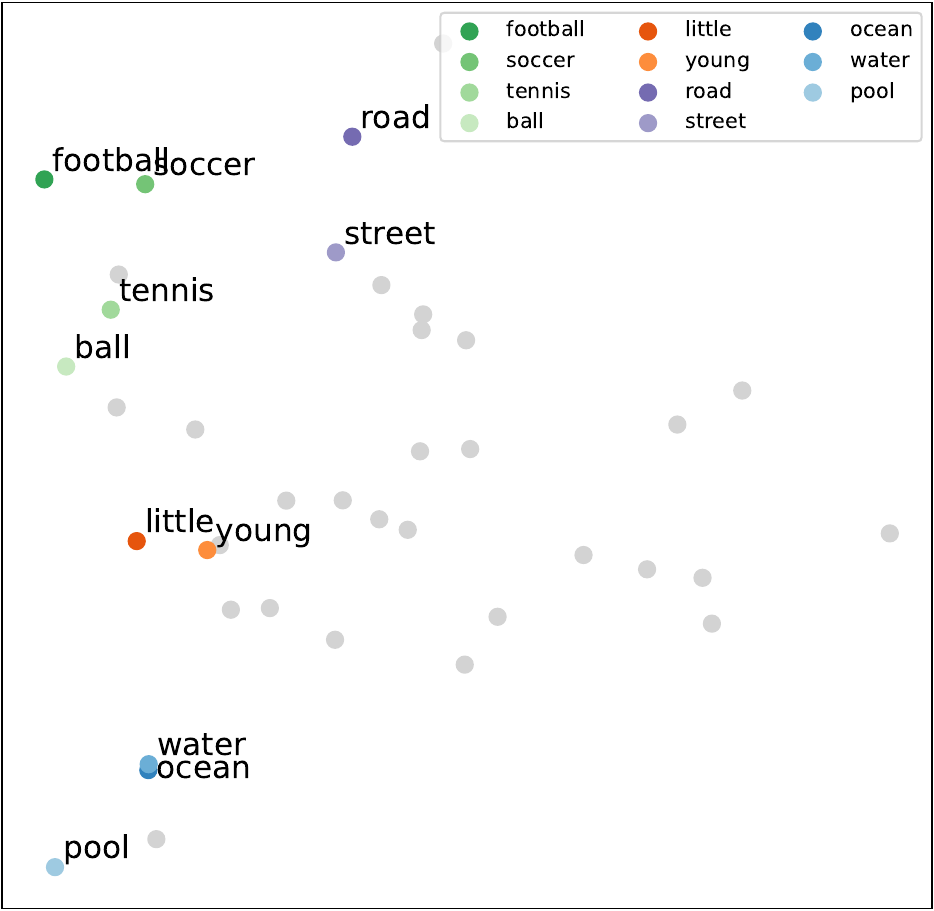}
		\caption{Textual word embeddings}
		\label{fig:intro_twe}
	\end{subfigure}
	\hfill
	\begin{subfigure}[c]{0.49\linewidth}
		\centering
		\includegraphics[width=0.99\linewidth]{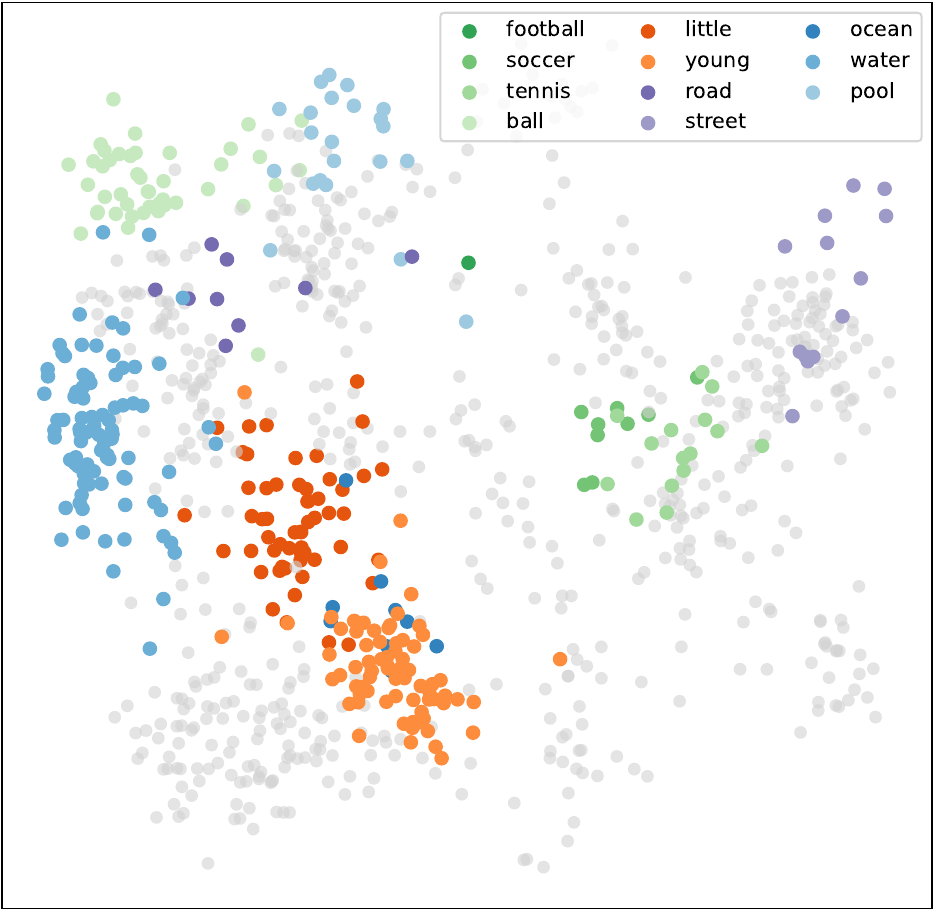}
		\caption{Acoustic word embeddings}
		\label{fig:intro_awe}
	\end{subfigure}
	\centering
	\caption{Illustrating the difference between (a) TWEs and (b) AWEs.
		In a TWE space, words sharing contextual meaning end up close to each other (different shades of the same colour).
		For AWEs, each spoken instance maps to a unique embedding (multiple of the same colour) with acoustically similar segments positioned close to each other.
	}
	
	\label{fig:intro_twe_awe}
\end{figure}

In Figure~\ref{fig:intro_twe_awe} we illustrate the fundamental differences between TWEs and AWEs.
In Figure~\ref{fig:intro_twe}, multiple embeddings are displayed in a two-dimensional vector space, obtained through training a TWE model and then projecting the embeddings to two dimensions using principal components analysis.
Note that each word type is assigned to a single embedding. 
Here we see embeddings of words that share contextual meaning are positioned close to each other.
For example, ``little'' and ``young'' (orange), and ``road'' and ``street'' (purple).

Figure~\ref{fig:intro_awe} shows the AWEs derived from the corresponding audio of the text data used for training the TWEs shown in Figure~\ref{fig:intro_twe}.
Here, spoken word segments are embedded such that each instance has a unique embedding that appears close to embeddings of other instances from the same word type.
For example, all the acoustic realisations of ``road'' (dark purple) and ``street'' (light purple) are close to each other. 
We also see how phonetic similarities are captured, for example, instances of ``ball'' and ``pool'' are also close, given that they end with the same phone.
It is clear that AWEs only store acoustic similarity and no information related to word meaning.
For example, instances of ``water'', ``pool'' and ``ocean'', although semantically similar (Figure~\ref{fig:intro_twe}), are spread out in the AWE space.


The main difference between AWEs and TWEs is summarised as follows: TWE models focus on learning embeddings that reflect the semantic relationship among words where instances of the same word are represented by the same embedding vector.
Conversely, AWE models focus on learning embeddings that encapsulate the acoustic properties of spoken words where instances of the same spoken word are represented by unique (but similar) embeddings. 
Throughout this thesis we only consider AWEs (in Section~\ref{sec:intro_aswe} we introduce a different form of AWEs that share similarities with TWEs, but they are also derived from spoken word segments without using word labels).



\subsection[Acoustic word embeddings in a zero-resource setting]{Acoustic word embeddings in a zero-resource setting}


In this section we present an overview of AWE modelling for zero-resource languages. This is necessary to understand the contributions of this thesis, which we start to present at the end of this section.

In the zero-resource setting, we face the challenge of developing AWE models for languages without labelled training data.
Two training strategies have been considered for AWE modelling in this scenario (more detail in Section~\ref{sec:background_settings}).
One approach is to use unlabelled audio data available in the target language~\cite{holzenberger_learning_2018, chung_unsupervised_2016, kamper_deep_2016, kamper_truly_2019}.
In the absence of labelled data, training word pairs are obtained from an unsupervised term discovery (UTD) system, which automatically finds recurring word-like patterns in an unlabelled speech collection.
Using discovered pairs from the target language enables the development of \emph{unsupervised monolingual} AWE models.

A recent alternative for obtaining embeddings on a zero-resource language is to use multilingual transfer learning~\cite{kamper_multilingual_2020, kamper_improved_2021, hu_acoustic_2021, hu_multilingual_2020}.
The idea is to train a \emph{supervised multilingual} AWE model jointly on a number of well-resourced languages for which labelled data is available, but to then apply the model to an unseen zero-resource language.
This multilingual transfer approach was found to outperform monolingual unsupervised learning approaches in~\cite{ruder_neural_2019, hu_multilingual_2020, kamper_improved_2021, hu_acoustic_2021, jacobs_acoustic_2021}

Various neural-based AWE models have been explored to learn latent representations from frame-level features.
The most effective ones leverage recurrent neural networks (RNNs)~\cite{chen_query-by-example_2015, chung_unsupervised_2016, settle_discriminative_2016, kamper_truly_2019, hu_multilingual_2020, kamper_improved_2021}.
We briefly introduce two existing AWE models which have been successfully applied to both the unsupervised monolingual and supervised multilingual training strategies (more detail in Chapter~\ref{chap:contrastive}): the \emph{correspondence autoencoder} (\system{CAE-RNN})~\cite{kamper_truly_2019} and \system{SiameseRNN}~\cite{settle_discriminative_2016}.
The \system{CAE-RNN} optimise an autoencoder-like reconstruction loss through an encoder-decoder RNN structure. 
Instead of reconstructing an input segment directly from the final encoder RNN hidden state, the \system{CAE-RNN} attempts to reconstruct another speech segment of the same type as the input.
Unlike the reconstruction loss used in the \system{CAE-RNN}, the \system{SiameseRNN} model explicitly optimises relative distances between embeddings~\cite{settle_discriminative_2016}.
The objective is to minimise the distance between the embedding of an input word and a word of the same type while at the same time maximising the distance between the input and an embedding from a different type~\cite{settle_discriminative_2016, kamper_improved_2021}.


Based on the overview provided, we identify three key elements to consider for potentially improving AWEs in a zero-resource setting: the learning objective of the AWE model, the training strategy taking into account the scarcity of labelled data, and the choice of languages used during multilingual training. We now concretely lay out our approach to addressing these aspects.


\textbf{AWE models and learning objectives.}
Recently, self-supervised contrastive learning has gained significant attention.
This approach involves using proxy tasks to automatically obtain target labels from the data~\cite{doersch_multi-task_2017, asano_critical_2020}.
Originally proposed for vision problems~\cite{doersch_unsupervised_2015, noroozi_unsupervised_2016, gidaris_unsupervised_2018}, it has since also been used as an effective pre-training step for supervised speech recognition~\cite{pascual_learning_2019, synnaeve_end--end_2020, baevski_vq-wav2vec_2020, baevski_effectiveness_2020, wang_unsupervised_2020}.
A number of loss functions have been introduced in the context of self-supervised learning which have not been considered for AWEs.
We specifically consider the contrastive loss of~\cite{chen_simple_2020, sohn_improved_2016} in a new AWE model which we call the \system{ContrastiveRNN}.
While a Siamese AWE model~\cite{kamper_deep_2016, settle_discriminative_2016} optimises the relative distance between one positive and one negative pair, our contrastive AWE model jointly embeds a number of speech segments and then attempts to select a positive item among several negative items.
\textbf{AWE training strategies.} 
While a monolingual AWE model is designed to capture language-specific nuances, a multilingual AWE model learns universal linguistic properties that are invariant across languages. 
We now ask whether unsupervised learning and multilingual transfer are complementary.
More specifically, can multilingual transfer further benefit from incorporating unsupervised learning? 
To answer this question we propose to adapt a multilingual AWE model to a target zero-resource language.
This involves fine-tuning a pre-trained multilingual model’s parameters using discovered word pairs.

\textbf{Language choice in multilingual transfer.}
Although there is a clear benefit in applying multilingual AWE models to an unseen zero-resource language, it is still unclear how the particular choice of training languages affects subsequent performance.
A careful selection of training languages, based on language family, proved successful in language identification~\cite{van_der_merwe_triplet_2020} and automatic speech recognition (ASR)~\cite{yi_language-adversarial_2019, van_der_westhuizen_multilingual_2021}, but this has not been considered in AWE modelling.
Preliminary experiments~\cite{kamper_improved_2021} show improved scores when training a monolingual model on one language and applying it to another from the same family.
But this has not been investigated systematically and there are still several unanswered questions:
Does the benefit of training on related languages diminish as we train on more languages (which might or might not come from the same family as the target zero-resource language)? When training exclusively on related languages, does performance suffer when adding an unrelated language? Should we prioritise data set size or language diversity when collecting data for multilingual AWE transfer? 
We address these questions by performing several experiments where we add data from different language families, and also control for the amount of data per language. 

Above we established three aspects this thesis considers aimed at improving the quality of AWEs for zero-resource language.
We now turn our attention to a practical downstream speech task that benefits from the application of these AWEs.

\subsection[Downstream application]{Downstream application}
\label{sec:intro_down}
For zero-resource languages, speech applications are developed without the need for labelled training data.
Among these applications is \emph{query-by-example} (QbE), a speech retrieval task.
In QbE, the goal is to use a spoken query to search an unlabelled audio corpus, aiming to retrieve utterances that contain instances of the query type.

DTW is typically used to match the speech features of a spoken query to the search utterances~\cite{park_unsupervised_2008, hazen_query-by-example_2009, zhang_unsupervised_2009, jansen_indexing_2012}.
However, DTW is slow and has some limitations~\cite{rabiner_considerations_1978, anastasopoulos_unsupervised_2016}.
AWEs have been proposed as an alternative for matching the query segment and search segments by jointly mapping them to the same vector space~\cite{levin_segmental_2015, settle_query-by-example_2017, yuan_learning_2018, hu_acoustic_2021, yuan_query-by-example_2019, ram_neural_2019}.
This allows fast comparison between the query segment and search segments.
Although AWE-based QbE systems have proven successful in controlled experiments, there has been limited work investigating the effectiveness of these systems beyond the experimental environment, where training and testing data come from different domains.

As far as we know, only the works of~\citet{saeb_very_2017} and~\citet{menon_radio-browsing_2017, menon_fast_2018, menon_feature_2019} consider speech retrieval performance on real in-the-wild data, applied to radio broadcast audio.
These approaches rely on labelled data from the target language.
Annotating audio (even in small quantities) is time-consuming and requires specialised linguistic knowledge which is not feasible when these systems need to be rapidly deployed.


In this thesis, we explore a real-world speech retrieval task of significant importance: hate speech detection in radio broadcasts~\cite{noauthor_united_nodate} through keyword spotting (KWS).
KWS typically relies on an ASR model trained on labelled data from the target language~\cite{larson_spoken_2012, mandal_recent_nodate}. 
To address this issue of label scarcity, we instead propose to extend QbE for KWS.
To perform KWS with QbE only requires a small number of spoken templates to serve as queries for the keywords of interest~\cite{menon_fast_2018, van_der_westhuizen_feature_2022}.


We approach this problem by first performing experiments in a controlled environment where training and test data come from the same domain. 
Then, we put our systems to a real-life test by evaluating performance on radio broadcast audio, specifically, Swahili audio (a low-resource language) scraped from radio stations in Kenya, a country in Sub-Saharan Africa.

\begin{figure}[!b]
	\centering
	\includegraphics[width=0.7\linewidth]{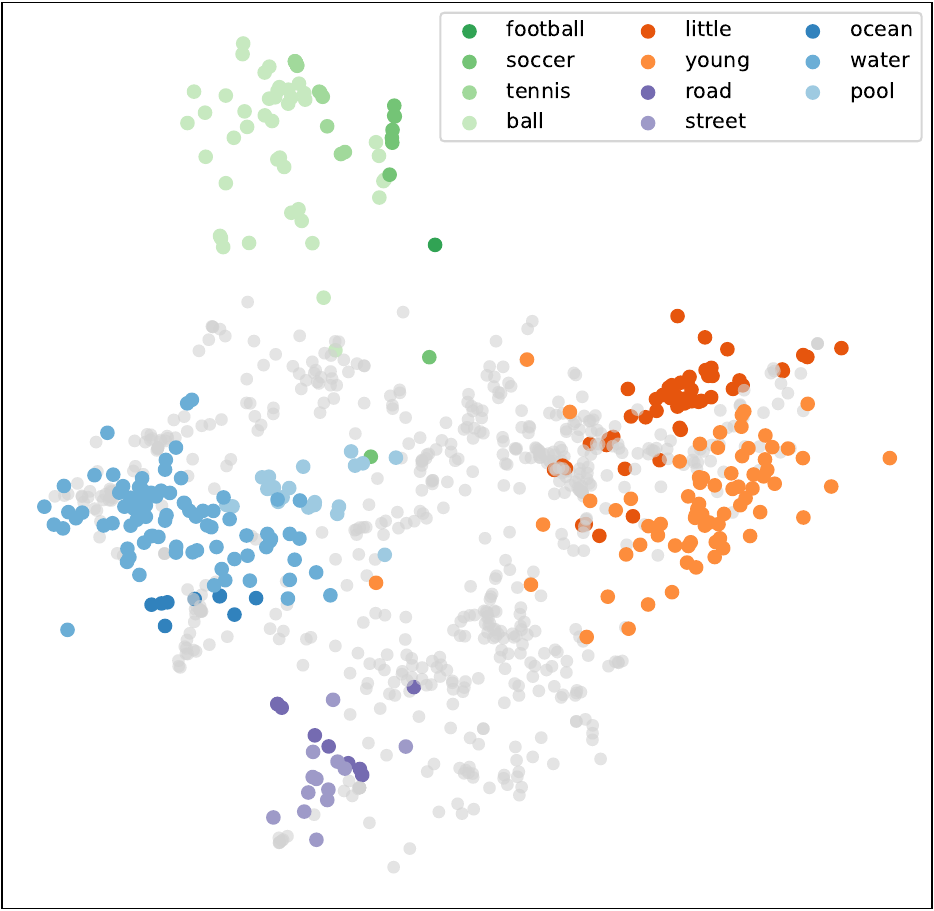}
	\caption{
	Semantic AWEs capture acoustic similarity among instances of the same word type (represented by the same colour) while also preserving word meaning (reflected in the shade of the same colour).
	}
	
	\label{fig:intro_sem}
\end{figure}

\subsection[Semantic acoustic word embeddings]{Semantic acoustic word embeddings}
\label{sec:intro_aswe}

In AWE modelling, the goal is to map variable-duration speech segments to fixed-dimensional vectors such that different acoustic realisations of the same word type have similar embeddings (Figure~\ref{fig:intro_awe}).
These embeddings have proven to be useful in applications requiring matching word segments from the same type (Section~\ref{sec:intro_down}). The TWEs described in Section~\ref{sec:intro_awe} have a distinct property where written words are mapped to similar representations if they share contextual meaning.

We now introduce a new form of representation learning called \emph{semantic} AWEs that should uphold both the properties of the AWEs and TWEs.
In semantic AWE modelling the goal would be to map speech segments to vector representations that not only capture whether two segments are instances of the same word, but also the semantic relationship between words.
Figure~\ref{fig:intro_sem} shows such a desired acoustic-semantic space.
Here we see all instances of the same type end up close to each other (as for the AWEs in Figure~\ref{fig:intro_awe}), for example ``road'' (dark purple) and ``street'' (light purple).
But now, the instances of ``road'' and ``street'' also end up close to each other (as the TWEs in Figure~\ref{fig:intro_twe}).

Only a limited number of studies have tried to address the problem of learning semantic AWEs from unlabelled speech alone~\cite{chung_speech2vec_2018, chen_phonetic-and-semantic_2019}.
These methodologies centre around training exclusively on unlabelled audio data from the target language.
In doing so, a model is challenged to distinguish between acoustic properties related to word type while simultaneously untangling word semantics.

To overcome this challenge, we propose leveraging the recent improvements in multilingual transfer in AWEs.
Specifically, we propose using a pre-trained multilingual AWE model as an additional signal to improve semantic AWEs in a target language where we only have unlabelled speech. 
Since the multilingual model already captures acoustic properties, this should simplify the semantic learning problem. 
In this thesis, we introduce three new semantic AWE models that incorporate multilingual AWEs.

\mysection{Contributions}{Contributions}
\label{sec:scope_and_contributions}



To take the above together, this thesis makes the following specific contributions to advancing zero-resource speech application via AWEs:

\begin{itemize}
	
	\item We propose a new AWE model, the \system{ContrastiveRNN}.
	The model architecture is similar to existing AWE models but optimises a different objective function, not previously considered for AWE modelling.
	We show this model outperforms existing AWE models in the unsupervised monolingual training strategy on six evaluation languages; the \system{ContrastiveRNN} performs on par with existing models in the multilingual transfer setting.
	
	\item To our knowledge, we present the first unsupervised adaptation of multilingual AWE models.
	Previous work performed adaptation of multilingual AWE models using speech segments with class labels~\cite{hu_multilingual_2020}.
	Instead of using true word segments, we use unknown word-like pairs.
	These are obtained from applying a UTD system---itself unsupervised---to an unlabelled speech corpus in the target language.
	The discovered word pairs are then used to fine-tune the multilingual model to the target language.
	We show that unsupervised adaptation is beneficial, with the new model we propose (\system{ContrastiveRNN}) giving the best performance after adaptation.
	
	\item To our knowledge, we are the first to perform extensive analysis on the choice of training languages in the multilingual transfer setting, specifically for AWEs.
	We experiment with different training languages and amounts of data on real low-resource languages. 
	We show the benefit of using languages that belong to the same language family as the target zero-resource language when training a multilingual AWE model. 
	
	\item 
		To our knowledge, we are the first to develop a keyword spotting system for hate speech detection in radio broadcasts using multilingual AWEs.
		We perform our main experiment using real broadcast in a low-resource language, Swahili, scraped from radio stations in Kenya.
		We show our AWE-based KWS system is more robust to a domain mismatch compared to existing ASR-based KWS systems, especially when large amounts of labelled data from the target language are unavailable.
		
%

	\item
	For the first time, we propose leveraging multilingual AWEs to learn representations of whole-word speech segments that capture the meaning of words (rather than acoustic properties) in a low-resource setting.
	Only a handful of studies have looked at this problem~\cite{chung_speech2vec_2018, chen_phonetic-and-semantic_2019}.
	We introduce three new semantic models, with one of them showing large improvements over previous approaches.
	We are also the first to apply these semantic acoustic embeddings in a downstream semantic speech retrieval task.

\end{itemize}

\subsection{Publications}
\label{ssec:intro_pubs}

The contributions above is summarised in the following publications.

\begin{tcolorbox}[width=\linewidth, colback=white!95!black, boxrule=0.5pt]
	\small
	\textit{Research paper 1: (Conference)} \\
	C. Jacobs, Y. Matusevych, and H. Kamper, ``Acoustic word embeddings for zero-resource languages using self-supervised contrastive learning and multilingual adaptation,'' \textit{IEEE Spoken Language Technology (SLT)}, 2021. \\
	
	\textit{Research paper 2: (Conference)} \\
	C. Jacobs and H. Kamper, ``Multilingual transfer of acoustic word embeddings improves when training on languages related to the target zero-resource language,'' \textit{Interspeech}, 2021. \\
	
	\textit{Research paper 3: (Conference)} \\
	C. Jacobs, N. C. Rakotonirina, E. A. Chimoto, B. A. Bassett, and H. Kamper, ``Towards hate speech detection in low-resource languages: Comparing ASR to acoustic word embeddings on Wolof and Swahili,'' \textit{Interspeech}, 2023. \\
	
	\textit{Research paper 4: (Journal)} \\
	C. Jacobs, and H. Kamper, ``Leveraging multilingual transfer for unsupervised semantic acoustic word embeddings,'' \textit{IEEE Signal Processing Letters}, 2023. \\
\end{tcolorbox}

%
%
%
%
%
%
%
%

\mysection{Thesis overview}{Thesis overview}
\label{sec:overview}

\ShortInTextTitle{Chapter~\ref{chap:background}: Background.}
The thesis starts by familiarising the reader with concepts to follow in subsequent chapters.
We describe various neural network models relevant to producing AWEs.
The processing of speech data is described.
A few zero-resource speech applications relevant to AWEs are introduced to the reader.
We provide a detailed description of AWE approaches in the zero-resource setting, including the training strategies we consider in this thesis.
Lastly, we discuss how we measure the quality of AWEs in both intrinsic and extrinsic evaluation tasks.
\\ 
\\
\ShortInTextTitle{Chapter~\ref{chap:contrastive}: {Contrastive learning for acoustic word embeddings.}}
This chapter presents a detailed description of existing AWE models that we reimplement as baselines for the rest of the thesis.
We then introduce a new embedding model we called the \system{ContrastiveRNN}.
We compare the \system{ContrastiveRNN} in both the unsupervised monolingual and supervised multilingual training strategies.
Following the unsupervised monolingual approach, the \system{ContrastiveRNN} outperforms the existing models on all six evaluation languages,  with an absolute increase in average precision ranging from 3.3\% to 17.8\%.
In the multilingual transfer setting, the \system{ContrastiveRNN} shows a marginal increase compared to existing models, mostly performing on par.
\\
\\
\ShortInTextTitle{Chapter~\ref{chap:adaptation}: {Multilingual adaptation.}}
This chapter introduces the unsupervised adaptation of multilingual AWE models, a new training strategy.
A multilingual AWE model is trained using labelled data from multiple-well-resourced languages; the model is then fine-tuned using unlabelled data from the target zero-resource language before applying it.
We perform unsupervised adaptation on all the AWE models considered in Chapter~\ref{chap:contrastive}.
The \system{ContrastiveRNN} model shows the highest performance increase after adaptation with roughly up to 5\% absolute increase in average precision on five out of six evaluation languages.
\\
\\
\ShortInTextTitle{Chapter~\ref{chap:related}: {Impact of language choice in a multilingual transfer setting.}}
This chapter investigates the impact the choice of training languages have in a transfer learning setting.
Here we assume a model trained on one set of languages will perform differently from a model trained on another set of languages.
We specifically consider the effect of training on languages that belong to the same language family as the target language.
We perform multiple experiments using data from South African languages.
We show the benefit of including related languages in multilingual modelling, even in small quantities.
This chapter also provides advice for practitioners who want to develop speech applications using AWEs.   
Here we also apply multilingual models in a downstream query-by-example speech search task.
\\
\\
{
\ShortInTextTitle{Chapter~\ref{chap:hatespeech}: {Hate speech detection using multilingual acoustic word embeddings.}}
This chapter addresses the problem of hate speech detection in low-resource languages through keyword spotting (KWS).
Our goal is to compare existing ASR-based KWS systems to a multilingual AWE-based approach on real radio broadcast audio.
We first develop and test our systems using experimental datasets: we use two low-resource languages, Swahili and Wolof.
In this in-domain setting, an ASR model fine-tuned using as little as five minutes of labelled data outperforms the AWE-based KWS system.
However, in a real-life scenario, when applying the systems to real radio broadcasts, the AWE system proves to be more robust by almost reaching the performance of an ASR model fine-tuned on 30 hours of data.
\\
\\
}
{
\ShortInTextTitle{Chapter~\ref{chap:semantic}: {Leveraging multilingual transfer for unsupervised semantic acoustic word embeddings.}}
This chapter introduces a novel approach to produce semantic AWEs.
Specifically, we propose using multilingual AWEs to assist in this learning task.
Our best semantic AWE approach involves clustering word segments using the multilingual AWE model, deriving soft pseudo-word labels from the cluster centroids, and then training a classifier model on the soft vectors.
In an intrinsic word similarity task measuring semantics, our multilingual transfer approach for semantic modelling outperforms all previous semantic AWE methods.
We also apply these semantic AWEs to a downstream semantic query-by-example search.
}
\\
\\
\ShortInTextTitle{Chapter~\ref{chap:conclusion}: {Summary and conclusion.}}
This chapter highlights the main findings of this thesis and provides recommendations for future work.


%


\mychapter{Background}{Background}
\label{chap:background}


{This chapter introduces concepts the reader needs to be familiar with to follow subsequent chapters. 
We give some background information on neural networks relevant to AWE modelling.
We describe the processing of speech data and different speech applications relevant to a zero-resource setting.
We then go on to describe different AWE models and how they are trained, and explain how we evaluate AWEs.
}

\mysection{Neural networks}{Neural networks}
\label{sec:background_networks}

In this section, we provide the reader with the necessary neural network fundamentals to follow subsequent chapters.

Feedforward neural networks (FFNNs) and convolutional neural networks (CNNs) are suitable for fixed-length inputs to create fixed-length outputs.
Recurrent neural networks (RNNs) are appropriate for handling variable-length sequences of inputs to produce fixed- or variable-length outputs.
Settle and Livescu~\cite{settle_discriminative_2016} showed that RNNs outperform FFNNs and CNNs in AWE modelling (more on this in Section~\ref{sec:background_models}); we therefore only consider AWE approaches built on RNNs.

We use the notation $g_\Phi(\mathbf{x})$ to represent a network where $\Phi$ are trainable weight parameters and $\mathbf{x}$ is an input to the network.
We use the notation $f_{\Theta}(\mathbf{x})$ to represent an entire model which can be composed of multiple network types for example:
\begin{equation}
	f_{\Theta_{Model}}(\mathbf{x}) = g_{\Phi_{FFNN}}(g_{\Phi_{RNN}}(\mathbf{x})),
\end{equation}
where $\Theta$ is the trainable parameters of all sub-networks.
In the following sections, we briefly explain the behaviour of RNNs and two network configurations used for AWE modelling.

\subsection[Recurrent neural network]{Recurrent neural network}
\label{ssec:background_rnn}
Recurrent neural networks (RNNs) are suitable for processing sequential data.
Unlike FFNNs and CNNs, RNNs contain feedback loops, allowing the network to predict an output using information from previous inputs and the current input~\cite{goodfellow_deep_2016}.

\begin{figure}[t]
	\centering
	\includegraphics[width=0.7\linewidth]{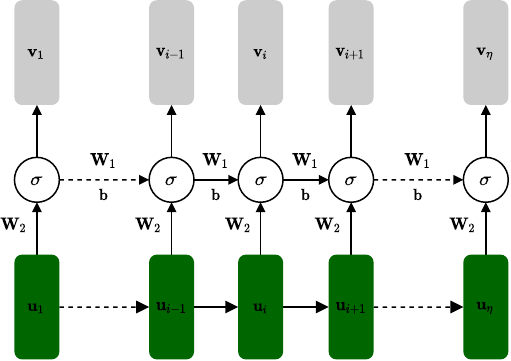}
	\caption{The structure of an RNN with one recurrent layer.}
	\label{fig:background_rnn}
\end{figure}

The network consists of one or more recurrent layers that share its parameters throughout.
Each recurrent layer has trainable weight parameters $\mathbf{W}_1 \in \mathbb{R}^{(N \times N)}$, $\mathbf{W}_2 \in \mathbb{R}^{(N \times M)}$ and $\mathbf{b} \in \mathbb{R}^{(N \times 1)}$. 
Given an input sequence $\mathbf{u}=(\mathbf{u}_1, \mathbf{u}_2, \ldots, \mathbf{u}_\eta)$, the recurrent layer produces an output sequence $\mathbf{v}=(\mathbf{v}_1, \mathbf{v}_2, \ldots, \mathbf{v}_\eta)$.
For a single input $\mathbf{u}_i \in \mathbb{R}^{M}$ the recurrent layer produces an output $\mathbf{v}_i \in \mathbb{R}^{N}$ using the following recursive function: 
\begin{equation}
	\mathbf{v}_i = \sigma(\mathbf{W}_2\mathbf{u}_i + \mathbf{W}_1\mathbf{v}_{i-1} + \mathbf{b}),
\end{equation}
where $\sigma$ is some activation function and $\mathbf{b}$ a bias parameter.
The network structure of an RNN is illustrated in Figure~\ref{fig:background_rnn}.
The weight parameter $\mathbf{W}_1$ is updated to preserve relevant information from the initial input $\mathbf{u}_1$ that is relevant to produce output $\mathbf{v}_i$ with current input $\mathbf{u}_i$.
Weight parameter $\mathbf{W}_2$ filters the relevant information from input $\mathbf{u}_i$ to produce $\mathbf{v}_i$.
These weight parameters control the state of the network over time depending on the inputs.

  
In practice RNNs do not learn long-term dependencies~\cite{bengio_learning_1994} due to exploding or vanishing gradients~\cite{shewalkar_performance_2019}.
A special type of recurrent unit, long short-term memory (LSTM), has been introduced to combat these difficulties~\cite{hochreiter_long_1997}.

\subsection[Autoencoder]{Autoencoder}
\label{ssec:background_ae}

An autoencoder (AE) neural network is a model that has a target output identical to the input~\cite{baldi_autoencoders_2012}.
The idea is that the model should learn a lower-dimensional representation of a given input vector, containing enough meaningful information, that would allow the model to reconstruct the input.
This model is useful for unsupervised learning since no knowledge regarding the input is needed.
The AE model consists of an encoder and decoder structure.
Given an input $\mathbf{x}$, the model's encoder $f_{\Phi_{\text{ENC}}}(.)$ processes the input to produce a latent embedding $\mathbf{z}$. 
The latent embedding is then given as input to the model's decoder $f_{\Phi_{\text{DEC}}}(.)$ to produce an output $\mathbf{\hat{y}}$, illustrated in Figure~\ref{fig:background_ae_cae}

The trainable parameters of the encoder and decoder together create the AE model $f_{\Theta_{AE}}(.)$.
With input $\mathbf{x}$, the model $f_{\Theta_{AE}}(.)$ is trained to minimise the reconstruction loss between the desired output $\mathbf{y}$ and the models output $\mathbf{\hat{y}}$:
\begin{figure}[t]
	\centering
	\includegraphics[width=0.7\linewidth]{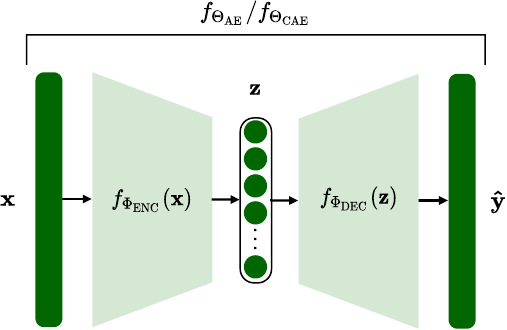}
	\caption{The AE and CAE  structure.}
	\label{fig:background_ae_cae}
\end{figure}
\begin{align}
	\begin{split}
	\ell(\mathbf{x}, \mathbf{y}) &= \lVert \mathbf{y} - f_{\Theta_{AE}}(\mathbf{x}) \rVert^2 \\
	&=
	\lVert \mathbf{y} - \mathbf{\hat{y}} \rVert^2
	\label{eqn:ae_general_loss}
	\end{split}
\end{align}
For the AE the target output $\mathbf{y}$ is equal to the input $\mathbf{x}$, the loss function can be written as: 
\begin{align}
	\begin{split}
	\ell(\mathbf{x}) &= \lVert \mathbf{x} - f_{\Theta_{AE}}(\mathbf{x}) \rVert^2 \\
	&=
	\lVert \mathbf{x} - \mathbf{\hat{y}} \rVert^2
	\end{split}
\end{align}

\subsection[Correspondence autoencoder]{Correspondence autoencoder}
\label{ssec:background_cae}

The correspondence autoencoder (CAE) network structure is identical to the AE as illustrated in Figure~\ref{fig:background_ae_cae}.
The only difference is that the target output of the CAE is not exactly the input.
Instead, given an input the CAE tries to produce an output that belongs to the same class as the input~\cite{kamper_unsupervised_2015}.

The CAE loss is given by setting the target output $\mathbf{y}$ in Equation~\ref{eqn:ae_general_loss} to $\mathbf{x_{\text{p}}}$:
\begin{align}
	\begin{split}
	\ell(\mathbf{x}, \mathbf{x}_{\text{p}}) &= \lVert \mathbf{x}_{\text{p}} - f_{\Theta_{CAE}}(\mathbf{x}) \rVert^2 \\
	&=
	\lVert \mathbf{x}_{\text{p}} - \mathbf{\hat{y}} \rVert^2,
	\label{eqn:cae_loss}
	\end{split}
\end{align}
where $\mathbf{x}_p$ represents an instance from the the same class as input $\mathbf{x}$. 

\subsection[Siamese neural network]{Siamese neural network}
\label{ssec:background_siamese}
\begin{figure}[!t]
	\centering
	{\includegraphics[width=0.6\linewidth]{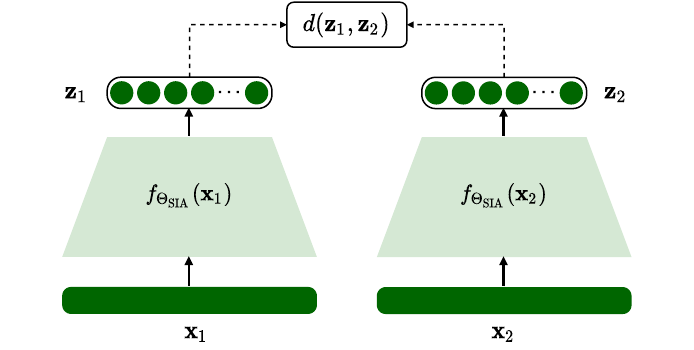}}
	\caption{
		The Siamese network structure.
	}
	\label{fig:background_siamese}
\end{figure}

Instead of reconstructing an input representation, the Siamese network measures the distance between representations of two inputs~\cite{bromley_signature_1993}.
The network consists of two identical sub-networks with tied parameters.
The model is trained to optimise the relative distance between representations of two data instances by mirroring parameter updates across both sub-networks.
Formally, the model $f_{\Theta_{SIA}}(.)$ produces representations $\mathbf{z}_1$ and $\mathbf{z}_2$ from inputs $\mathbf{x}_1$ and $\mathbf{x}_2$ respectively, illustrated in Figure~\ref{fig:background_siamese}.
Ideally, embeddings $\mathbf{z}_1$ and $\mathbf{z}_2$ from same class instances $\mathbf{x}_1$ an $\mathbf{x}_2$ should be similar, and embeddings $\mathbf{z}_1$ and $\mathbf{z}_2$ from different class instances $\mathbf{x}_1$ an $\mathbf{x}_2$ should be unalike.


\begin{figure}[!t]	
	\begin{minipage}[a]{0.45\linewidth}
		\centering
		\centerline{\includegraphics[width=0.95\linewidth]{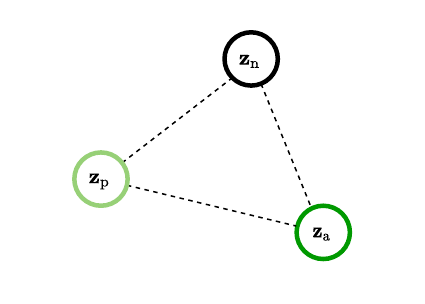}}
		\centerline{(a) Before training.}
	\end{minipage}
	\hfill
	\begin{minipage}[a]{0.45\linewidth}
		\centering
		\centerline{\includegraphics[width=0.95\linewidth]{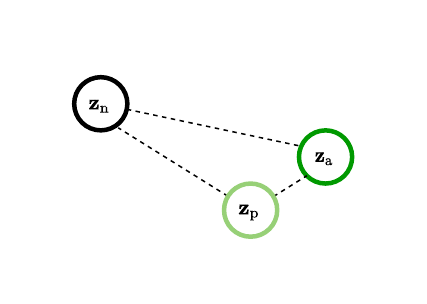}}
		\centerline{(b) After training.}
	\end{minipage}
	\vspace*{1.5mm}
	\caption{Visualisation of the relative distance between embeddings (a) before training and (b) after training, showing how representations from the same class $(\mathbf{z}_{\text{a}}, \mathbf{z}_{\text{p}})$ are pulled together and pushed away from a different class $\mathbf{z}_{\text{n}}$ by optimising the triplet loss.
	} 
	\label{fig:background_triplet}
\end{figure}
In other words, embeddings from the same class should be close to each other in vector space and far away from embeddings from other classes.
By introducing negative samples during training we can better accomplish this goal~\cite{wang_learning_2014, chechik_large_2010, hermans_defense_2017, hoffer_deep_2015}.
For a single training instance the model needs to minimise the distance between an input $\mathbf{x}_{\text{a}}$ and another instance from the same class $\mathbf{x}_{\text{p}}$ while simultaneously maximising the distance between the input $\mathbf{x}_{\text{a}}$ and an instance from a different class $\mathbf{x}_{\text{n}}$ (subscripts refer to anchor, positive and negative) as illustrated in Figure~\ref{fig:background_triplet}.
To optimise the relative distance, the Siamese network optimises a triplet hinge-like loss:
\begin{equation}
	l_{\text{triplet}}(\mathbf{z}_{\text{a}}, \mathbf{z}_{\text{p}}, \mathbf{z}_{\text{n}}) = \text{max}\{0, m + d(\mathbf{z}_{\text{a}}, \mathbf{z}_{\text{p}}) - d(\mathbf{z}_{\text{a}}, \mathbf{z}_{\text{n}})\},
	\label{eqn:background_triplet}
\end{equation}
where 
\begin{equation}
	d(\mathbf{z}_1, \mathbf{z}_2) = 1 - \frac{\mathbf{z}_1 \cdot \mathbf{z}_2}{\lVert \mathbf{z}_1 \rVert \lVert \mathbf{z}_2 \rVert}  \medskip
\end{equation}

is the cosine distance between embeddings $\mathbf{z}_1$ and $\mathbf{z}_2$ extracted from  inputs $\mathbf{x}_1$ and $\mathbf{x}_2$ with $m$ a margin parameter.
The triplet hinge-like loss is at a minimum when all embedding pairs $(\mathbf{z}_{\text{a}}, \mathbf{z}_{\text{p}})$ of the same type are a distance $m$ closer than embedding pairs $(\mathbf{z}_{\text{a}}, \mathbf{z}_{\text{n}})$ of different types.
Various studies showed the benefit of optimising the relative distances instead of only minimising the absolute distance between data instances from the same class~\cite{wang_learning_2014, chechik_large_2010}. 

\mysection{Speech processing}{Speech processing}
\label{sec:background_speech_processing}


We now describe the feature representation of speech segments that are presented to the neural networks (described above in Section~\ref{sec:background_networks}) as inputs and outputs.
We then go on to describe speech tasks relevant to AWEs in a zero-resource setting.

\subsection[Speech features]{Speech features}
\label{ssec:background_features}

Throughout this thesis we use mel-frequency cepstrum coefficients (MFCCs) to make a reliable comparative study between AWE models considered in previous work.  
All speech segments are parametrised as 13-dimensional MFCCs using a window size of 25 ms and a 10 ms frame shift.
Additionally, speaker normalisation is performed per utterance.

All AWE models are trained on MFCC features from isolated word segments.
In the supervised setting, word boundaries are obtained from forced alignments.
In the unsupervised setting, we use an unsupervised term discovery system (Section~\ref{ssec:background_utd}) to obtain unknown word-like segments from unlabelled data in the target language.  

For dynamic time warping (described below in Section~\ref{ssec:background_dtw}) baseline experiments, delta and double-delta MFCCs are included.

{Throughout the course of this thesis, significant advancements have been made in the field of self-supervised speech representation learning (SSRL) using transformer networks~\cite{baevski_wav2vec_2020, conneau_unsupervised_2021, hsu_hubert_2021, chen_wavlm_2022}.
The primary objective of SSRL is to train a network using unlabelled speech data to generate frame-level representations that offer more contextual information compared to traditional rule-based MFCCs.
In our experiments, we use the Wav2Vec2.0 XLSR model, which has been trained on unlabelled audio from multiple languages~\cite{conneau_unsupervised_2021}.
Using the XLSR model, we extract speech features from 25 ms of speech with a frame shift of 20 ms.
We use the output of the 12th transformer layer producing input features of 1024 dimensions.
From Chapter~\ref{chap:hatespeech} we include these features in our experiments.}

\subsection[Dynamic time warping]{Dynamic time warping}
\label{ssec:background_dtw}

\begin{figure}[!t]
	\centering
	\begin{minipage}[a]{0.45\linewidth}
		\centering
		\centerline{\includegraphics[width=0.7\linewidth]{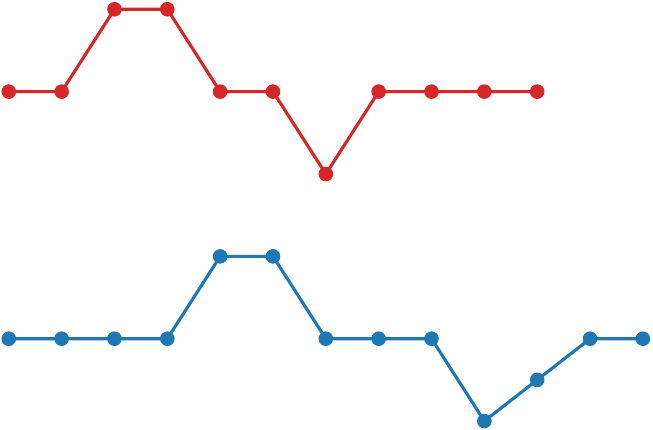}}
		\centerline{(a)}\medskip
	\end{minipage}
	\begin{minipage}{0.45\linewidth}
		\centering
		\includegraphics[width=0.7\linewidth]{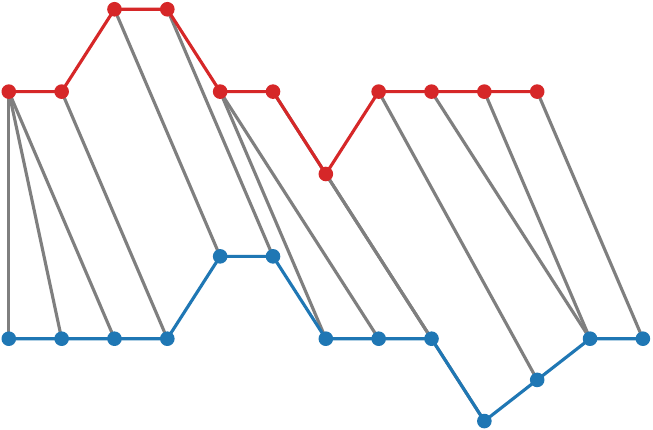}
		\centerline{(b)}\medskip
	\end{minipage}
	
	\caption{Illustration of the mapping between two signals of different durations that yield the lowest alignment cost using DTW.}
	\label{fig:background_dtw}
\end{figure}

Dynamic time warping (DTW) is used to calculate the dissimilarity between two temporal sequences.
This method is closely related to the \textit{edit distance} used in natural language processing.
Edit distance is a method used to quantify the difference between two strings by identifying the minimum number of operations required to transform one of the strings into the other.\footnote{\footnotesize{\url{https://en.wikipedia.org/wiki/Edit_distance}}}
Assume there are three possible operations to transform one string into the other: insertion, deletion and substitution.
Now consider the following two words, ``time'' and ``camel''. To transform ``time'' into ``camel'' we need to substitute the `t' with a `c', substitute the `i' with an `a' and insert a `l' at the end. 
This transformation requires three operations, therefore the edit distance between the two words is three.

DTW is used to calculate the dissimilarity between two temporal sequences similar in the way edit distance is used to calculate the dissimilarity between strings.
DTW is an algorithm that finds the optimal alignment between two time series through a dynamic programming method.
Consider the example signals in Figure~\ref{fig:background_dtw}~(a).
DTW can be used to calculate how similar the blue signal is to the red signal by calculating the distance between each point from the blue signal to every point in the red signal.
This results in a matrix of distances known as a cost matrix from which we can identify a particular mapping of points from the blue signal to the red signal that results in the lowest alignment cost.
The mapping of points from the blue signal to points in the red signal that yield the lowest alignment cost is displayed in Figure~\ref{fig:background_dtw}~(b).

\begin{figure}[!t]
	\centering\	
	\includegraphics[width=0.30\linewidth]{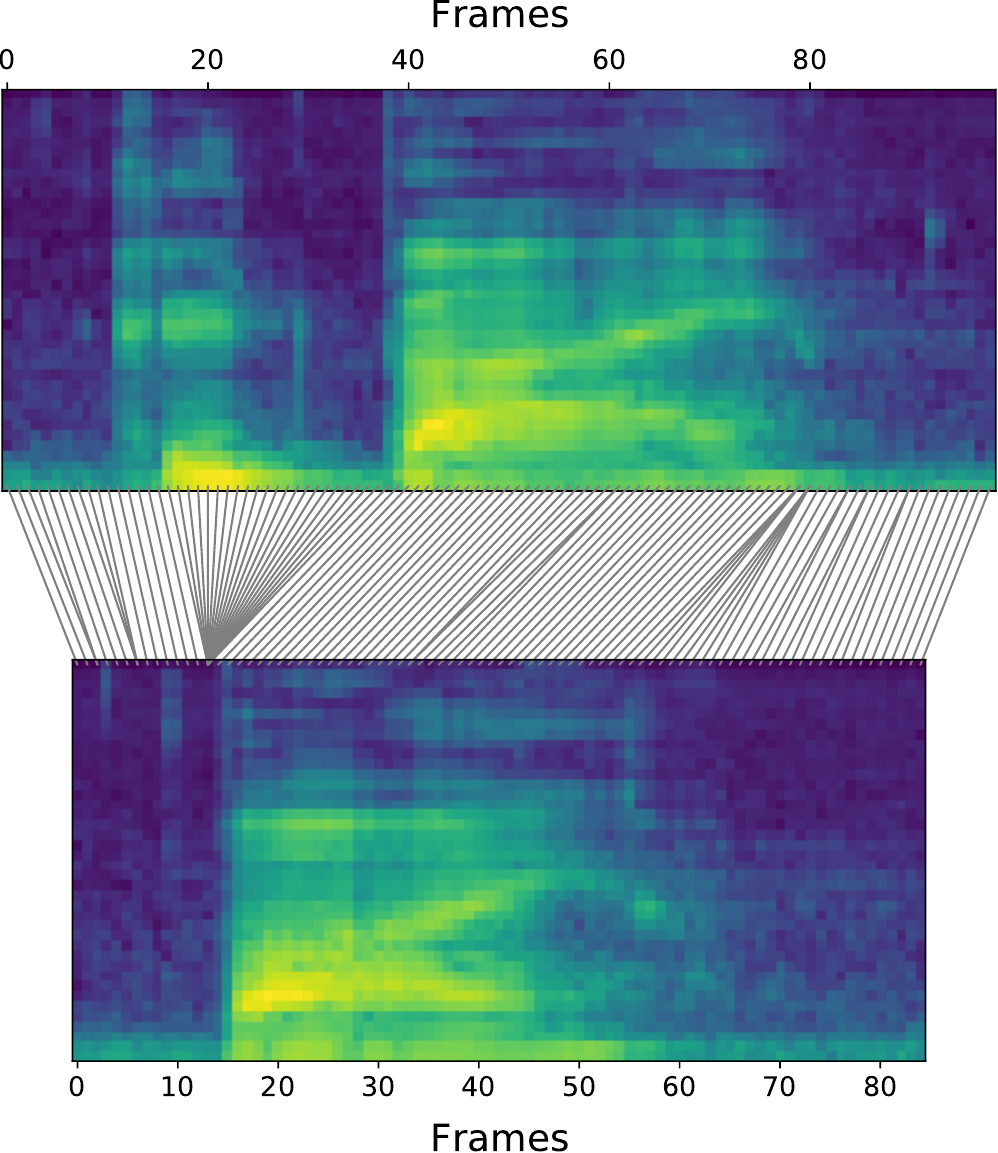}
	\caption{Illustration of the optimal alignment of MFCCs extracted from two speech segments using DTW.}
	\label{fig:background_dtw_mfcc}
\end{figure}

For speech signals, DTW can be applied to a set of extracted speech features for example MFCCs.
Similarity between two ``points'' is calculated as some distance between the MFCC vectors such as the cosine distance.
Figure~\ref{fig:background_dtw_mfcc} shows an example of the mapping of one speech segment to another using extracted MFCC frames.

DTW is found in most state-of-the-art unsupervised term discovery (Section~\ref{ssec:background_utd}) and query-by-example (Section~\ref{ssec:background_qbe}) systems as described below.
AWEs were introduced as an alignment-free method to  measure the similarity between word segments operating at linear time complexity compared to the polynomial time complexity of DTW.
We use DTW as a baseline to compare AWEs in a word discrimination task (Section~\ref{sec:background_evaluation}) and in a query-by-example speech search (Section~\ref{ssec:background_qbe}) task.

\subsection[Query-by-example speech search]{Query-by-example speech search}
\label{ssec:background_qbe}
Query-by-example (QbE) speech search is the task of retrieving utterances from a speech collection related to a given a spoken query.
Early QbE systems implemented with large vocabulary continuous speech recognition (LVCR) systems proved to be successful~\cite{miller_rapid_2007, saraclar_lattice-based_2004}.
However, training an LVCR requires large amounts of labelled data and has limitations to handling out-of-vocabulary 
words.
Studies focussing on implementing systems following a phonetic approach alleviate these impediments~\cite{yu_vocabulary-independent_2005, ng_subword-based_2000}.
These systems still rely on language-specific knowledge and labelled data which is impractical for the zero-resource setting.

Recently, several studies focused on developing QbE speech search systems, specifically for the zero-resource setting~\cite{rajput_spoken_2011, barnard_spoken_2012, metze_spoken_2013, anguera_query_2014, szoke_query_2015}.
In the zero-resource setting, a typical approach is to learn efficient frame-level feature representations for both the query and all possible segments from the speech collection then perform DTW to find matching representations~\cite{hazen_query-by-example_2009, zhang_unsupervised_2009}.
However, DTW is computationally expensive, restricting QbE on large-scale speech collections.
Several studies considered less expensive DTW QbE alignment methods~\cite{mantena_speed_2013, zhang_piecewise_2011}.
Moreover, Jansen van Durme~\cite{jansen_indexing_2012} introduced a QbE system where raw speech frames QbE hashed to bit signatures followed by applying a QbE nearest neighbour similarity search algorithm, reducing the computational time to logarithmic time.

Levin et al.~\cite{levin_segmental_2015} introduced AWEs to replace the frame-based representations used to perform QbE in~\cite{jansen_indexing_2012}, to alternatively use whole-word segments representations, using Laplacian eigenmaps.
This approach showed a reduction in computation time and improved accuracies given that the AWEs have better lexical discrimination (more in Section~\ref{sec:background_models}).

Settle et al.~\cite{settle_query-by-example_2017} replaced the Laplacian eigenmaps with deep neural network AWEs, specifically, they implement a \system{SiamseseRNN} (Section~\ref{ssec:background_siamese}) AWE model  to perform QbE search which showed significant improvement.
Recently, several studies proposed performing QbE using AWEs with a neural approach~\cite{yuan_learning_2018, yuan_query-by-example_2019, hu_acoustic_2021}.

In Chapter~\ref{chap:related} we discuss the implementation of a simplified QbE approach used by Kamper et al.~\cite{kamper_semantic_2019}  for measuring AWE performance in a downstream application.

\subsection[Unsupervised term discovery]{Unsupervised term discovery}
\label{ssec:background_utd}

{The goal of QbE is to retrieve one or more utterances within an unlabelled speech collection given a spoken query. Unsupervised term discovery (UTD) instead aims to extract and group unknown word-like segments from an unlabelled speech collection (i.e. no query is given).}
{Throughout this thesis, we use UTD to find word-like speech segments from an unlabelled speech collection that allows us to train AWE models in a fully unsupervised way.}

Park and Glass~\cite{park_unsupervised_2008} were the first to introduce a UTD system.
The UTD system aims to identify repeating subsequences within a speech signal using only the signal itself. 
The system of Park and Glass tries to find recurring speech patterns by applying a segmental DTW algorithm to audio data.
The segmental DTW algorithm tries to find subsequence alignments between the acoustic features of pairs of continuous spoken utterances. 

Jansen and van Durme~\cite{jansen_efficient_2011} presented a UTD system that reduces the computation complexity from quadratic in~\cite{park_unsupervised_2008} to linearithmic.
They did this by applying a random projection algorithm that maps sequences in the acoustic feature space to fixed-length bit signatures, allowing efficient similarity calculation through an approximate nearest neighbour search algorithm.

This task is of great interest as it allows matching speech segments predicted to be of the same word type from untranscribed speech, which is useful in the zero-resource setting.
Many unsupervised AWEs have been trained using word pairs obtained from UTD systems~\cite{kamper_truly_2019, kamper_improved_2021, jacobs_acoustic_2021}.

%

\mysection{Extracting acoustic word embeddings}{Extracting acoustic word embeddings}
\label{sec:background_models}

We now explore previous work addressing AWEs, focussing on zero-resource languages for which transcribed audio data is not available. 

Levin et al.~\cite{levin_fixed-dimensional_2013} were the first to propose representing whole word speech segments as fixed-dimensional vectors that contain linguistic meaning.
They explored various embedding techniques.
For each, different levels of information are assumed to be available.
In a setting where the only available data is unknown isolated speech segments (word boundaries are available), they perform downsampling on extracted speech features to create fixed-dimensional representations.
They considered uniform and non-uniform downsampling (using hidden Markov models), both methods yielding fixed-dimensional embeddings.
Using downsampling as a means to create AWEs was also used as a baseline for comparing more sophisticated approaches considered in ~\cite{kamper_truly_2019, holzenberger_learning_2018}.

Another approach Levin et al.~\cite{levin_fixed-dimensional_2013} presented involves reference vectors and Laplacian eigenmaps. Assuming a reference set of distinct unknown speech segments is available, the DTW cost between a speech segment and each reference segment is concatenated to create a reference vector. The reference set should contain enough segments to form a basis for all possible speech segments, consequently, producing very high-dimensional reference vectors.
They apply a linear dimensionality reduction technique, principal component analysis, to reduce the dimensionality of the reference vectors, showing a minimal decrease in performance.
They also proposed an AWE approach using a non-linear graph embedding technique.
They evaluated all their methods on a word discrimination task where only the Laplacian eigenmaps approach reached the performance of DTW.
Although, some of these unsupervised approaches were successfully implemented in ~\cite{settle_query-by-example_2017, kamper_embedded_2017, kamper_segmental_2017, kamper_unsupervised_2016} they are still far from reaching the performance of the supervised approaches (also considered by Levin et al.) where word labels are available.

Many studies followed, aiming to improve AWEs in the zero-resource setting, mostly using deep neural networks.
Kamper et al.~\cite{kamper_deep_2016} follow a CNN-based approach that makes use of weak supervision.
Here speech segments are grouped into unknown word types using available word class labels. 
They use these word pairs to train an AWE model called the \system{SiameseCNN} using a hinge-like contrastive loss function, where the model tries to minimise the distance between the unknown word pairs while maximising the distance between words from different unknown types.
This loss is described in Section~\ref{sec:background_networks}.
They show that the \system{SiameseCNN} outperforms the Laplacian eigenmaps approach of Levin et al.~\cite{levin_fixed-dimensional_2013} and, more importantly, the \system{SiameseCNN}, trained with weak supervision, performs on par with a supervised classification model (\system{ClassifierCNN}) where word labels are available.

Chung et al.~\cite{chung_unsupervised_2016} were the first to apply an RNN to produce fixed-dimensional representations from variable-duration speech segments without any supervision.
They present a sequence-to-sequence autoencoder AWE model. 
An encoder RNN processes an input sequence
, a decoder RNN then tries to reconstruct the input sequence taking the final layer of the encoder RNN as input.
This reconstruction loss is described in Section~\ref{sec:background_networks}.
After training the model, the final hidden layer of the encoder RNN is taken as the embedding of an input sequence.
They did not explicitly compare embedding quality to previous methods, however, they show applying the autoencoder AWE outperforms DTW in a QbE speech search task.

Settle and Livescu~\cite{settle_discriminative_2016} reimplemented the CNN-based models of Kamper et al.~\cite{kamper_deep_2016} using RNNs. 
In a similar experimental setup as Kamper et al.~\cite{kamper_deep_2016} they show that the Siamese network build on RNNs (\system{{SiameseRNN}}) outperform the \system{SiameseCNN} of \citet{kamper_deep_2016}.

Finally, Kamper~\cite{kamper_truly_2019} propose a method for embeddings in a truly zero-resource setting.
In this unsupervised setting, no labelled speech data or word boundaries are available.
A UTD (Section~\ref{ssec:background_utd}) system---itself unsupervised---is used to find word-like pairs predicted to be of the same unknown type.   
These pairs are presented to an autoencoder-like network where the target to reconstruct is not identical to the input (like the autoencoder from Chung et al.~\cite{chung_unsupervised_2016}) but rather an instance predicted to be of the same type.
This model, the correspondence autoencoder RNN (\system{CAE-RNN}), outperforms all previous unsupervised AWE approaches and delivers similar results as DTW executing at a much lower run-time.

Although these unsupervised methods are useful for the zero-resource setting, there still exists a large performance gap between the supervised AWE methods~\cite{levin_fixed-dimensional_2013, kamper_truly_2019}.
Recently, Kamper et al.~\cite{kamper_improved_2021} exploits labelled data from multiple well-resourced languages to train a single supervised multilingual AWE model and then apply the model to an unseen zero-resource language.
In this transfer learning setting, they train multilingual versions for three existing models, \system{ClassifierRNN}~\cite{settle_discriminative_2016}, \system{SiameseRNN}~\cite{settle_discriminative_2016} and \system{CAE-RNN}~\cite{kamper_truly_2019}.
They compare these multilingual models to their monolingual unsupervised counterparts, where the multilingual models outperform all of them.
Others have since also considered a multilingual approach to AWE modelling~\cite{hu_multilingual_2020, hu_acoustic_2021, van_staden_improving_2021}.
The multilingual transfer approach is currently the best approach.
However, the performance compared to the supervised models still lack by a large margin.

In the following section we summarise some of the training strategies identified in this section.

\mysection{Unsupervised acoustic word embedding training strategies}{{Unsupervised acoustic word embedding training strategies}}
\label{sec:background_settings}

In this section we explicitly mention the training strategies used for unsupervised AWE modelling that we use in subsequent chapters.
Firstly, we summarise two training strategies identified in the preceding section.
One option is to train unsupervised monolingual models directly on unlabelled data (Section~\ref{ssec:monolingual}), introduced by \citet{kamper_truly_2019}.
Another option is to train a supervised multilingual model on labelled data from well-resourced languages and then apply the model to a zero-resource language (Section~\ref{ssec:multilingual}), introduced by \citet{kamper_multilingual_2020}.
In Section~\ref{ssec:multilingual_adapt} we describe a new AWE training strategy, which we include here in the background chapter for the sake of completeness---we really propose this new setting only in Chapter~\ref{chap:adaptation}.
In this strategy, multilingual models are fine-tuned to a zero-resource language using unsupervised adaptation.
All the AWE models considered throughout this thesis can be used in all three settings, as explained below.

\subsection{Unsupervised monolingual models}
\label{ssec:monolingual} 
For any of the AWE models in Chapter~\ref{chap:contrastive}, we need pairs of segments containing words of the same type; for the \system{SiameseRNN} and \system{ContrastiveRNN} we additionally need negative examples.
In a zero-resource setting there is no transcribed speech to construct such pairs.  
But pairs can be obtained automatically 
by applying a UTD (Section~\ref{ssec:background_utd}) system 
to an unlabelled speech collection from the target zero-resource language. 
This system discovers pairs of word-like segments, predicted to be of the same unknown type. 

The discovered pairs can be used to sample positive and negative examples for any of the three models in Chapter~\ref{chap:contrastive}.
Since the UTD system has no prior knowledge of the language or word boundaries within the unlabelled speech data, the entire process can be considered unsupervised.
Using this methodology, we consider purely unsupervised monolingual versions of each AWE model in Chapter~\ref{chap:contrastive}.

\subsection{Supervised multilingual models}
\label{ssec:multilingual}

Instead of relying on discovered words from the target zero-resource language, we can exploit labelled data from well-resourced languages to train a single multilingual supervised AWE model~\cite{kamper_improved_2021, hu_multilingual_2020}.
This model can then be applied to an unseen zero-resource language.
Since a supervised model is trained for one task and applied to another, this can be seen as a form of \textit{transfer learning}~\cite{pan_survey_2009, ruder_neural_2019}

Experiments in~\cite{kamper_improved_2021} showed that multilingual versions of the \system{CAE-RNN} and \system{SiameseRNN} outperform unsupervised monolingual variants.
A multilingual \system{ContrastiveRNN} hasn't been considered in a previous study, as far as we know.
We consider supervised multilingual variants of all three models in Chapter~\ref{chap:contrastive}.

\subsection{Unsupervised adaptation of multilingual models}
\label{ssec:multilingual_adapt}

While previous studies have found that multilingual AWE models (Section~\ref{ssec:multilingual}) are superior to unsupervised AWE models (Section~\ref{ssec:monolingual}), one question is whether multilingual models could be tailored to a particular zero-resource language in an unsupervised way.

We propose to adapt a multilingual AWE model to a target zero-resource language by fine-tuning the multilingual model's parameters using discovered pairs from the target zero-resource language.   
These discovered segments are obtained by applying a UTD system to unlabelled data from the target zero-resource language.
The idea is that adapting the multilingual AWE model to the target language would allow the model to learn aspects unique to that language.
We consider the adaptation of multilingual versions of all three AWE models in Chapter~\ref{chap:adaptation}.

As far as we know, we are the first to perform unsupervised adaptation of multilingual AWE models for the zero-resource setting.

\mysection{Evaluation of acoustic word embeddings}{Evaluation of acoustic word embeddings}
\label{sec:background_evaluation}

Ultimately, we want to evaluate the performance of AWEs in downstream speech applications.
However, we would like to measure the quality of the embeddings without being tied to a specific system architecture for easy and fast comparison between different embedding approaches.
We therefore use a word discrimination task that measures the intrinsic quality of the embeddings.
The \textit{same-different} task, introduced by Carlin et al.~\cite{carlin_rapid_2011}, is specifically designed to quantify the separability of same-word and different-word representations across a range of word types and speakers. 

In the same-different task, we are given a pair of acoustic segments, each a true word, and we must decide whether the segments are examples of the same or different words.
To evaluate a particular embedding method, a set of isolated test words are first embedded.
Two words can then be classified as being of the same or different type based on some distance threshold, and a precision-recall curve is obtained by varying the threshold.
The area under this curve is used as the final evaluation metric, referred to as the average precision (AP).
\citet{carlin_rapid_2011} show that this AP correlates with the phone recognition accuracy in a spoken term discovery system, indicating its potential for predicting performance in downstream tasks.

In our implementation, we use a cosine distance to measure the similarity between two vectors that Levin et al.~\cite{levin_fixed-dimensional_2013} found to generally perform better than Euclidean distance. 
Concretely, let's consider two test word embeddings, $\mathbf{z}_i$ and $\mathbf{z}_j$.
Two words are declared to be the same if their cosine distance is less or equal to some distance threshold $\tau$, as follows:
\begin{equation}
	1 - \frac{\textbf{z}_i \cdot \textbf{z}_j}{\lVert \textbf{z}_i \rVert \lVert \textbf{z}_j \rVert} \le \tau
\end{equation}
%

Moreover, we are particularly interested in obtaining embeddings that are speaker-invariant. 
We therefore calculate AP by only taking the recall over instances of the same word spoken by different speakers.
In a test set of all possible word pairs, there exist a subset $C_{SWDP}$ containing pairs of words that are the same but said by different speakers (SWDP).
If the number of word pairs declared to be the same based on a threshold $\tau$ from this subset is $N_{SWDP}(\tau)$, we calculate the recall as follows:

\begin{equation}
	R_{SWDP}(\tau) = \frac{N_{SWDP}(\tau)}{|C_{SWDP}|}
\end{equation}
In other words, we consider the more difficult setting where a model does not get credit for recalling the same word if it is said by the same speaker.

However, Algayres et al.~\cite{algayres_evaluating_2020} have shown that this same-different evaluation task is not always indicative of downstream system performance specifically for AWEs.
Here they compared the results of the intrinsic word-discrimination task against a downstream task: the unsupervised estimation of frequencies of speech segments in a given corpus.
In general, they found the AP to correlate with frequency estimation.
However, some inconsistencies appear in fine-grained distinctions across embedding approaches and languages.  
{Moreover, \citet{abdullah_acoustic_2021} analyse the correlation between the word discrimination task and word phonological similarity.
In their experiments, they show that AWEs optimising contrastive objectives (such as the \system{SiameseRNN} and \system{ContrastiveRNN}) yield strong discriminative embeddings but for the most part, fail to reflect the phonological distance between word forms.}

{Given these inconsistencies in the evaluation of the quality of AWEs, we perform QbE speech search (Section~\ref{ssec:background_qbe}) as an additional evaluation of AWEs.
In contrast to the word discrimination task, this test does not assume a set of isolated words, but instead operates on full unsegmented utterances.}
\mysection{Chapter summary}{Chapter summary}
\label{sec:background_summary}

We looked at 
AWEs for zero-resource languages and identified the multilingual transfer strategy as state-of-the-art, specifically when implemented using the \system{CAE-RNN} and \system{SiameseRNN} models of \cite{kamper_truly_2019} and \cite{settle_discriminative_2016}, respectively.
We introduced two zero-resource speech applications, QbE and UTD.
We use QbE as a downstream evaluation task to evaluate the quality of AWEs in Chapter~\ref{chap:related}, \ref{chap:hatespeech}, and~\ref{chap:semantic} (semantic QbE is performed and explained in the latter).
We use UTD to obtain word pairs for training unsupervised monolingual AWEs in Chapter~\ref{chap:contrastive}.
Details regarding the implementation of AWEs were covered by exploring different neural network models, as well as the preprocessing of audio and how the quality of AWEs is measured.


\mychapter{Contrastive learning for acoustic word embeddings}{Contrastive Learning for Acoustic Word Embeddings}
\label{chap:contrastive}

In this chapter we first describe two state-of-the-art AWE models previously mentioned in Section~\ref{sec:background_models}, namely the \system{CAE-RNN} and \system{SiameseRNN}.
We then present a new model that has not been used for AWE modelling, which relies on contrastive learning.
We call this model the \system{ContrastiveRNN}.
A direct comparison is made between the two existing models and the new model by reproducing an experimental setup used in previous work.
We evaluate each AWE model following both the unsupervised monolingual approach (Section~\ref{ssec:monolingual}) and the multilingual transfer approach (Section~\ref{ssec:multilingual}).
In the first, an unsupervised AWE model is trained on unlabelled data in the target language.
In the second, a supervised multilingual model is trained on labelled data from multiple well-resourced languages and then only applied to an unseen target language.
Here we describe the training setup and discuss the results obtained for each of the three AWE models.

\begin{tcolorbox}[width=\linewidth, colback=white!95!black, boxrule=0.5pt]
	\small
	\textit{Parts of this chapter were presented at IEEE Spoken Language Technology (SLT):} \\
	C. Jacobs, Y. Matusevych, and H. Kamper, ``Acoustic word embeddings for zero-resource languages using self-supervised contrastive learning and multilingual adaptation,'' in \textit{Proceedings of SLT}, 2021. 
\end{tcolorbox}

\mysection{Related work}{Related work}
\label{sec:contrastive_background}

Many unsupervised AWE modelling approaches have been proposed as an alternative to DTW (Section~\ref{ssec:background_dtw}) for calculating the similarity between speech segments of different durations. 
Here we remind the reader of the state-of-the-art approaches identified in Section~\ref{sec:background_models}, all of which employ deep neural networks built on RNNs (Section~\ref{ssec:background_rnn}).

An RNN-based AWE model sequentially processes a word segment $X = (\mathbf{x}_1, \mathbf{x}_2, \mathbf{x}_3, \ldots, \mathbf{x}_T)$, where each $\mathbf{x}_t$ is a frame-level acoustic feature, to produce a fixed-dimensional vector that represents segment $X$.
The \system{AE-RNN} of \citet{chung_unsupervised_2016} consists  of an encoder RNN and decoder RNN structure.
A single recurrent layer processes a word segment $X$. 
The output of the recurrent layer (encoder RNN), after processing the final acoustic feature $\mathbf{x}_T$, is given as input to another single recurrent layer (decoder RNN) to reconstruct the original input sequence.
After training, the final output of the recurrent layer is taken as the AWE of an input segment.
In their experiments, the \system{AE-RNN} is trained on unknown isolated speech segments. The speech segments are obtained by segmenting unlabelled speech data in the zero-resource language using forced-alignments. 
Although the \system{AE-RNN} models itself is trained in an unsupervised fashion, using word boundaries for segmentation relies on transcriptions of the zero-resource language.

The \system{SiameseRNN} of \citet{settle_discriminative_2016} explicitly optimise for the distance between embeddings instead of using a reconstruction loss as the \system{AE-RNN}. 
They show the \system{SiameseRNN} is especially better at organising embeddings of words that were not seen during training compared to other methods they considered.
Here they implement an encoder RNN with multiple recurrent layers, instead of a single recurrent layer.
Additionally, they apply a set of fully connected layers to the output of the last recurrent layer, projecting the output to a lower-dimensional vector.
The output of the RNN might be larger than the desired learned representation needed to learn intermediate information while processing the sequential input sequence.
Therefore, the final layer of the encoder RNN might contain redundant information not contributing to the discriminative characteristics of the learned representation.
By adding a set of fully connected layers, the representation can be transformed to be more discriminative.
They train the \system{SiameseRNN} using a weak form of supervision where word segments are grouped into unknown word types. 
Again, this approach is not completely unsupervised since word boundaries are used for segmentation and words are grouped using class labels.

The \system{CAE-RNN} model of \citet{kamper_truly_2019} is an extension of the \system{AE-RNN}.
In the \system{CAE-RNN}, unlike the \system{AE-RNN}, the target output is not identical to the input, but rather an instance of the same word type.
They were the first to use this correspondence learning technique in an encoder-decoder structure operating on whole-word speech segments.
Similar to Settle and Livescu~\cite{settle_discriminative_2016}, they extended the encoder RNN structure by adding a single fully connected feedforward layer to the output of the final output of the last recurrent layer.
Most importantly, the \system{CAE-RNN} is trained in a complete zero-resource setting.
In this setting, no word boundaries or class labels are assumed to be available.
Isolated word-like pairs are obtained by applying a UTD system (Section~\ref{ssec:background_utd}) to the unlabelled speech data.
The UTD system is in itself unsupervised making the whole process unsupervised.
This training setting is the unsupervised monolingual setting as we described in Section~\ref{ssec:monolingual}.
They showed the \system{CAE-RNN} outperforms the \system{AE-RNN} trained in the fully zero-resource setting.

Recently, Kamper et al.~\cite{kamper_improved_2021} presented the multilingual transfer training strategy as described in Section~\ref{ssec:multilingual}.
Here they implement multilingual variants of the \system{CAE-RNN} and \system{SiameseRNN}.
They show that applying a multilingual model to an unseen target language, outperforms an unsupervised monolingual model trained on unlabelled data from the target language. 

\mysection{Baseline acoustic word embedding models}{Baseline acoustic word embedding models}
\label{sec:contrastive_baselines}

In this section we describe the \system{CAE-RNN} and \system{SiameseRNN} AWE models as implemented in Kamper et al.~\cite{kamper_improved_2021} which we reimplement.
We compare these two models to the \system{ContrastiveRNN}, a new model we introduce in Section~\ref{sec:contrastive_rnn}.

\subsection{CAE-RNN}
\label{ssec:contrastive_cae_rnn}


\begin{figure}[!b]
	\centering
	{\includegraphics[width=0.95\linewidth]{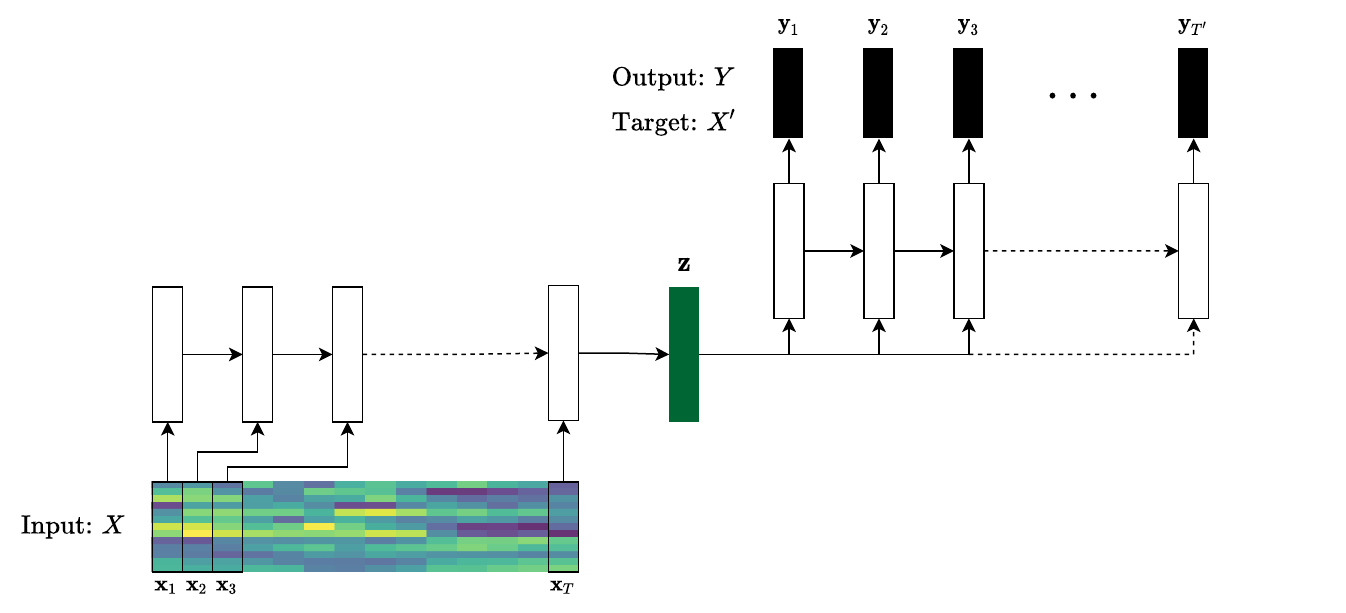}}
	\caption{
		The \system{CAE-RNN} is trained to reconstruct an instance $X^\prime$ of the same word type as the 
		input sequence $X$.
		$T^\prime$ and $T$ are the lengths of $X^\prime$ and $X$, respectively. 
	}
	\label{fig:cae_rnn}
\end{figure}

The \system{CAE-RNN} AWE model of \citet{kamper_truly_2019, kamper_improved_2021} 
uses the CAE encoder-decoder network structure (Section~\ref{ssec:background_cae}) built on RNNs.  
The \system{CAE-RNN} is trained on pairs of speech segments $(X, X^\prime)$, with $X = \mathbf{x}_1, \ldots, \mathbf{x}_T$ and $X^\prime = \mathbf{x}_1, \ldots, \mathbf{x}_{T^{\prime}}$, containing different instances of the same word type, with each $\mathbf{x}_t$ an acoustic feature vector.
Given an input segment $X$, the encoder RNN produces a latent embedding $\mathbf{z}$ from projecting the output of the final hidden layer to a lower-dimensional representation.
The embedding $\mathbf{z}$ is given as input to the decoder RNN to reconstruct an output  sequence $Y = (\mathbf{y}_1, \mathbf{y}_2, \mathbf{y}_3, \ldots, \mathbf{y}_{T^{\prime}})$, where $\mathbf{y}_t$ is the projection of the finale hidden layer at time $t$ conditioned on $\mathbf{z}$.
The \system{CAE-RNN} model is trained using the reconstruction loss,
\begin{align}
\begin{split}
	\ell(X, X') &= \lVert X' - f_{\Theta_{CAE}}(X) \rVert^2 \\
	&=
	\lVert X' - Y \rVert^2,
	\label{eqn:contrastive_cae_loss}\end{split}\end{align} 
with $X^{\prime}$ and $Y$ the target and output, respectively.  
Figure~\ref{fig:cae_rnn} illustrates this model.
After training, the projection $\mathbf{z}$ is taken as the AWE of input segment $X$.

By forcing the model to reconstruct a word segment from a different instance of the same word type, the AWEs should be invariant to properties not common to two segments such as speaker, gender and channel, but capture the aspects that are, for instance, word type.
As in~\cite{kamper_truly_2019}, we first pretrain the \system{CAE-RNN} as an autoencoder using the loss in Equation~\ref{eqn:ae_general_loss} and then switch to the loss function for correspondence training.


\subsection{SiameseRNN}
\label{ssec:contrastive_siamese_rnn}

\begin{figure}[!b]
	\centering
	{\includegraphics[width=0.9\linewidth]{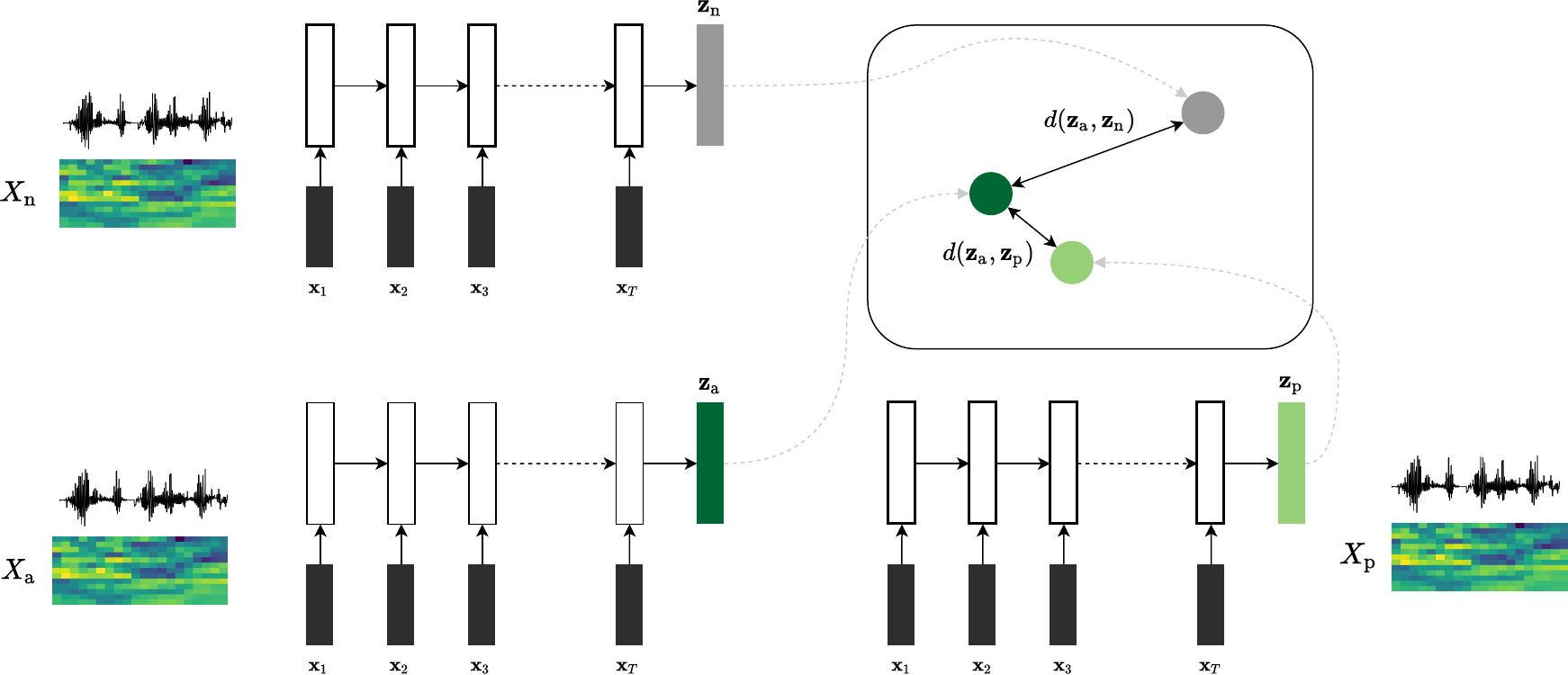}}
	\caption{In the \system{SiameseRNN}, three encoder RNNs use the same set of parameters to produce embeddings $\mathbf{z}_{\text{a}}$, $\mathbf{z}_{\text{p}}$, $\mathbf{z}_{\text{n}}$ from input segments $X_{\text{a}}$, $X_{\text{p}}$, $X_{\text{n}}$. The model is trained to minimise the distance between the anchor and the positive item while maximising the distance between the anchor and negative item.}
	\label{fig:siamese_rnn}
\end{figure}

The \system{SiameseRNN} of \citet{settle_discriminative_2016, kamper_improved_2021}, uses the Siamese network configuration (Section~\ref{ssec:background_siamese}) built on RNNs. 
The model consists of an encoder RNN that processes a word segment $X$ to produce an embedding $\mathbf{z}$, similar to the encoder of the \system{CAE-RNN}.
Instead of reconstructing a target output segment, the \system{SiameseRNN} model explicitly optimises the relative distances between embeddings. 

Given input sequences $X_{\text{a}}$, $X_{\text{p}}$, $X_{\text{n}}$, the model produces embeddings $\mathbf{z}_{\text{a}}$, $\mathbf{z}_{\text{p}}$, $\mathbf{z}_{\text{n}}$, as illustrated in Figure~\ref{fig:siamese_rnn}.
Inputs $X_{\text{a}}$ and $X_{\text{p}}$ are from the same word type (subscripts indicate anchor and positive) and $X_{\text{n}}$ is from a different word type (negative).
For a single triplet of inputs, the model is trained using the triplet loss:
\begin{equation}
	l_{\text{triplet}}(\mathbf{z}_{\text{a}}, \mathbf{z}_{\text{p}}, \mathbf{z}_{\text{n}}) = \text{max}\{0, m + d(\mathbf{z}_{\text{a}}, \mathbf{z}_{\text{p}}) - d(\mathbf{z}_{\text{a}}, \mathbf{z}_{\text{n}})\},
	\label{eqn:contrastive_triplet_loss}
\end{equation}
with $m$ a margin parameter and $d(\textbf{u}, \textbf{v}) = 1 - \textbf{u}^\intercal \textbf{v} / \lVert\textbf{u}\rVert \Vert\textbf{v}\rVert$ denoting the cosine distance between two vectors $\textbf{u}$ and $\textbf{v}$.
This loss is at a minimum when all embedding pairs ($\mathbf{z}_{\text{a}}$, $\mathbf{z}_{\text{p}}$) of the same type are more similar by a margin $m$ than pairs ($\mathbf{z}_{\text{a}}$, $\mathbf{z}_{\text{n}}$) of different types.

Generating all possible triplets would lead to slow convergence given that many of them would easily satisfy the triplet constraint in Equation~\ref{eqn:contrastive_triplet_loss}.
Therefore, we select triplets that violate the triplet constraint.
The \system{SiameseRNN} of \citet{kamper_improved_2021} uses the online semi-hard mining strategy of \citet{schroff_facenet_2015} to sample negative examples.
They sample hard negatives and positives from a single training batch.
In a single training batch, they sample all anchor-positive pairs and select the hardest negative for each. 
In our implementation we use an online hard mining strategy~\cite{hermans_defense_2017}: for each item (anchor) in the batch we select the hardest positive and hardest negative example. 
For each item in the batch, the hardest positive is an instance of the same word type whose embedding is currently the furthest away out of all the other positive items in the batch. Conversely, the hardest negative is an item from a different word type whose embedding is the closest out of all the negative items in the batch.

\mysection{ContrastiveRNN}{ContrastiveRNN}
\label{sec:contrastive_rnn}
This section introduces a new AWE model, namely the \system{ContrastiveRNN}.
We will compare this model to two existing state-of-the-art AWE models, namely the \system{CAE-RNN} and \system{SiameseRNN}.
We first describe the concept of contrastive learning.
We then describe how we implement contrastive learning in AWE modelling.

\subsection[Contrastive learning]{Contrastive learning}
In broad terms, contrastive learning is a technique to learn representations from data inputs such that representations from the same class are close in vector space and representations from different classes are separated.  
A contrastive loss function calculates the distance between two outputs from the same class and contrasts that with the distance between the output of one or more different classes.
The triplet loss optimised by the \system{SiameseRNN} is a form of contrastive learning, but in this thesis, we refer to contrastive loss where a positive example is compared to multiple negative examples. 

Distance metric learning frameworks, such as the triplet loss, often suffer from slow convergence and poor local optima~\cite{sohn_improved_2016, medela_constellation_2020}.
The triplet loss only compares an input example to one negative example without considering examples from the remaining negative classes.
Consequently, for a single training instance, an input example is only guaranteed to be pushed away from one negative example. 
Hence, samples are only separated from limited negative examples, still appearing close to several other classes in vector space.
In practice, after enough iterations of randomly selecting negative examples, the triplet loss should be optimised for optimal class separability.
However, even with curated negative example mining strategies, the triplet loss often struggles to organise the embedding space adequately.




To alleviate this problem, \citet{sohn_improved_2016} proposed a \textit{multi-class $N$-pair} loss that optimises to identify a positive example from $N-1$ negative examples.
This loss extends the triplet loss by incorporating multiple negative examples for each positive pair: an input example is being compared to a positive example and negative examples from multiple classes that it needs to discriminate from at the same time.

\citet{sohn_improved_2016} formalised the $N$-pair loss as follows: given a $(N\text{+1})$-tuplet training example $\{X_a, X_p, X_{n_{1}}, X_{n_{2}}, \cdots, X_{n_{N-1}}\}$, with $X_a$ the input example, $X_p$ a positive example and $\{X_{i}\}_{i=1}^{N-1}$ negative examples, the $(N\text{+1})$-tuplet loss is defined as follows:
\begin{equation}
	\small
	l_{(N+1)\text{-tuplet}}(\{X_a, X_p, X_{n_{1}}, X_{n_{2}}, \cdots, X_{n_{N-1}}\}; f_{\Theta}) = \text{log}\left(1 + \sum_{i=1}^{N-1} \text{exp}(\mathbf{z}_a^T\mathbf{z}_{n_{i}} - \mathbf{z}_a^T\mathbf{z}_p) \right)
	\label{eqn:N_tuplet}
\end{equation}
where $f_{\Theta}$ is an embedding function producing embeddings $\mathbf{z}$ from segments $X$.
Sohn showed that for the scenario when $N=2$, the (2+1)-tuplet loss closely resembles the triplet loss in Equation~\ref{eqn:background_triplet}: 
\begin{equation}
	\small
	l_{(2+1)\text{-tuplet}}(\{X_a, X_p, X_n\}; f_{\Theta}) = \text{log}(1 + \text{exp}(\mathbf{z}_a^T\mathbf{z}_n - \mathbf{z}_a^T\mathbf{z}_p))
\end{equation}
where the embedding function $f_{\Theta}$ minimises the $l_{(2+1)\text{-tuplet}}$ if and only if it minimises $l_\text{triplet}$.
Rewriting the loss function in Equation~\ref{eqn:N_tuplet} to 
\begin{equation}
	\small
	\text{log}\left(1 + \sum_{i=1}^{N-1} \text{exp}(\mathbf{z}_a^T\mathbf{z}_{n_{i}} - \mathbf{z}_a^T\mathbf{z}_p) \right) = - \text{log}\frac{\text{exp}(\mathbf{z}_a^T\mathbf{z}_p)}{\text{exp}(\mathbf{z}_a^T\mathbf{z}_p) + \sum_{i=1}^{L-1}\text{exp}(\mathbf{z}_a^T\mathbf{z}_{n_i})}
	\label{eqn:tuplet_softmax}
\end{equation}
looks similar to the multi-class logistic loss where including more negative examples gives a better approximation.
This shows the advantage of the $(N\text{+1})$-tuplet loss over the triplet loss.


Previous work successfully implemented this multi-class negative mining loss for visual representation learning~\cite{bachman_learning_2019, chen_simple_2020}.
To the best of our knowledge, this loss has not been implemented in whole-word speech representation learning in AWE modelling. 
In the following section we describe how we apply contrastive learning to AWE modelling.

\subsection[Contrastive learning in acoustic word embeddings]{Contrastive learning in acoustic word embeddings}

We specifically implement the contrastive loss of~\cite{chen_simple_2020} using the same embedding function  as the \system{SiameseRNN} (Section~\ref{ssec:contrastive_siamese_rnn}); we call this model the \system{ContrastiveRNN}.
Concretely, given inputs $X_a$ and $X_p$ and multiple negative examples $X_{n_{1}}, \ldots, X_{n_{K}}$, the \system{ContrastiveRNN} produces embeddings $\mathbf{z}_a, \mathbf{z}_p, \mathbf{z}_{n_{1}}, \ldots, \mathbf{z}_{n_{K}}$.
Let $\text{sim}(\mathbf{u}, \mathbf{v}) = \mathbf{u}^{\top}\mathbf{v}/\lVert\mathbf{u}\rVert \Vert\mathbf{v}\rVert$ denote the cosine similarity between two vectors $\mathbf{u}$ and $\mathbf{v}$.
The loss given a positive pair $(X_a, X_p)$ and the set of negative examples is then defined as~\cite{chen_simple_2020}:

\begin{equation}
	\ell(\mathbf{z}_a, \mathbf{z}_p, \mathbf{z}_{n_{1}}, \ldots, \mathbf{z}_{n_{K}}) = -\text{log}\frac{\text{exp}\big\{\text{sim}(\mathbf{z}_a, \mathbf{z}_p)/\tau\big\}}{\sum_{j \in \{p, n_1, \hdots, n_K\}}^{}\text{exp}\big\{\text{sim}(\mathbf{z}_a, \mathbf{z}_j)/\tau\big\}}\,\text{,}
	\label{eqn:contrastive_loss}
\end{equation} 
where $\tau$ is a temperature parameter.
The temperature parameter does not directly affect the accuracy but helps gradients to be propagated more easily.

\begin{figure}[t]	
	\vspace{3mm}
	\begin{minipage}[a]{0.45\linewidth}
		\centering
		\centerline{\includegraphics[width=0.95\linewidth]{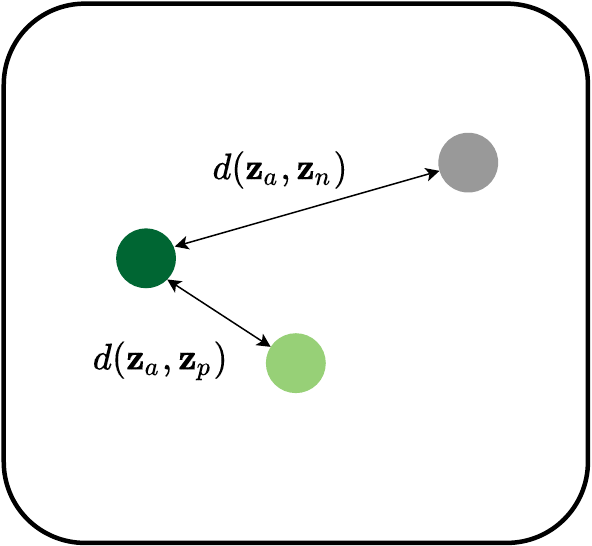}}
		\centerline{(a) Single negative example.}\medskip
	\end{minipage}
	\hfill
	\begin{minipage}[a]{0.45\linewidth}
		\centering
		\centerline{\includegraphics[width=0.95\linewidth]{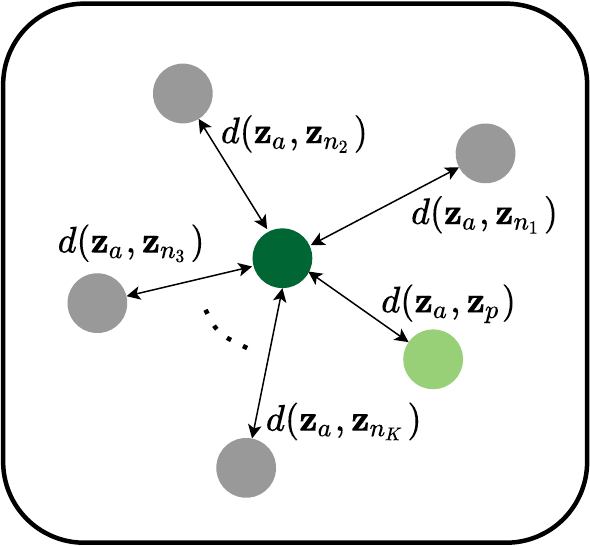}}
		\centerline{(b) Multiple negative examples.}\medskip
	\end{minipage}
	
	\caption{A visualisation of the difference in the optimisation of (a) the \system{SiameseRNN} and (b) the \system{ContrastiveRNN} for a single positive pair $(\mathbf{z}_a, \mathbf{z}_p)$ in the embedding space.
	} 
	\label{fig:contrastive_rnn}
\end{figure}
The difference between this loss and the triplet loss used in the \system{SiameseRNN} is illustrated in Figure~\ref{fig:contrastive_rnn}.
Ideally, we would like to include a negative example of each class for an input example. However, this is not computationally efficient and could lead to poor generalisation with class imbalances. 
We follow an offline batch construction process where we randomly choose $N$ distinct positive pairs.
Given a positive pair $(X_a, X_p)$, the remaining $2(N-1)$ items are then treated as negative examples. 
The final loss is calculated as the sum of the loss over all $N$ positive pairs within the batch. 


The \system{ContrastiveRNN} can be trained in both the unsupervised monolingual (Section~\ref{ssec:monolingual}) and supervised multilingual setting (Section~\ref{ssec:multilingual}).

\mysection{Experimental setup}{Experimental setup}
\label{sec:contrastive_experiment_setup}

{We follow the same setup as Kamper et al.~\cite{kamper_improved_2021} to train AWE models in the monolingual unsupervised (Section~\ref{ssec:monolingual}) and supervised multilingual (Section~\ref{ssec:multilingual}) settings.
In the first, an unsupervised AWE model is trained on unlabelled data in the target language.
In the second, a supervised multilingual model is trained on labelled data from multiple well-resourced languages and then only applied to an unseen target language.
Here we describe the training and evaluation data, and details regarding the configuration of the AWE models (\system{CAE-RNN}, \system{SiameseRNN}, \system{ContrastiveRNN}). }

\subsection[Data]{Data}
\label{ssec:contrastive_setup_data}

We perform experiments using the GlobalPhone corpus of read speech~\cite{schultz_globalphone_2013}.
This corpus contains 20 languages covering a variety of speech peculiarities. 
For each language, 100 adult native speakers were recorded reading approximately 100 sentences selected from national newspaper articles available on the web.
As in 
~\cite{kamper_improved_2021}, we treat six languages as our target zero-resource languages: Spanish~(ES), Hausa~(HA), Croatian~(HR), Swedish~(SV), Turkish~(TR) and Mandarin~(ZH).
Each language has on average 16 hours of training, 2 hours of development and 2 hours of test data.
We apply the UTD system of~\cite{jansen_efficient_2011} to the training set of each zero-resource language and use the discovered pairs to train unsupervised monolingual embedding models (Section~\ref{ssec:monolingual}).
The UTD system discovers around 36k pairs for each language,
where pair-wise matching precisions vary between $32\%$ (SV) and $79\%$ (ZH).
Training conditions for the unsupervised monolingual \system{CAE-RNN}, \system{SiameseRNN} and \system{ContrastiveRNN} models are determined by doing validation on the Spanish
development data.
The same hyperparameters are then used for the five remaining zero-resource languages.

For training supervised multilingual embedding models (Section~\ref{ssec:multilingual}), six other GlobalPhone languages are chosen as well-resourced languages: Czech, French, Polish, Portuguese, Russian and Thai.
Each well-resourced language has on average 21 hours of labelled training data.
We pool the data from all six well-resourced languages and train a multilingual \system{CAE-RNN}, a \system{SiameseRNN} and a \system{ContrastiveRNN}.
Instead of using the development data from one of the zero-resource languages, we use another well-resourced language, German, for validation of each model before applying it to the zero-resource languages. 
We only use 300k positive word pairs for each model, as further increasing the number of pairs did not give improvements on the German validation data.

All speech audio is parametrised as $D = 13$ dimensional static Mel-frequency cepstral coefficients (MFCCs).

\subsection[Acoustic word embedding models]{Acoustic word embedding models}
{All our models use a similar architecture as implemented in~\cite{kamper_improved_2021}}: encoders and decoders consist of three unidirectional RNNs with 400-dimensional hidden vectors, and all models use an embedding size of 130 dimensions.
Models are optimised using Adam optimisation~\cite{kingma_adam_2017}, with a learning rate of $0.001$ for the \system{CAE-RNN} and \system{ContrastiveRNN}, and $5\cdot10^{-4}$ for the \system{SiameseRNN}.
The margin parameter $m$ in Section \ref{ssec:contrastive_siamese_rnn} and temperature parameter $\tau$ in Section \ref{ssec:contrastive_cae_rnn} are set to $0.25$ and $0.1$, respectively.
We train the \system{CAE-RNN} and \system{SiameseRNN} with a batch size of $300$, and set the batch size for the \system{ContrastiveRNN} to $600$.

{Learning rates and hyperparameters are set by experimenting on Spanish development data for the unsupervised monolingual models, and German development data for the supervised multilingual models.}



\mysection{Experiments}{Experiments}
\label{sec:contrastive_experiments}

We start in Section~\ref{ssec:word_discrimination} by evaluating the different AWE models using the word discrimination task described in Section~\ref{sec:background_evaluation}.
{Instead of only looking at word discrimination results, it is useful to also use other methods to try and better understand the organisation of AWE spaces~\cite{matusevych_analyzing_2020}, especially in light of recent results showing that AP has limitations as described in Section~\ref{sec:background_evaluation}.}
We therefore also look at speaker classification performance in Section~\ref{ssec:speaker_classification}.
{Speaker classification is not necessarily a task in which we want our models to be good at.
Since we are mostly interested in producing AWEs that are speaker invariant, a lower speaker classification is desirable.
This probing experiment is useful to get insight on the AWEs produced by different AWE models.
For example, two AWEs might show similar AP in the word discrimination task, while one may capture more speaker information.}



\subsection[Word discrimination]{Word discrimination}
\label{ssec:word_discrimination}

We first consider purely unsupervised monolingual models (Section~\ref{ssec:monolingual}).
We are particularly interested in the performance of the \system{ContrastiveRNN}, which has not been considered in previous work.
The top section in Table~\ref{tbl:multi} shows the performance for the unsupervised monolingual AWE models applied to the test data from the six zero-resource languages.\footnote{We note that the results for the \system{CAE-RNN} and \system{SiameseRNN} here are slightly different to that of~\cite{kamper_multilingual_2020, kamper_improved_2021}, despite using the same test and training setup. We believe this is due to the different negative sampling scheme for the \system{SiameseRNN} and other small differences in our implementation.}
As a baseline, we also give the results where DTW is used directly on the MFCCs to perform the word discrimination task.
We see that the \system{ContrastiveRNN} consistently outperforms the \system{CAE-RNN} and \system{SiameseRNN} approaches on all six zero-resource languages.
The \system{ContrastiveRNN} is also the only model to perform better than DTW on all six zero-resource languages, which is noteworthy since DTW has access to the full sequences for discriminating between words.

\begin{table}[!t]
	\mytable
	\caption{AP (\%) on test data for the six zero-resource languages.
		The purely unsupervised monolingual models are trained on discovered word segments (top).
		The multilingual models are trained on true words by pooling labelled training data from six well-resourced languages (bottom).
		The best approach in each subsection is shown in bold.}
	
	\begin{tabularx}{1.0\linewidth}{@{}lCCCCCC@{}}
		\toprule
		\addlinespace
		Model & {ES} & HA & HR & SV & TR & ZH \\
		\midrule
		\addlinespace
		\underline{\textit{Unsupervised models:}} \\[2pt]        
		\tablesystem{DTW} &36.2 &23.8 &17.0 &27.8 &16.2 &35.9 \\
		\tablesystem{CAE-RNN} &52.7 &18.6 &24.5 &28.0 &14.2 &33.7 \\
		\tablesystem{SiameseRNN} &56.6 &16.8 &21.1 &31.8 &22.8 &52.0 \\
		\tablesystem{ContrastiveRNN} &\textbf{70.6} &\textbf{36.4} &\textbf{27.8} &\textbf{37.9} &\textbf{31.3} &\textbf{57.1} \\[2pt]
		\underline{\textit{Multilingual models:}} \\[2pt]
		\tablesystem{CAE-RNN} &72.4 &49.3 &44.5 &\textbf{52.7} &34.4 &\textbf{53.9} \\
		\tablesystem{SiameseRNN} &70.3 &45.3 &40.6 &47.5 &27.7 &49.9 \\
		\tablesystem{ContrastiveRNN} &\textbf{73.3} &\textbf{50.6} &\textbf{45.1} &46.4 &\textbf{34.6} &53.2 \\[2pt]
		\bottomrule
		\addlinespace
	\end{tabularx}

	\label{tbl:multi}
\end{table}

Next, we consider the supervised multilingual models (Section~\ref{ssec:multilingual}). 
The bottom section of Table~\ref{tbl:multi} shows the performance for the supervised multilingual models applied to the six zero-resource languages. 
By comparing these supervised multilingual models to the unsupervised monolingual models (top), we see that in almost all cases the multilingual models outperform the purely unsupervised monolingual models, as also in~\cite{kamper_multilingual_2020, kamper_improved_2021}.
However, on Mandarin (ZH), the unsupervised monolingual \system{ContrastiveRNN} model outperforms all three multilingual models.
Comparing the three multilingual models, we do not see a consistent winner between the \system{ContrastiveRNN} and \system{CAE-RNN}, with one performing better on some languages while the other performs better on others.
The multilingual \system{SiameseRNN} generally performs worst, although it outperforms the \system{ContrastiveRNN} on Swedish (SV).

\subsection[Speaker classification]{Speaker classification}
\label{ssec:speaker_classification}

{We now perform a probing experiment by considering the extent to which different AWE models capture speaker information.
Performing well in a speaker classification task is not always desirable.
A lower speaker classification score indicates that the representations capture little speaker information, which is what we want.}

To measure speaker invariance, we use a linear classifier to predict a word's speaker identity from its AWE.
Specifically, we train a multi-class logistic regression model on 80\% of the development data and test it on the remaining 20\%.

The top section of Table~\ref{tbl:mono_speaker} shows speaker classification results on development data for the three types of monolingual unsupervised models (Section~\ref{ssec:monolingual}).
Since we are interested in how well models abstract away from speaker information, we consider lower accuracy as better (shown in bold).
The \system{ContrastiveRNN} achieves the lowest speaker classification performance across all languages, except on Croatian where it performs very similarly to the \system{SiameseRNN}.
This suggests that among the unsupervised monolingual models, the \system{ContrastiveRNN} is the best at abstracting away from speaker identity (at the surface level captured by a linear classifier).

Next, we consider speaker classification performance for the multilingual models (Section~\ref{ssec:multilingual}).
Comparing the middle and top sections of Table~\ref{tbl:mono_speaker}, we see that for each multilingual model (bottom) the speaker classification performance drops from its corresponding unsupervised monolingual version (top) across all six languages, again indicating an improvement in speaker invariance. 
Comparing the three multilingual models to each other (bottom), the \system{ContrastiveRNN} has the lowest speaker classification performance on four out of the six evaluation languages. 

{We conclude that the multilingual models are better at producing speaker-invariant AWEs compared to their monolingual variants. Also, the \system{ContrastiveRNN} is better at abstracting away from speaker information in both the unsupervised monolingual and multilingual transfer setting compared to the \system{CAE-RNN} and \system{SiameseRNN}.}

\begin{table}[!t]
	\mytable
	\caption{
		Speaker classification accuracy (\%) on development data for the six zero-resource languages using purely unsupervised monolingual models (top) and multilingual models (bottom).
	}$  $
	
	\begin{tabularx}{1\linewidth}{@{}lCCCCCC@{}}
		\toprule
		\addlinespace
		Model & {ES} & HA & HR & SV & TR & ZH \\
		\midrule
		\addlinespace
		\underline{\textit{Unsupervised models:}} \\[2pt]        
		\tablesystem{CAE-RNN} & 52.8 & 49.5 & 53.8 & 47.8 & 48.9 & 62.0 \\
		\tablesystem{SiameseRNN} & 38.9 & 38.4 & \textbf{37.2} & 36.4 & 38.6 & 44.3 \\
		\tablesystem{ContrastiveRNN} & \textbf{34.6} & \textbf{33.1} & 37.8 & \textbf{33.6} & \textbf{35.5} & \textbf{39.9} \\[2pt]
		\underline{\textit{Multilingual models:}} \\[2pt]        
		\tablesystem{CAE-RNN} & 40.6 & 44.7 & 45.0 & 42.3 & 42.0 & 49.4 \\
		\tablesystem{SiameseRNN} & 32.6 & 30.3 & 31.9 & \textbf{28.6} & \textbf{30.3} & 32.5 \\ 
		\tablesystem{ContrastiveRNN} & \textbf{27.6} & \textbf{28.7} & \textbf{26.9} & 29.2 & 32.1 & \textbf{31.9} \\[2pt]     
		\bottomrule
		\addlinespace
	\end{tabularx}
	
	\label{tbl:mono_speaker}
\end{table}

\mysection{Chapter summary}{Chapter summary}
\label{sec:contrastive_summary}

In this chapter we reimplemented two existing AWE models (\system{CAE-RNN} and \system{SiameseRNN}).
We reproduced an experimental setup to compare these two models to a new model we presented (\system{ContrastiveRNN}), in both the unsupervised monolingual and supervised multilingual settings.  
We show that the \system{ContrastiveRNN} outperforms the \system{CAE-RNN} and \system{SiameseRNN} in the unsupervised monolingual setting and performs on par with the \system{CAE-RNN} in the supervised multilingual setting. 
In the following chapter, we determine whether some available data (unlabelled) in the zero-resource setting can be used to adapt a multilingual model to cater for a specific zero-resource language.

\mychapter{Multilingual adaptation}{Multilingual Adaptation}
\label{chap:adaptation}


In the previous chapter, we concluded that applying a supervised multilingual AWE model, trained on labelled data from multiple languages, to an unseen target zero-resource language outperforms an unsupervised monolingual AWE model trained on unlabelled data from the target language.
In this chapter, we explore whether the performance of a multilingual AWE model for a specific zero-resource language can be improved by adapting the model using unlabelled data available in that particular language. 

\begin{tcolorbox}[width=\linewidth, colback=white!95!black, boxrule=0.5pt]
	\small
	\textit{Parts of this chapter were presented at IEEE Spoken Language Technology (SLT):} \\
	C. Jacobs, Y. Matusevych, and H. Kamper, ``Acoustic word embeddings for zero-resource languages using self-supervised contrastive learning and multilingual adaptation,'' in \textit{Proceedings of SLT}, 2021.
\end{tcolorbox}


\mysection{Intuition and related work}{Intuition and related work}
\label{sec:adapt_related}

{The supervised multilingual transfer approach of \citet{kamper_improved_2021} as described in Section~\ref{ssec:multilingual} and implemented in Chapter~\ref{chap:contrastive}, is a form of transfer learning categorised as \textit{transductive transfer learning}.}
\newline

\begin{adjustwidth}{1.5em}{0em}
	\begin{flushright}
		\textbf{Transductive transfer learning}~\cite{pan_survey_2009}: Given a source domain $\mathcal{D}_{\text{S}}$ and a corresponding learning task $\mathcal{T}_{\text{S}}$, a target domain $\mathcal{D}_{\text{T}}$  and a corresponding learning task $\mathcal{T}_{\text{T}}$, transductive transfer learning aims to improve the learning of the target predictive function $f_{\text{T}(.)}$ in $\mathcal{D}_{\text{T}}$ using the knowledge in $\mathcal{D}_{\text{S}}$ and $\mathcal{T}_{\text{S}}$, where $\mathcal{D}_{\text{S}}$ $\neq$ $\mathcal{D}_{\text{T}}$ and $\mathcal{T}_{\text{S}}$ $=$ $\mathcal{T}_{\text{T}}$. In addition, some unlabelled target domain data must be available at training time.
		\newline
	\end{flushright}
\end{adjustwidth}
The scenario of training a model in one domain and then transferring it to a domain of interest, is especially useful in the unsupervised setting where a sufficient amount of labelled data is available in the source domain and only unlabelled data in the target domain are assumed to be available. 
In the supervised multilingual approach, we identify the source domain $\mathcal{D}_{\text{S}}$ as the setting in which labelled training data from multiple well-resourced languages are available and the target domain $\mathcal{D}_{\text{T}}$ as the zero-resource setting where only unlabelled speech data from an unseen target language is available.
We identify the target predictive function as an embedding function $f_{\Theta}(\mathbf{x})$ that produces a fixed-dimensional embedding given an arbitrary-length speech segment $\mathbf{x}$.
The task (producing fixed-dimensional embeddings from variable-length speech segments) across both domains $\mathcal{T}_{\text{S}}$ and $\mathcal{T}_{\text{T}}$ are the same, irrespective of the language.
We confirm the multilingual transfer approach fits the definition of transductive transfer learning as given in the definition.
Note, the transfer learning method implemented by Kamper et al.~\cite{kamper_improved_2021} (Section~\ref{ssec:multilingual}) only partially performs transductive transfer learning by not considering any data from the target domain.
An AWE model was simply trained on labelled data from multiple languages (source domain) and then directly applied to an unseen zero-resource language (target domain). 


In this chapter, we extend this transfer learning technique by using available unlabelled data from a zero-resource language to modify the embedding function.
This transfer learning technique can simply be treated as a regular semi-supervised problem, ignoring the domain difference by considering the source and target data as labelled and unlabelled, respectively.

Recently, Hu et al.~\cite{hu_multilingual_2020} presented adaptation of acoustically grounded word embeddings (AWEs are trained jointly on speech and text) by using labelled data (limited) in the target domain, showing improved results on the target language. However, to the best of our knowledge, we are the first to perform adaptation of AWEs with unlabelled data in the target domain.

\mysection{Unsupervised adaptation of multilingual models}{Unsupervised adaptation of multilingual models}
\label{sec:adapt_models}




We aim to improve the performance of a multilingual AWE model on a specific zero-resource language by manipulating the learned parameters of a multilingual model using unlabelled data from that language. 
In other words, we adapt a multilingual model by fine-tuning the pre-trained multilingual model directly on the available unlabelled data in the target domain.
The idea is that adapting the multilingual AWE model to the target language would allow the model to learn aspects unique to the language.

We consider the unsupervised adaptation of all three multilingual AWE models (\system{CAE-RNN}, \system{SiameseRNN}, \system{ContrastiveRNN}) in Chapter~\ref{chap:contrastive}.
We again use the UTD system of \cite{jansen_efficient_2011} to obtain discovered word pairs from the target language.
We then use these discovered pairs to adapt the multilingual models in the same way we trained the unsupervised monolingual models in Chapter~\ref{chap:contrastive}.
We need to update a multilingual model's parameters in such a way that it doesn't unlearn its current ``knowledge".
Using a learning rate lower than the learning rate used for training the source multilingual model is one way to prevent parameters from deviating too much.
When fine-tuning a model, instead of allowing all model parameters to update, freezing some parameters is another way to preserve some of the learned information.

\mysection{Experimental setup}{Experimental setup}
\label{sec:adapt_setup}

\begin{figure}[b]
	\centering
	\includegraphics[width=0.65\linewidth]{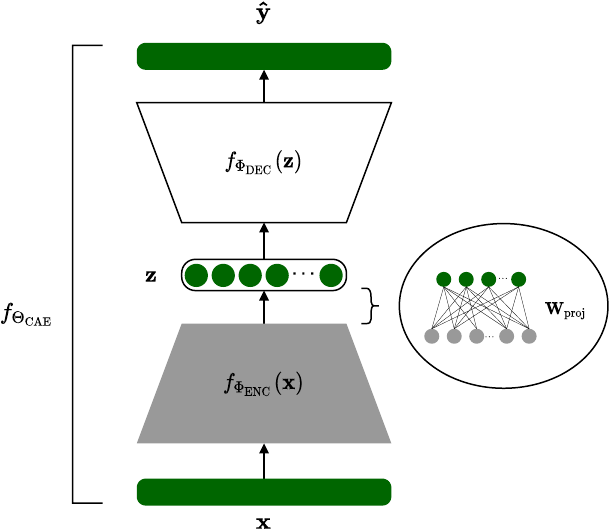}
	\caption{The same \system{CAE-RNN} network architecture as displayed in Figure~\ref{fig:background_ae_cae}.
		When adapting the \system{CAE-RNN} we freeze the encoder RNN weights (highlighted in grey) only allowing the projection weights $\mathbf{W}_{\text{proj}}$ between the final hidden layer of the encoder RNN and latent embedding $\mathbf{z}$ to update. The decoder RNN is re-initialised with random weights before training on the target data (highlighted in white).}
	\label{fig:freeze_cae}
\end{figure}

The experimental setup is the same as described in Section~\ref{ssec:contrastive_setup_data}, except for the differences we describe here.

We adapt each multilingual model to every zero-resource language using the same discovered word pairs as for the unsupervised monolingual models in Chapter~\ref{chap:contrastive}. 
We performed validation experiments to determine which parameters to update and which to keep fixed from the source multilingual models using Spanish development data.
The same hyperparameters are used to adapt a multilingual model to each of the five remaining zero-resource languages.
For the \system{CAE-RNN} (described in Section~\ref{ssec:contrastive_cae_rnn}), we found that it is best to freeze the multilingual encoder RNN weights and only update the weights between the final encoder RNN hidden state and the embedding $\mathbf{z}$, as illustrated in Figure~\ref{fig:freeze_cae}.
We also found that it is best to re-initialise the decoder RNN weights randomly before training on the target data.

For the \system{SiameseRNN} (Section~\ref{ssec:contrastive_siamese_rnn}) and \system{ContrastiveRNN} (Section~\ref{sec:contrastive_rnn}) we update all weights during adaptation.
The learning rate for the \system{CAE-RNN}, \system{SiameseRNN} and \system{ContrastiveRNN} were adjusted to $1{\cdot}10^{-4}$, $1{\cdot}10^{-5}$ and $1{\cdot}10^{-4}$, respectively.



\mysection{Experiments}{Experiments}
\label{sec:adapt_experiments}

We repeat the experiments in Chapter~\ref{chap:contrastive}, this time evaluating the multilingual adapted models on the word discrimination and speaker classification evaluation tasks.
We specifically, compare the results of the adapted multilingual models to the original multilingual models in Table~\ref{tbl:multi}.
In addition to the quantitative evaluations, we also provide some qualitative analysis to illustrate the effect of adaptation in vector space.

\subsection[Word discrimination]{Word discrimination} 
\label{ssec:adapt_discrimiation}

\begin{table}[!b]
	\mytable
	\caption{The AP (\%) on test data for all six zero-resource languages.
		The top section is a duplicate of the multilingual models in Table~\ref{tbl:multi}.
		The bottom section is the results achieved by adapting each multilingual model (\tablesystem{CAE-RNN}, \tablesystem{SiameseRNN}, \tablesystem{ContrastiveRNN}) to every zero-resource language using discovered words from the target zero-resource language.
		The model delivering the highest score per language is highlighted in bold.}
	
	\begin{tabularx}{1.0\linewidth}{@{}lCCCCCC@{}}
		\toprule
		\addlinespace
		Model & ES & HA & HR & SV & TR & ZH \\
		\midrule
		\addlinespace
		\underline{\textit{Multilingual models:}} \\[2pt]
		\tablesystem{CAE-RNN} &72.4 &49.3 &44.5 &\textbf{52.7} &34.4 &\textbf{53.9} \\
		\tablesystem{SiameseRNN} &70.3 &45.3 &40.6 &47.5 &27.7 &49.9 \\
		\tablesystem{ContrastiveRNN} &\textbf{73.3} &\textbf{50.6} &\textbf{45.1} &46.4 &\textbf{34.6} &53.2 \\[2pt]
		\underline{\textit{Multilingual adapted:} 
		} \\[2pt]
		\tablesystem{CAE-RNN} &74.2 &49.4 &\textbf{45.9} &53.4 &34.5 &53.9 \\
		\tablesystem{SiameseRNN} &74.5 &44.7 &37.6 &50.3 &30.3 &57.3 \\
		\tablesystem{ContrastiveRNN} &\textbf{76.6} &\textbf{56.7} &44.4 &\textbf{54.4} &\textbf{40.5} &\textbf{60.4} \\
		\bottomrule
		\addlinespace		
	\end{tabularx}
	
	\label{tbl:multi_adapt}
\end{table}
The results after adapting each multilingual model (\system{CAE-RNN}, \system{SiameseRNN}, \system{ContrastiveRNN}) to each of the zero-resource languages are shown in the bottom section of Table~\ref{tbl:multi_adapt}.
In the top section of Table~\ref{tbl:multi_adapt}, we again show the performance of the source multilingual models on each language (bottom section of Table~\ref{tbl:multi}). 
Comparing the adapted models (bottom) to the original multilingual models (top) in Table~\ref{tbl:multi_adapt}, we see that most of the adapted models outperform their corresponding source multilingual models, with the \system{ContrastiveRNN} and \system{SiameseRNN} improving substantially after adaptation on some of the languages.
The adapted \system{ContrastiveRNN} models outperform the adapted \system{CAE-RNN} and \system{SiameseRNN} models on five out of the six zero-resource languages.
The results in Table~\ref{tbl:multi_adapt} are some of the best-reported results on these data from the same-different AP.
\begin{table}[!b]
	\mytable
	\caption{AP (\%) on development data.
		The supervised monolingual models are trained on ground truth words from the target languages to determine an upper bound on performance.
		The adapted models are the same as those in Table~\ref{tbl:multi_adapt} but applied to development data here for the purpose of analysis.
	}
	
	\begin{tabularx}{1.0\linewidth}{@{}lCCCCCC@{}}
		\toprule
		\addlinespace
		Model & {ES} & HA & HR & SV & TR & ZH \\
		\midrule
		\addlinespace
		\underline{\textit{Supervised monolingual:}} \\[2pt]
		\tablesystem{CAE-RNN} &70.2 &79.7 &63.0 &55.8 &65.6 &84.2  \\
		\tablesystem{SiameseRNN} &78.6 &85.2 &79.3 &68.6 &77.6 &93.1 \\
		\tablesystem{ContrastiveRNN} &81.8 &82.4 &80.3 &70.9 &80.5 &92.1 \\[2pt]
		\underline{\textit{Multilingual adapted:}} \\[2pt]
		\tablesystem{CAE-RNN} &51.7 &59.5 &44.3 &38.6 &40.9 &52.2 \\
		\tablesystem{SiameseRNN} &51.5 &52.7 &38.4 &40.3 &33.0 &56.5 \\
		\tablesystem{ContrastiveRNN} &58.0 &60.5 &40.5 &43.8 &46.9 &60.2 \\
		\bottomrule
		\addlinespace
	\end{tabularx}
	\label{tbl:top_line}
\end{table}
From the results in Table~\ref{tbl:multi_adapt} we conclude that unsupervised adaptation of multilingual models to a target zero-resource language is an effective AWE approach, especially when coupled with the self-supervised contrastive loss.

One question is whether adapted models close the gap between the zero-resource setting and the best-case scenario where we have labelled data available in a target language.
To answer this, Table~\ref{tbl:top_line} compares adapted multilingual models (bottom) to ``oracle'' supervised monolingual models trained on labelled data from the six evaluation languages (top) on development data.
Although Table~\ref{tbl:multi_adapt} shows that adaptation greatly improves performance in the zero-resource setting, Table~\ref{tbl:top_line} shows that multilingual adaptation still does not reach the performance of supervised monolingual models.

\subsection[Speaker classification]{Speaker classification}
\label{ssec:adapt_classifiction}

\begin{table}[!t]
	\mytable
	\caption{Speaker classification accuracy (\%) on development data for the zero-resource languages using multilingual models after adaptation.}
	
	\begin{tabularx}{1\linewidth}{@{}lCCCCCC@{}}
		\toprule
		\addlinespace
		Model & {ES} & HA & HR & SV & TR & ZH \\
		\midrule
		\addlinespace
		\underline{\textit{Multilingual models:}} \\[2pt]        
		\tablesystem{CAE-RNN} & 40.6 & 44.7 & 45.0 & 42.3 & 42.0 & 49.4 \\
		\tablesystem{SiameseRNN} & 32.6 & 30.3 & 31.9 & \textbf{28.6} & \textbf{30.3} & 32.5 \\ 
		\tablesystem{ContrastiveRNN} & \textbf{27.6} & \textbf{28.7} & \textbf{26.9} & 29.2 & 32.1 & \textbf{31.9} \\[2pt]     
		\underline{\textit{Multilingual adapted:}} \\[2pt]
		\tablesystem{CAE-RNN} & 40.3 & 45.3 & 45.0 & 43.1 & 41.0 & 49.7 \\
		\tablesystem{SiameseRNN} & 33.6 & 33.9 & 30.6 & \textbf{29.4} & \textbf{33.7} & 37.9 \\
		\tablesystem{ContrastiveRNN} & \textbf{30.3} & \textbf{32.1} & \textbf{26.9} & 29.5 & 36.5 & \textbf{35.1} \\        
		\bottomrule
		\addlinespace
	\end{tabularx}
	
	\label{tbl:adapt_speaker}
\end{table}

{As in Section~\ref{ssec:speaker_classification}, we again consider speaker classification probing experiments to see to what extent the resulting features normalise out speaker characteristics after adaptation.
As a reminder, the resulting AWEs should avoid capturing speaker information, in other words, a lower speaker classification score is desired.}

The impact of unsupervised adaptation (Section~\ref{ssec:multilingual_adapt}) on speaker invariance, is shown at the bottom of Table~\ref{tbl:adapt_speaker}.
We again show the original multilingual speaker classification results (bottom section of Table~\ref{tbl:mono_speaker}) at the top of Table~\ref{tbl:adapt_speaker}.  
After adaptation (bottom) we see that speaker classification results improve consistently compared to their corresponding source multilingual model (top).
Although this seems to indicate that the adapted AWEs capture more speaker information, these embeddings still lead to better word discrimination performance (Table~\ref{tbl:multi_adapt}).
A similar trend was observed in~\cite{kamper_improved_2021}: a model leading to better (linear) speaker classification performance does not necessarily give worse AP.
Note, in Chapter~\ref{chap:contrastive} we emphasised the desire for producing AWEs that capture less speaker information.
Although this is the case for many tasks, capturing speaker information may be favourable for some downstream tasks.

{We conclude that that unsupervised adaptation using unlabelled data in a target zero-resource language leads to representations allowing better word discrimination while retaining more speaker information. Capturing speaker information might be beneficial for some downstream tasks.
For practical implementation, the choice of applying adaptation will depend on the downstream application at hand.}


\subsection[Qualitative analysis]{Qualitative analysis}
\label{ssec:qualitative}

\begin{figure}[!t]	
	\begin{minipage}[a]{.49\linewidth}
		\centering
		\centerline{\includegraphics[width=\linewidth]{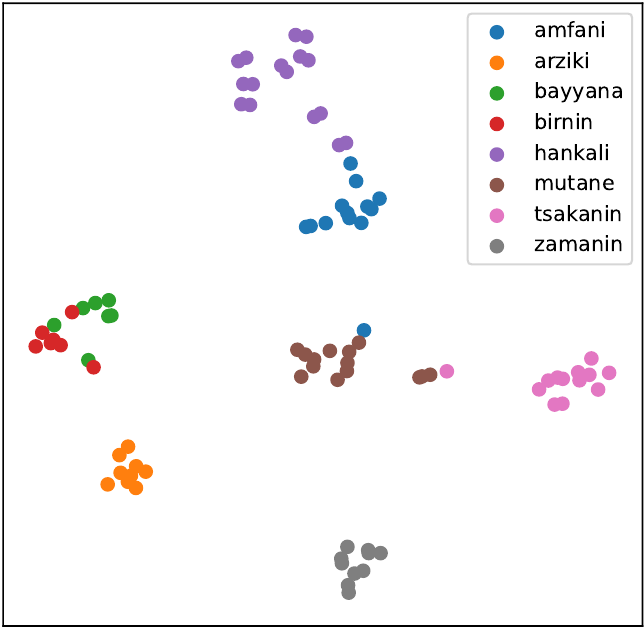}}
		\centerline{(a) Before adaptation.}\medskip
	\end{minipage}
	\hfill
	\begin{minipage}[a]{0.49\linewidth}
		\centering
		\centerline{\includegraphics[width=\linewidth]{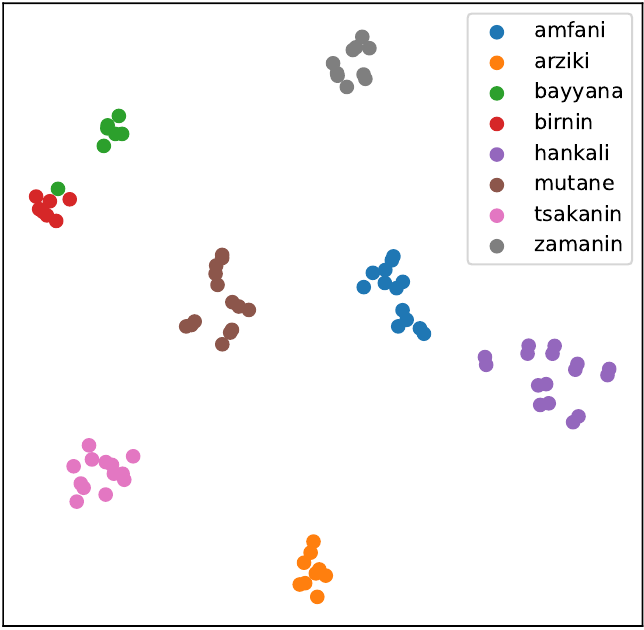}}
		\centerline{(b) After adaptation.}\medskip
	\end{minipage}
	
	\caption{t-SNE visualisations of acoustic embeddings for the most frequent words in the Hausa test data, produced by (a) the multilingual \system{ContrastiveRNN} model (Table~\ref{tbl:multi}) and (b) the multilingual \system{ContrastiveRNN} model adapted to Hausa (Table~\ref{tbl:multi_adapt}).} 
	\label{fig:tsne}
\end{figure}

Given the high-dimensional embedding space it is not easy to gain an intuitive understanding of how these embedding spaces are organised.
We can however apply dimensionality reduction techniques to visualise the high-dimensional data in a lower-dimensional space.
t-SNE~\cite{maaten_visualizing_2008} is a non-linear dimensionality reduction technique for the purpose of this task, specifically designed for data exploration and visualisation of high-dimensional data.
The goal is to find a lower-dimensional mapping of higher-dimensional data while preserving relative distances between data points.
Figure~\ref{fig:tsne} shows a t-SNE visualisation of the AWEs produced by the \tablesystem{ContrastiveRNN} on Hausa data before and after adaptation.
In this curated example, we see how some of the words that are clustered together by the multilingual model (e.g. “amfani” and “hankali”) are separated after adaptation.
This observation supports the results in Table~\ref{tbl:multi_adapt} that shows, exposing the multilingual model to the target zero-resource language benefits the model by learning some peculiarities from the specific language to improve the organisation of embeddings in vector space.

\mysection{Chapter summary}{Chapter summary}
\label{sec:summary}

In this chapter we explored the benefit of using some unlabelled speech data available from a zero-resource language to cater a pretrained multilingual AWE model to that specific language.
Our results indicate that the multilingual transfer approach (Section~\ref{ssec:multilingual}) along with unsupervised adaptation (proposed in this chapter) is beneficial, especially when implemented with the \system{ContrastiveRNN} AWE model which showed the highest performance increase after adaptation compared to the \system{CAE-RNN} and \system{SiameseRNN}.
In the following chapter, we further investigate multilingual models, where we investigate the choice of multilingual training language for specific zero-resource languages.

\mychapter{Impact of language choice in a multilingual transfer setting}{Impact of Language choice in a Multilingual Transfer Setting}
\label{chap:related}

In the previous chapter, we showed how the performance of an AWE model can improve on a zero-resource language by adapting the model to that particular language using unlabelled data from that language.
In this chapter, we again try to improve the performance of a multilingual model on a zero-resource language, this time considering the choice of training languages in the multilingual transfer setting.
{We perform our experiments on a speech corpus containing under-resourced languages spoken in South Africa.}
{We hypothesise that applying a multilingual model trained on one set of languages to a target language would achieve different results from a model trained on another set of languages.
Specifically, we investigate the difference in performance when training on languages that are more related to the target than others, and the amount of training data required to achieve competitive results.
With such knowledge, practitioners could save time and money on collecting resources when implementing a multilingual AWE model for a specific target zero-resource language.}

\begin{tcolorbox}[width=\linewidth, colback=white!95!black, boxrule=0.5pt]
	\small
	\textit{The work in this chapter was presented at Interspeech:} \\
	C. Jacobs and H. Kamper, ``Multilingual transfer of acoustic word embeddings improves when training on languages related to the target zero-resource language,'' in \textit{Proceedings of Interspeech}, 2021. 	
\end{tcolorbox}




\mysection{Related work}{Related work}
\label{sec:related_related}

The impact training languages have on a particular target language, in a multilingual transfer setting, has been observed by several studies in general speech processing (i.e. unrelated to AWEs).
Recent work in~\citet{van_der_westhuizen_multilingual_2021} considers the choice of training languages in a multilingual ASR model.
They perform their experiments in a similar setting to the multilingual transfer setting we explored in the previous chapter.
However, in contrast to having no available labelled data from the target zero-resource languages, they have access to a limited amount of transcribed speech data for each of the target languages.
With a total of six training languages, they train various combinations of multilingual models for each target language, always including the training data from the target language.
They show that any combination of two or more languages (where one is the target language) outperform a monolingual model trained on the target language alone.
This is similar to our findings in Table~\ref{tbl:multi}, showing a model trained on labelled data from multiple languages outperforms a model trained on unlabelled data (instead of limited labelled data) from the target language itself.
Importantly, although adding all languages improves over a monolingual model, they show a significant improved performance can be achieved by a more careful selection of training languages: best results are achieved for both target languages when only training a multilingual model with one additional language to the target language.
While some of the languages are phonetically more similar to the target languages than others, the impact thereof was not observed.
{In broad terms, languages are said to be phonetically similar when they share similar contrastive sound units (phonemes), which are the individual sounds that make up all words in a language.}   
Although the benefit of using related languages were not observed by \citet{van_der_westhuizen_multilingual_2021}, we identify a more careful selection of training languages in a multilingual transfer setting might be beneficial.

In contrast to the results of \citet{van_der_westhuizen_multilingual_2021}, not observing the benefit of related language in a multilingual ASR model, \citet{yi_language-adversarial_2019} however shows performance increase when adding languages related to the target in a multilingual ASR model.
They also show the benefit of adding data from more languages, where a multilingual ASR model trained with up to twelve languages outperform all models trained on fewer languages on all the target languages.
Importantly, different to \citet{van_der_westhuizen_multilingual_2021}, they do not include any data from the target languages when training the multilingual models.

In another task, spoken language identification (LID)---identifying the language spoken in a speech recording--- van der Merwe~\cite{van_der_merwe_triplet_2020} and \citet{bedyakin_low-resource_2021} show that more confusion occurs between the identity of languages that belong to the same language family.
Interestingly to us, the experiments in \citet{van_der_merwe_triplet_2020} are performed on the same dataset we use to perform the experiments in this chapter to train multilingual AWE models (see Section~\ref{ssec:related_data}).
Van der Merwe~\cite{van_der_merwe_triplet_2020} shows that most confusion occurs between languages from the same family, for example, Zulu and Xhosa, belonging to the Nguni language family, and Afrikaans and English, belonging to the West-Germanic family (see Section~\ref{sec:related_africa}), with the least confusion observed between language from far related languages, for example, Zulu and Afrikaans.
This might be related to words shared among the vocabularies of two closely related languages, but also the linguistic characteristics of related languages since for example, there doesn't exist much overlap in words between Afrikaans and English.
We further investigate this matter in Section~\ref{ssec:related_data}.

Although the impact of the family to which a language belong have been observed in other speech tasks, it has not been investigated extensively for AWEs.
Preliminary experiments in Kamper et al.~\cite{kamper_improved_2021} showed improved scores when training a monolingual AWE model on one language and applying it to another from the same language family.
But this has not been investigated thoroughly or with multilingual models, leaving several unanswered questions.


In this chapter, we answer some of the following questions:  Does the benefit of training on related languages diminish as we train on more languages (which might or might not come from the same family as the target zero-resource language)? When training exclusively on related languages, does performance suffer when adding an unrelated language? Should we prioritise data set size or language diversity when collecting data for multilingual AWE transfer?
To answer these questions we use a corpus of under-resourced languages spoken in South Africa.
In the following section, we describe these languages and how they can be used to benefit the performance of a specific language.

\mysection{Languages of South Africa}{Languages of South Africa}
\label{sec:related_africa}

\begin{figure}[!t]
	\centering
	\begin{subfigure}[b]{0.24\textwidth}
		\centering
		\includegraphics[width=\textwidth]{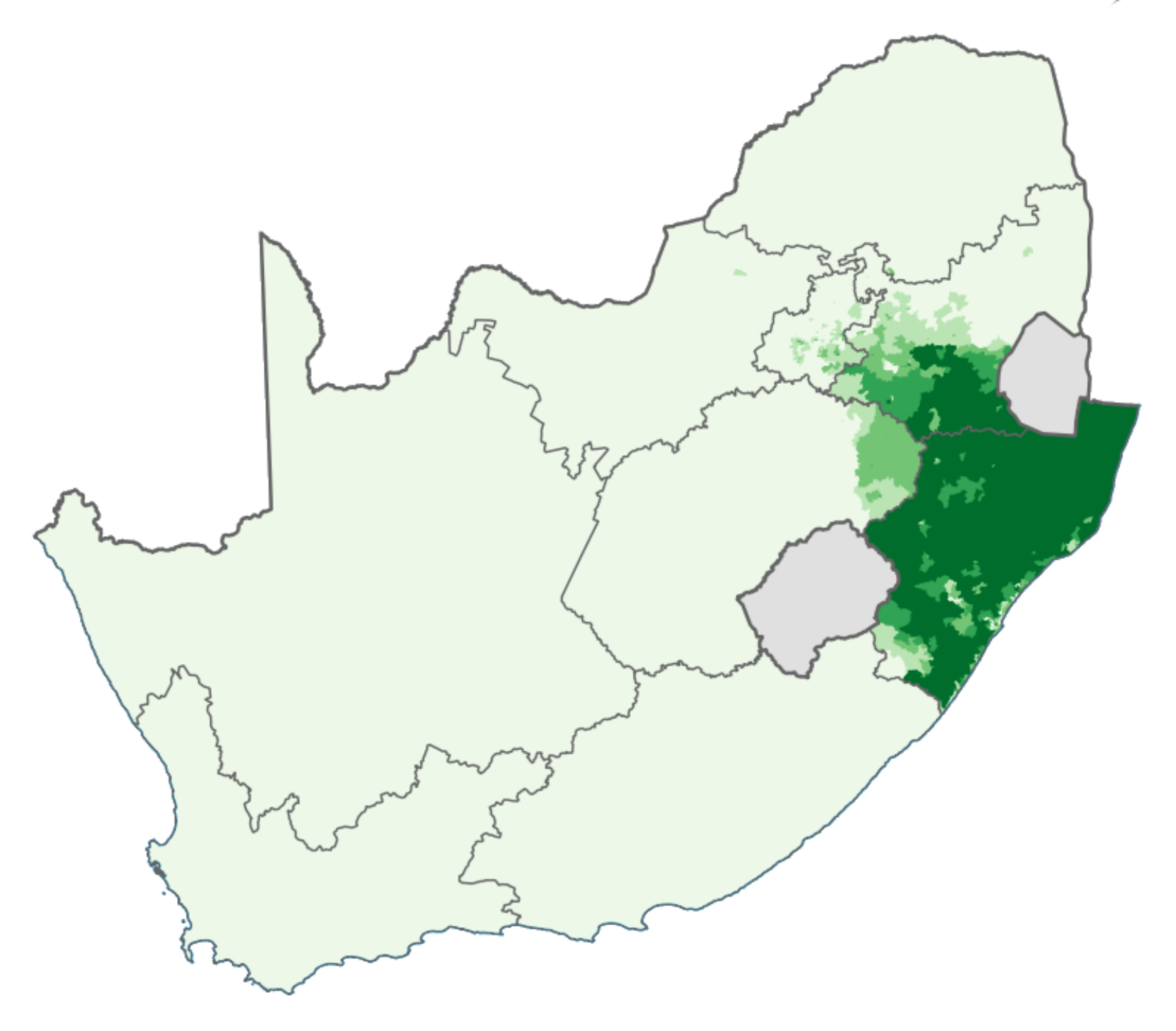}
		\caption{isiZulu}
		\label{fig:map_zul}
	\end{subfigure}
	\hfill
	\begin{subfigure}[b]{0.24\textwidth}
		\centering
		\includegraphics[width=\textwidth]{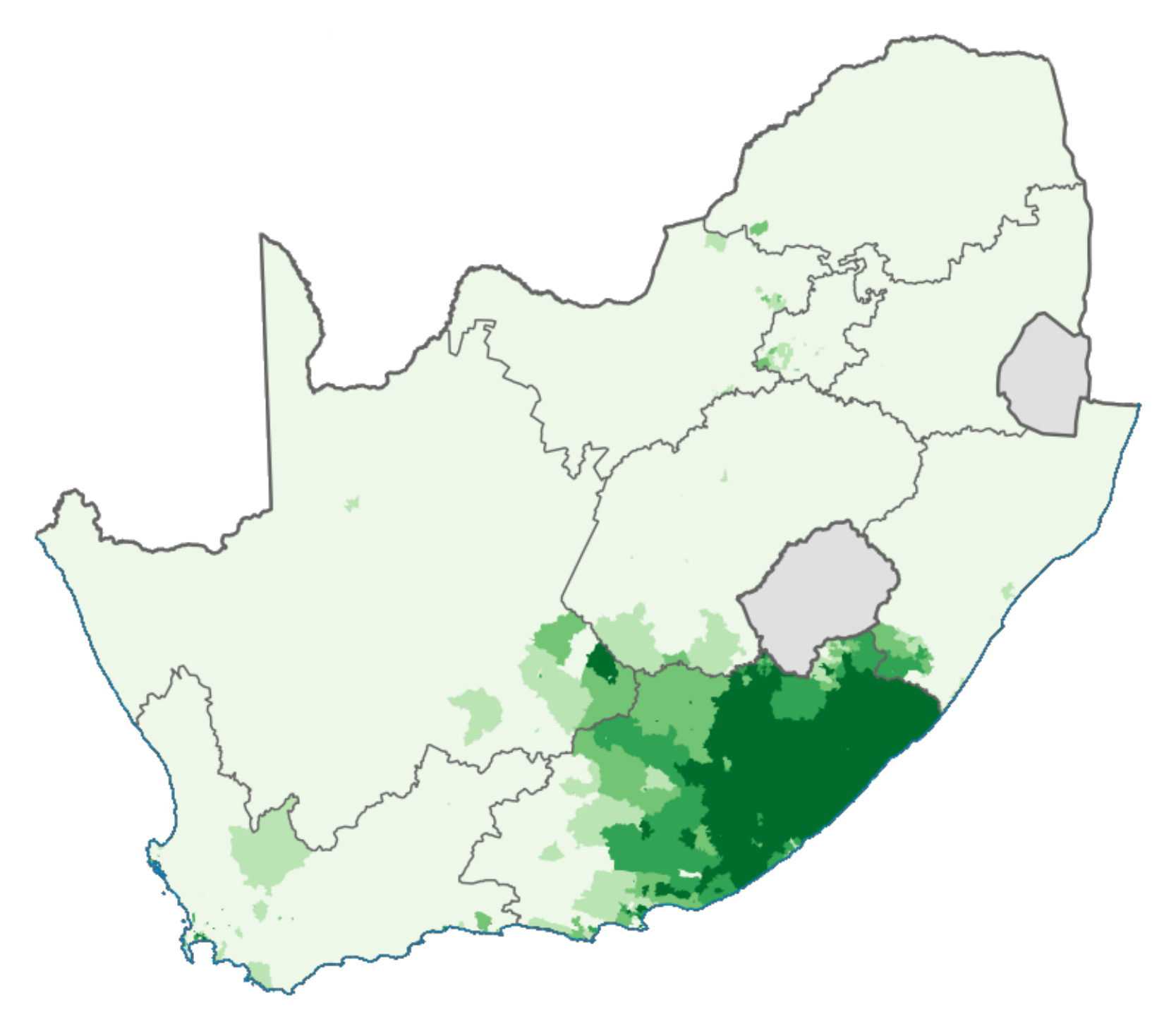}
		\caption{isiXhosa}
		\label{fig:map_xho}
	\end{subfigure}
	\hfill
	\begin{subfigure}[b]{0.24\textwidth}
		\centering
		\includegraphics[width=\textwidth]{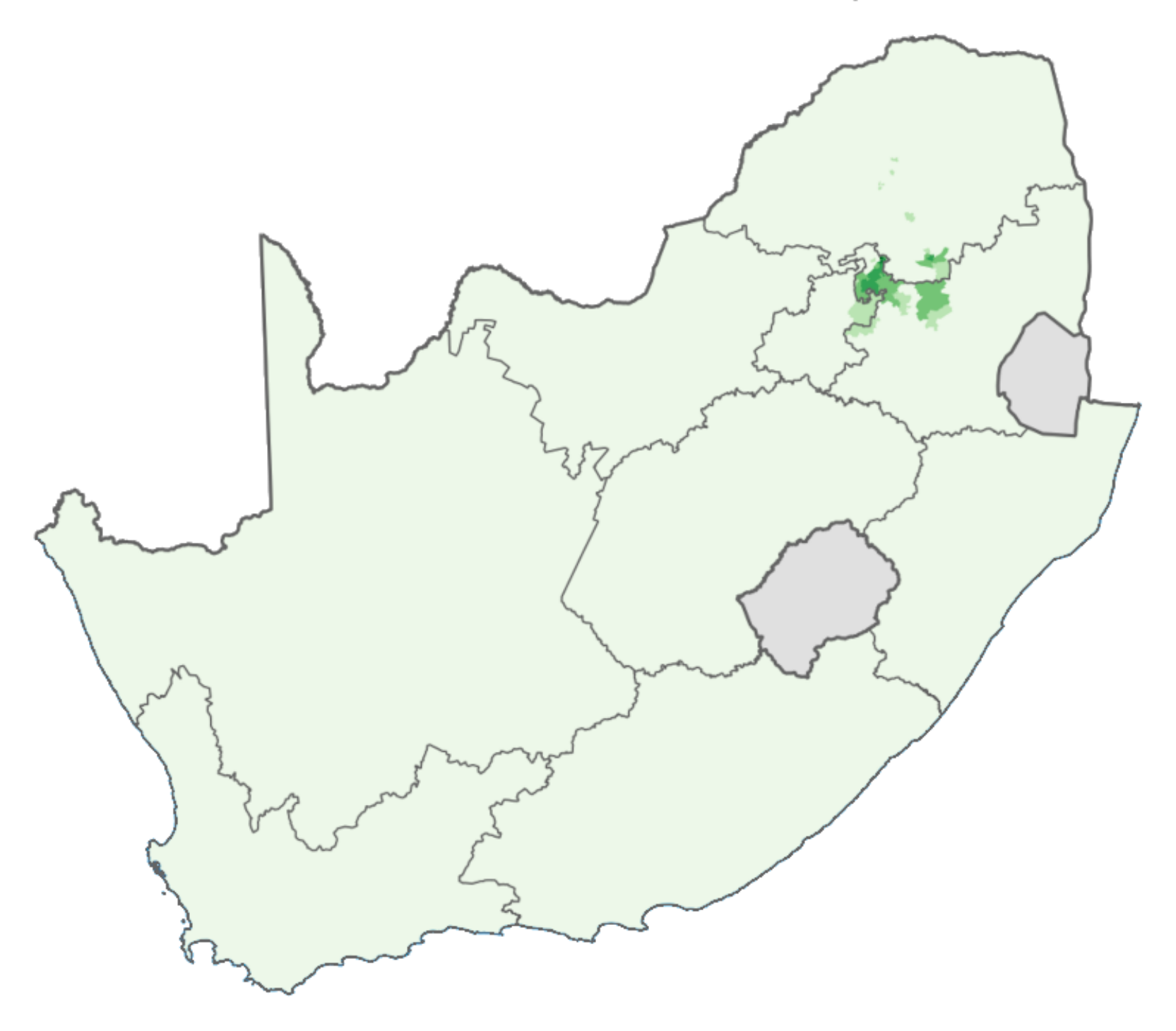}
		\caption{isiNdebele}
		\label{fig:map_nbl}
	\end{subfigure}
	\hfill	
	\begin{subfigure}[b]{0.24\textwidth}
		\centering
		\includegraphics[width=\textwidth]{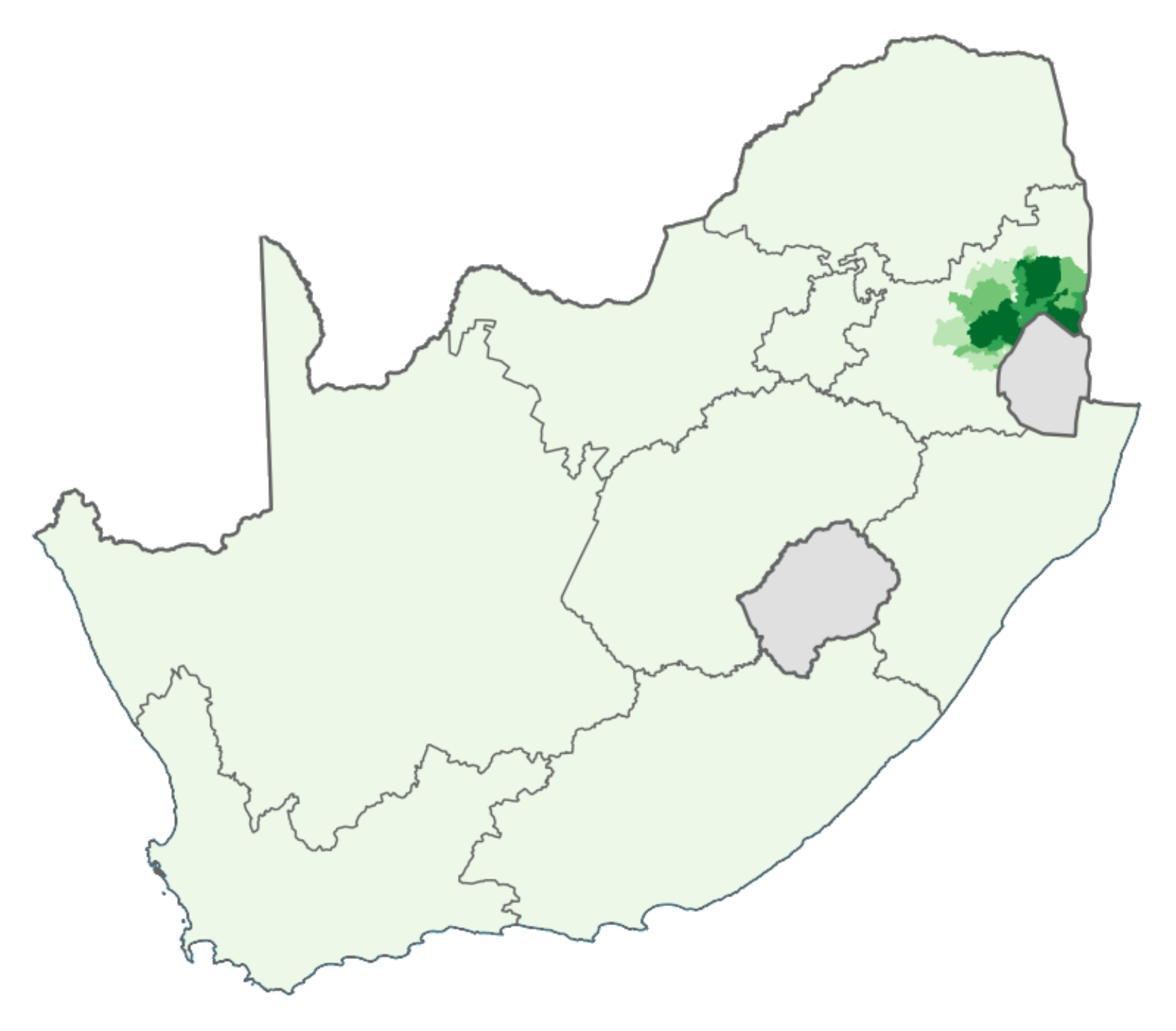}
		\caption{siSwati}
		\label{fig:map_ssw}
	\end{subfigure}
	\newline
	\vfill
	\begin{subfigure}[b]{0.24\textwidth}
		\centering
		\includegraphics[width=\textwidth]{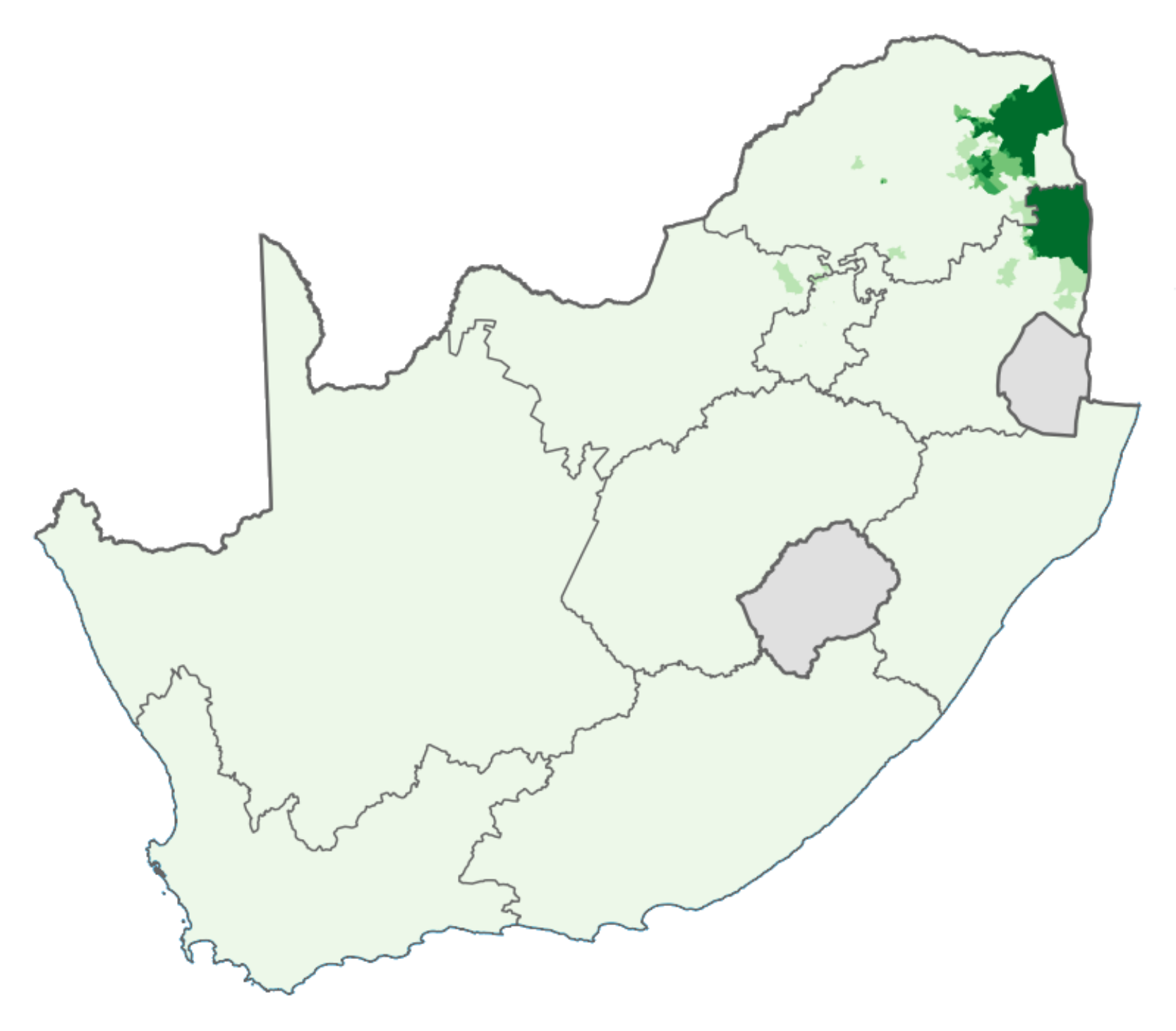}
		\caption{Xitsonga}
		\label{fig:map_tso}
	\end{subfigure}
	\hfill
	\begin{subfigure}[b]{0.24\textwidth}
		\centering
		\includegraphics[width=\textwidth]{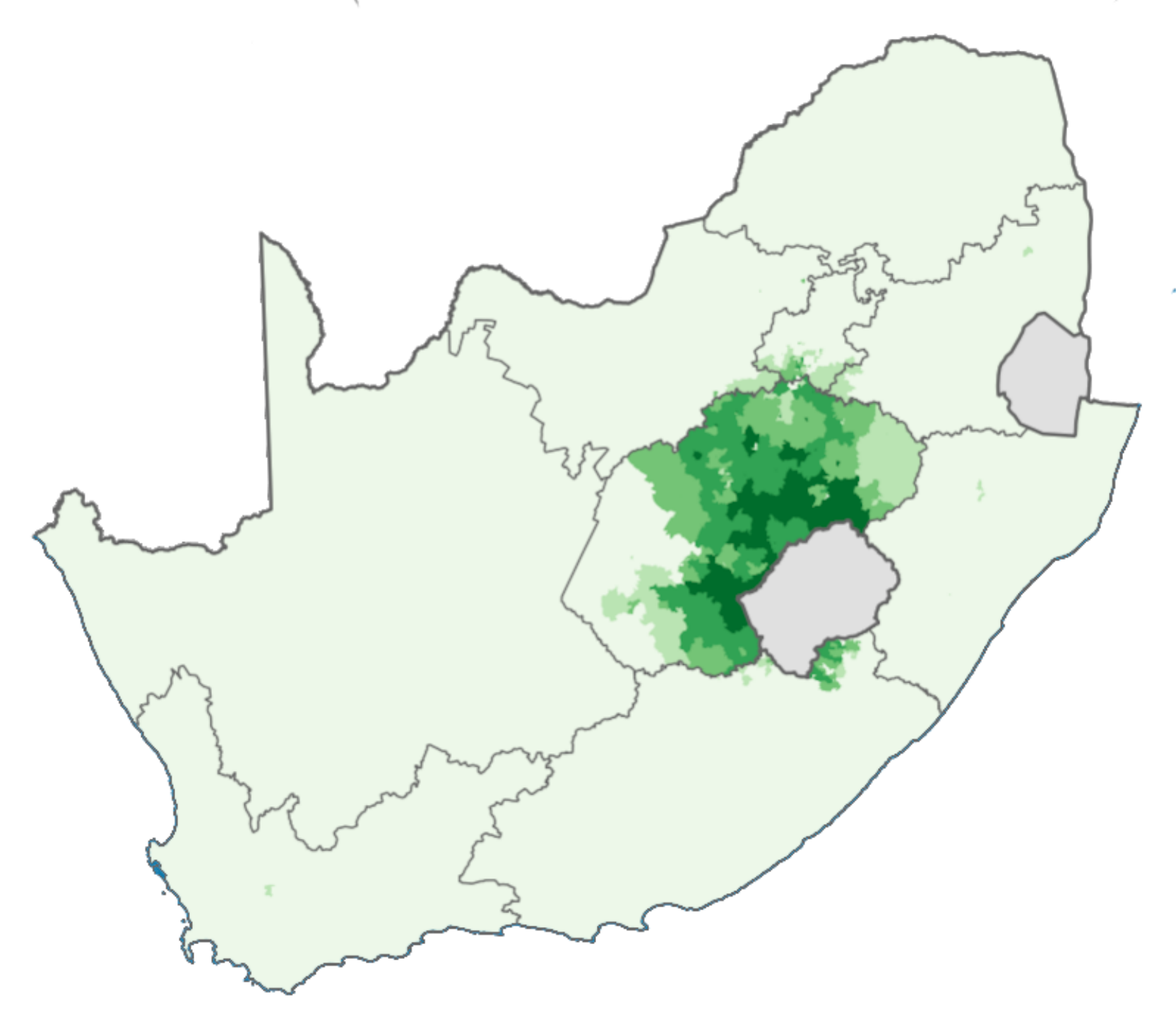}
		\caption{Sesotho}
		\label{fig:map_sot}
	\end{subfigure}
	\hfill
	\begin{subfigure}[b]{0.24\textwidth}
		\centering
		\includegraphics[width=\textwidth]{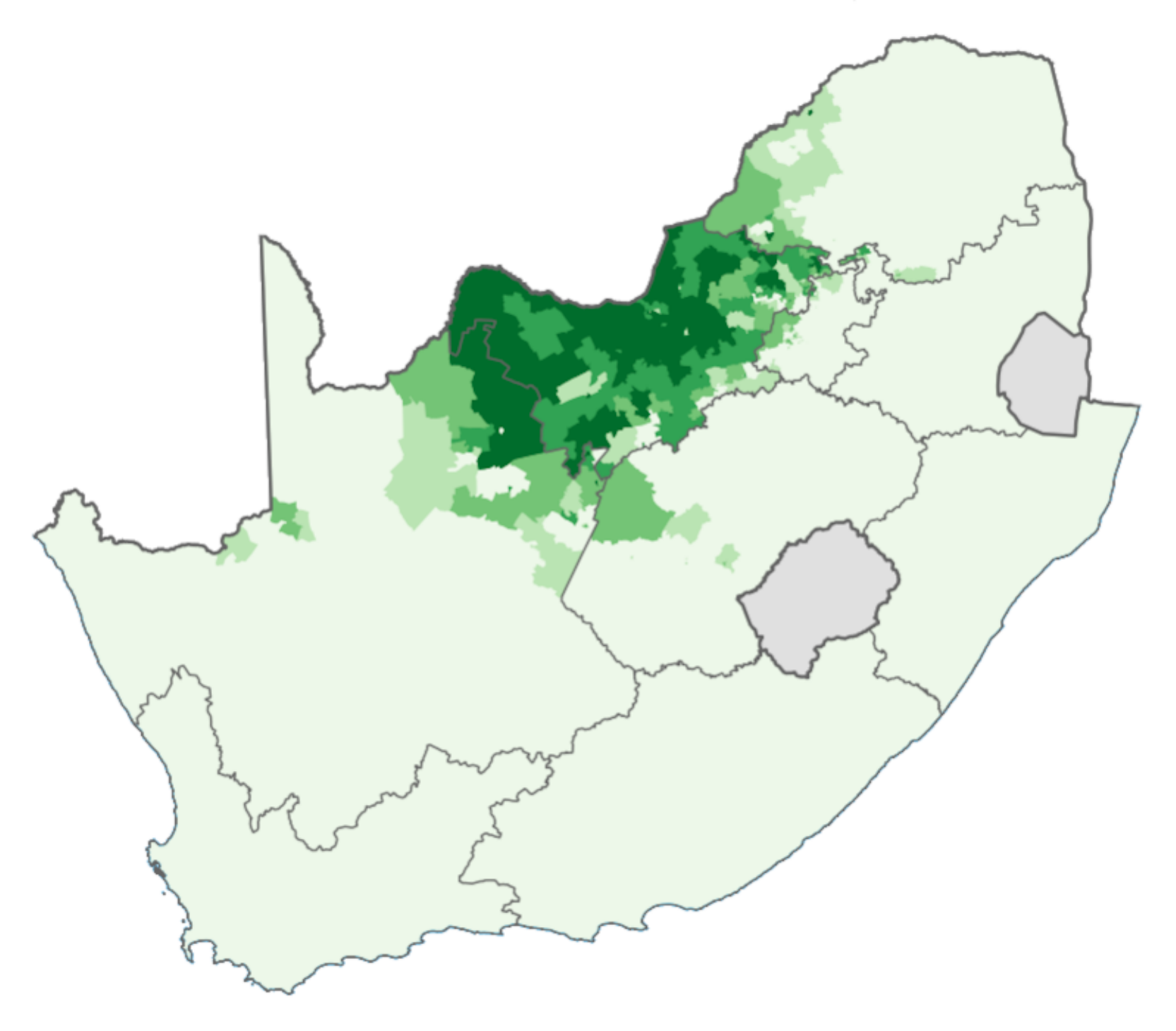}
		\caption{Setswana}
		\label{fig:map_tsn}
	\end{subfigure}
	\hfill
	\begin{subfigure}[b]{0.24\textwidth}
		\centering
		\includegraphics[width=\textwidth]{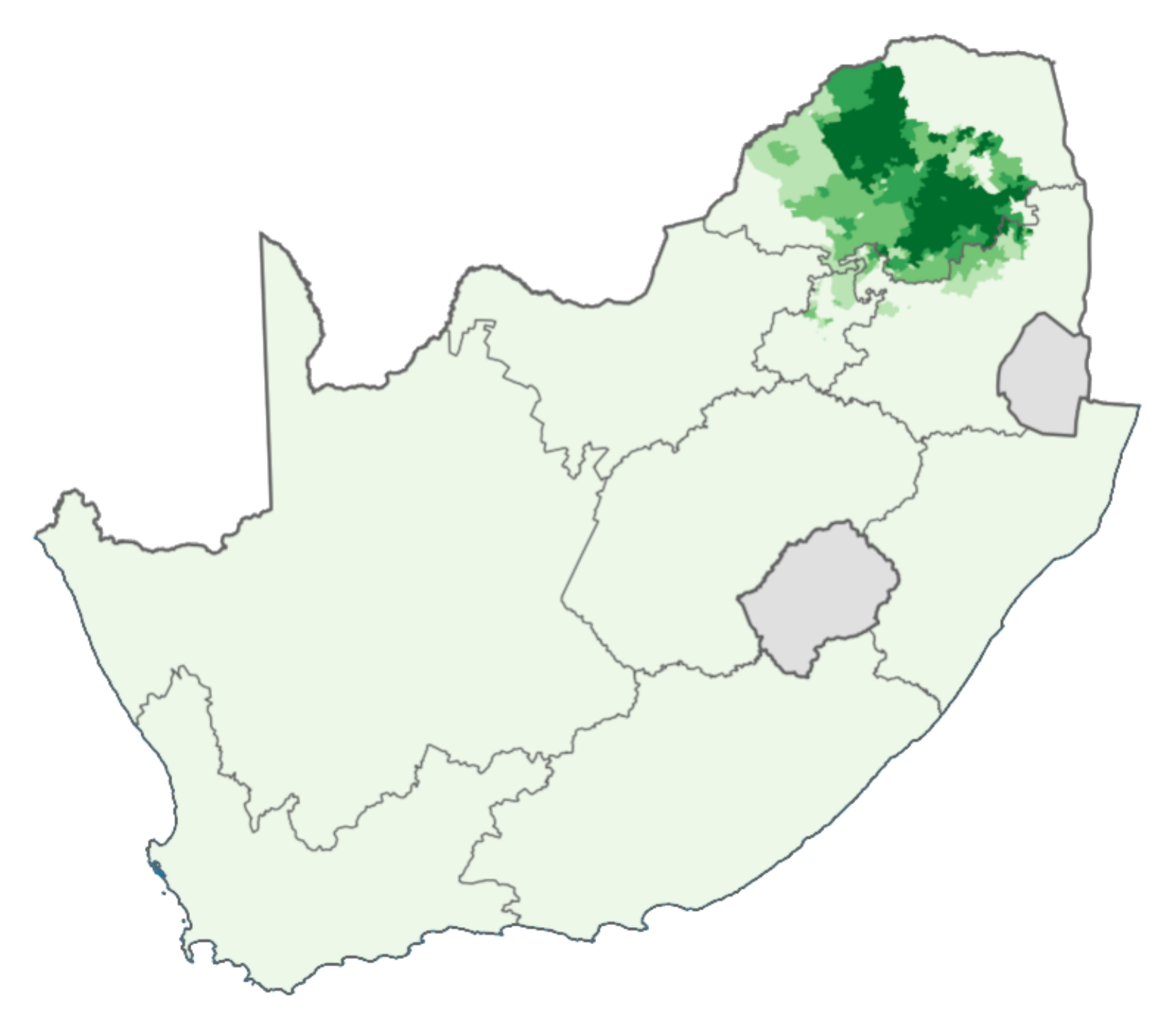}
		\caption{Sepedi}
		\label{fig:map_nso}
	\end{subfigure}
	\newline
	\vfill
	\begin{subfigure}[b]{0.24\textwidth}
		\centering
		\includegraphics[width=\textwidth]{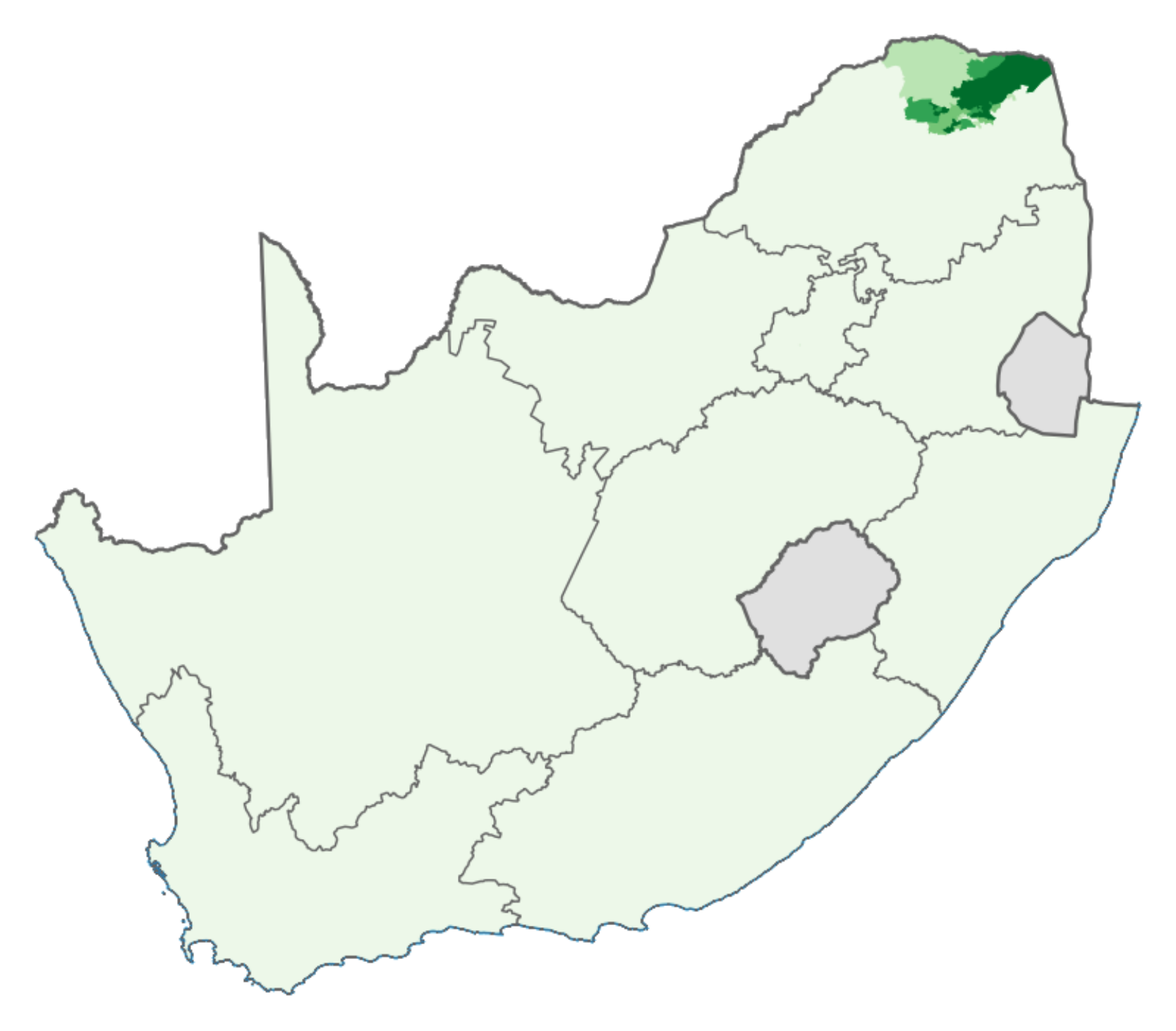}
		\caption{Tshivenda}
		\label{fig:map_ven}
	\end{subfigure}
	\hfill
	\begin{subfigure}[b]{0.24\textwidth}
		\centering
		\includegraphics[width=\textwidth]{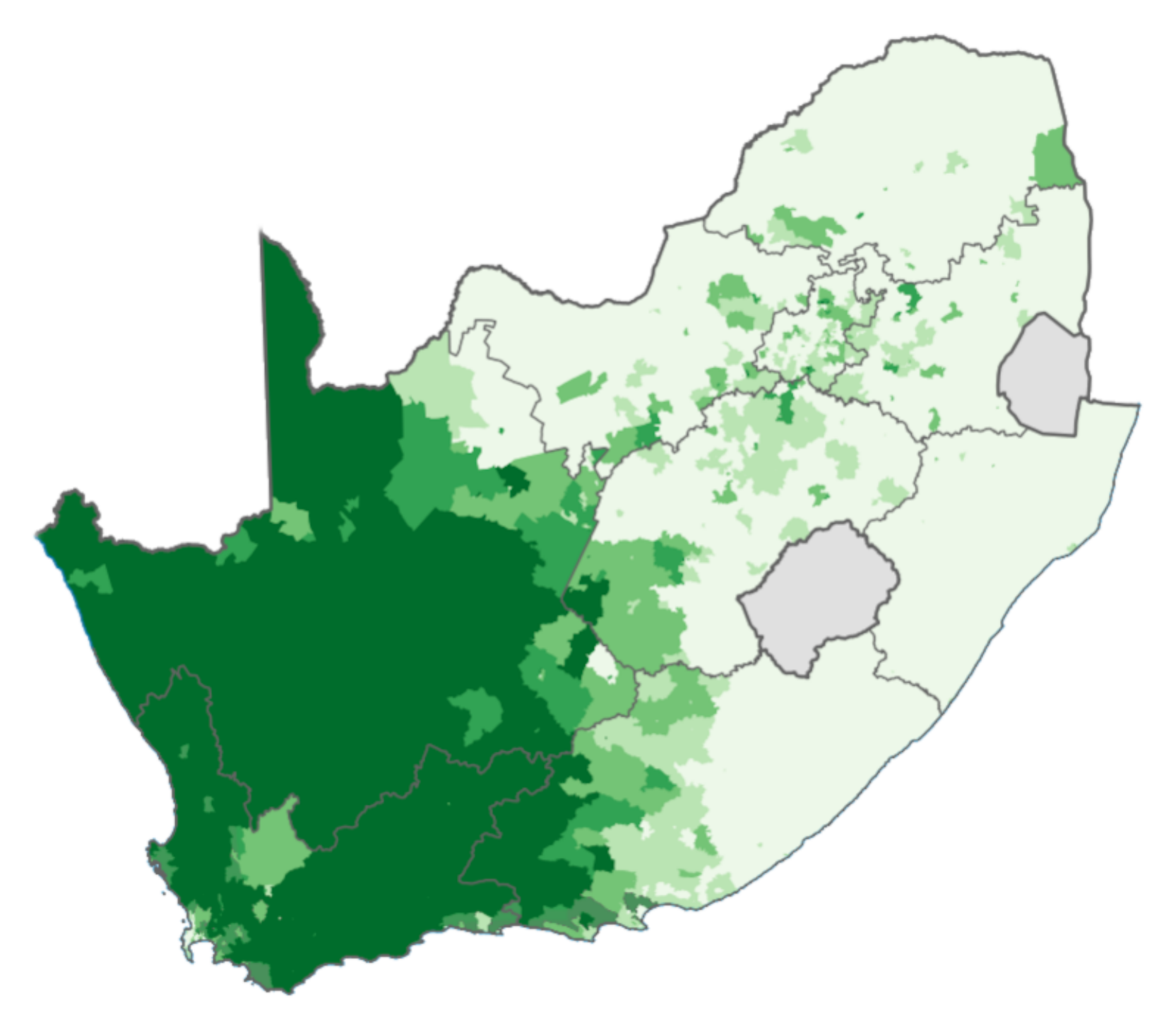}
		\caption{Afrikaans}
		\label{fig:map_afr}
	\end{subfigure}
	\hfill
	\begin{subfigure}[b]{0.24\textwidth}
		\centering
		\includegraphics[width=\textwidth]{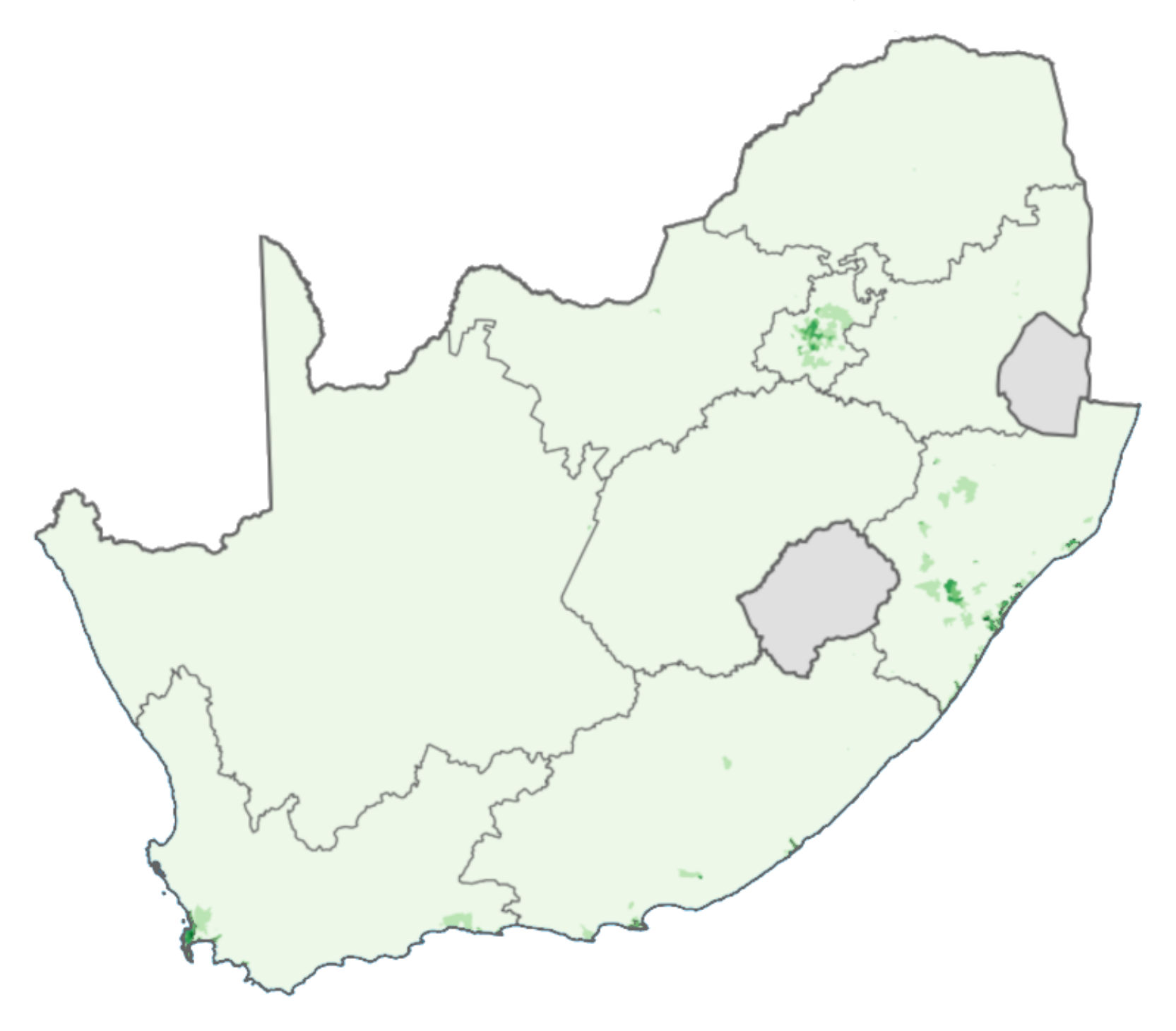}
		\caption{English}
		\label{fig:map_eng}
	\end{subfigure}
	\hfill
	\begin{subfigure}[b]{0.24\textwidth}
		\centering
		\includegraphics[width=\textwidth]{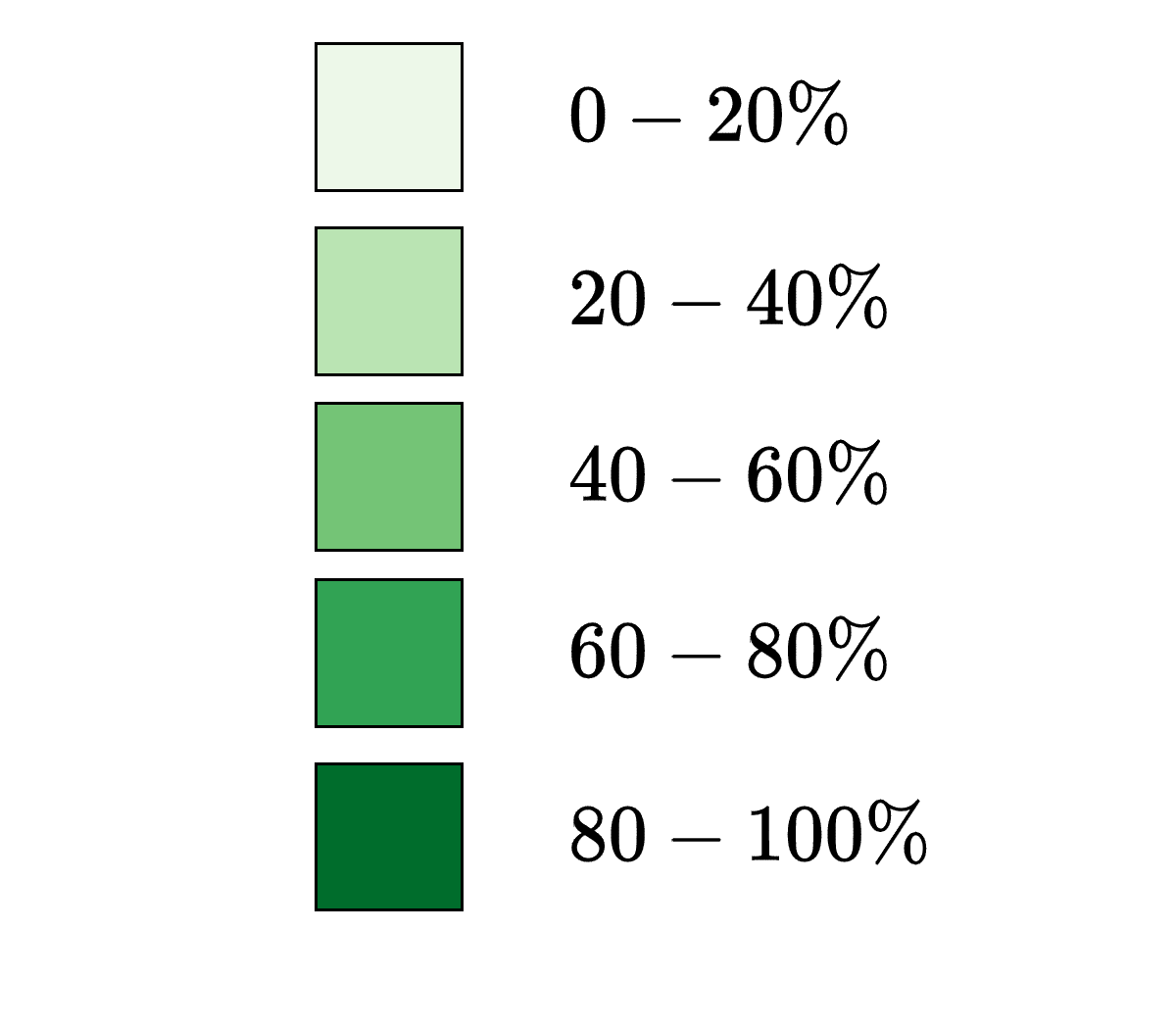}
		\vspace*{2mm}
	\end{subfigure}
	\vspace{2mm}
	\caption[Caption for LOF]{Geographical distributions showing the preferred language of communication in households in South Africa.\footnotemark}
	\label{fig:distr_maps}
\end{figure}

The majority of languages in Africa are considered under-resourced~\cite{orife_masakhane_2020}.
This includes the eleven official languages of South Africa, ten of which are indigenous to South Africa: isiZulu, isiXhosa, isiNdebele, siSwati, Sepedi, Setswana, Sesotho, Xitsonga, Tshivenda and Afrikaans.
The eleventh language, English, is in contrast, the most well-resourced language in the world.
It is the country’s lingua franca and the primary language of government, business, and commerce, bridging the language barrier between native languages among different groups in South Africa.
Figure~\ref{fig:distr_maps} shows the geographical regions where each language is predominantly spoken as first language.

\begin{figure}[!t]
	\centering
	\includegraphics[width=0.95\linewidth]{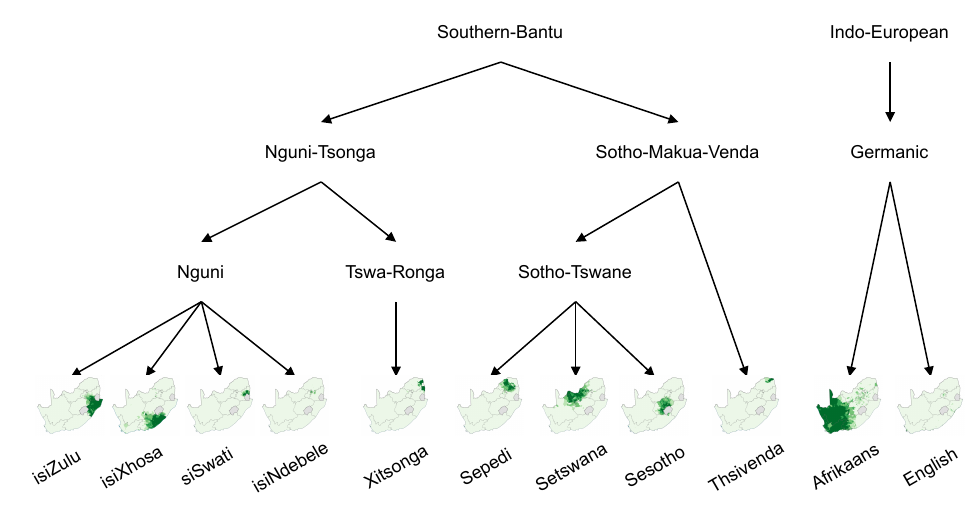}	
	\caption{A family tree for the official South African languages.}
	\label{fig:tree}
\end{figure}

These languages can be grouped into different language families based on the linguistic links they share.
A language family is a group of different languages related through descent from a common ancestral language or parental language, called the protolanguage~\cite{society_family_2020}.
Although some languages do not have a protolanguage
, most languages spoken throughout the world belong to a language family.
The eleven official languages of South Africa are grouped into two primary groups, Southern Bantu and Indo-European, from where the languages are further grouped into sub-families as illustrated in Figure~\ref{fig:tree}.
From the Southern Bantu family there exist two principal branches, Nguni-Tsonga and Sotho-Makua-Venda, to which nine of the languages belong.

\footnotetext{\scriptsize{Images show statistical data from the 2011 South Africa's Census retrieved and adapted from Wikipedia, distributed under the CC-BY 2.0 license.}}

The Nguni-Tsonga group further branches into two groups, Nguni and Tswa-Ronga where 
isiZulu, isiXhosa, siSwati, and isiNdebele belong to the former and Xitsonga to the latter. 
The Nguni languages use a five-vowel system and are characterised by click consonants.
Xitsonga shares similar properties to the Nguni languages but has very few words with click consonants.


From the Sotho-Makua-Venda family, three languages belong to the Sotho-Tswana group: Sepedi, Setswana and Sesotho.
In the Sotho-Makua-Venda subfamily, Venda is somewhat of a standalone.
They use a 9-vowel system and unlike the Nguni languages, Sepedi and Setswana do not have clicks but Sesotho utilises click consonants in some words.

All the languages from the Southern Bantu group are known as tonal languages, where a  word said with different tonal inflexions will convey different meanings.
Moreover, for the Southern Bantu branch, languages grouped into families at the lowest level would (to an extent) be intelligible to a native speaker of another language in the same group.



The other two languages, Afrikaans and English, belong to the Indo-European family and are grouped at a higher level to belong to the set of West-Germanic languages.
Although they are not as similar as some of the Southern Bantu languages, they are far less similar to each other than any of the languages in the Souther Bantu family.


\mysection{Query-by-example speech search}{Query-by-example speech search}
\label{sec:related_qbe}

In addition to the word discrimination task measuring the intrinsic quality of the AWEs (Section~\ref{sec:background_evaluation}), we apply AWE models to a speech task, Query-by-Example (QbE) (Section~\ref{ssec:background_qbe}), to measure an AWE model's performance in an actual zero-resource downstream application.   
In contrast to the word discrimination task, QbE speech search does not assume a test set of isolated words but instead operates on full unsegmented utterances.
Our goal is not to deliver competitive QbE results compared to existing work. Instead, we use this task as a supplementary evaluation task to support the word discrimination results as discussed in Section~\ref{sec:background_evaluation}.
We hope that the QbE results would correlate with the AP performance of the same-different word discrimination task (Section~\ref{sec:background_evaluation}).

\begin{figure}[!t]
	\centering
	\includegraphics[width=0.90\linewidth]{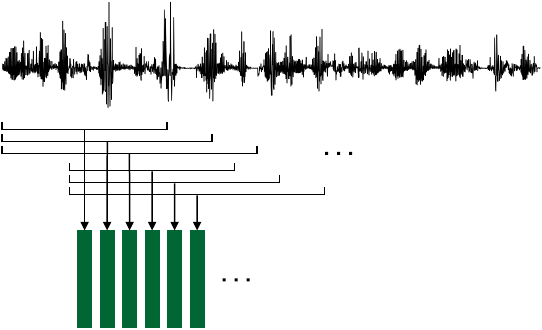}
	\vspace{2mm}
	\caption[Caption]{A search utterance is split into overlapping
		segments which are embedded individually using the same AWE model.}
	
	\label{fig:related_qbe}
\end{figure}

We implement a simplified QbE system used in Kamper et al.~\cite{kamper_semantic_2019}.
This system aims to retrieve utterances containing a given spoken query word from a speech collection containing multiple speech utterances.
If we knew the word boundaries in the search collection, we could embed each of the words in an utterance and simply look up the closest embeddings to the query. 
Instead, because we do not have word boundaries in the zero-resource setting, each utterance is split into overlapping segments from some minimum to some maximum duration as illustrated in Figure~\ref{fig:related_qbe}. 
Each segment from each utterance is then embedded separately using the same AWE model.
The spoken query word is also embedded using the same AWE model.
To do the QbE task, the query embedding is compared to each of the utterance sub-segment embeddings (using cosine distance), and the minimum distance over the utterance is then taken as the score for whether the utterance contains the given query. 

\mysection{Experimental setup}{Experimental setup}
\label{sec:setup}

Here we describe the setup for training multilingual AWE models for the purpose of investigating the impact of related languages.    
We also give the implementation details for the QbE system (described above in~Section~\ref{sec:related_qbe}) and how we evaluate the system retrieval accuracy. 


\subsection[Data]{Data}
\label{ssec:related_data}

We perform all our experiments on the NCHLT speech corpus~\cite{barnard_nchlt_2014}, which provides wide-band speech from each of the eleven official South African languages (Section~\ref{sec:related_africa}). 
We use a version of the corpus where all repeated utterances are removed, leaving roughly 56 hours of speech from around 200 speakers in each language with an equal split between males and females.
We use the default training, validation and test sets from which word segments are extracted using forced alignments.
After preliminary experiments and inspecting the extracted word segments, we added some constraints on the duration of word segments to be considered for this experiment.
Nguni languages follow a conjunctive writing system and the Sotho languages follow a disconjunctive writing system such that the average word-length of Bantu languages are much longer in duration than Sotho languages~\cite{probert_word_2016}. 
We modified all the sets to only contain words with a minimum number of characters and frames for each language as shown in Table~\ref{tbl:related_data_splits}.
We observe that, although the number of training items for each language is somewhat equal, the number of unique words in each language is quite different.
For example,  Xhosa and Afrikaans have similar number of training items, 95k and 98k, but the number of unique word types within the set is 26.4k compared to 7.5k. 
We will observe this effect in our experiments in the following section. 

\begin{figure}[!t]
	\centering
	\includegraphics[width=0.9\linewidth]{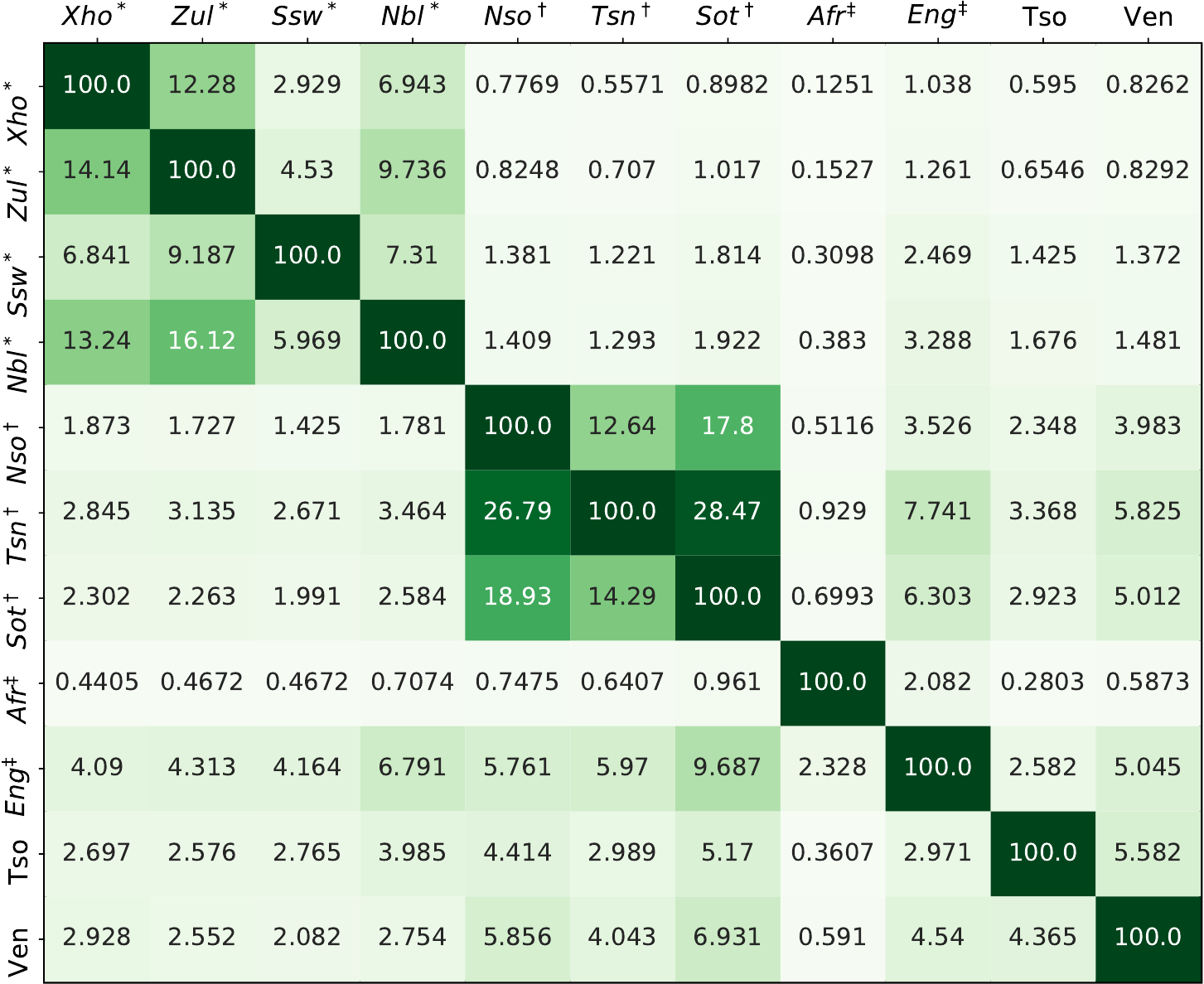}	
	\caption{The percentage of words in the training set of each language (row) that appear in the training set of every other language (column).}
	\label{fig:overlap}
\end{figure}

\begin{table}[!t]
	\mytable
	\caption{The number of word segments for each language in every set (train, development, test) after applying word constraints.}
	\footnotesize
	\begin{tabularx}{1\linewidth}{L c@{\extracolsep{3mm}}cccccccc}
		\toprule
		& \multicolumn{2}{c}{Constraints} & \multicolumn{3}{c}{No. items} & \multicolumn{3}{c}{No. types}\\
		\cline{2-3}		\cline{4-6} \cline {7-9}
		Language & Min. chars & Min. frames & Train & Val & Test & Train & Val & Test \\
		\midrule       
		{Xhosa} &5 &50 &95k &7k &6.4k &26.4k &4.9k &4.2k\\
		{isiZulu} &5 &50 &103k &8k &7.6k &22.9k &6.3k &4.9k \\
		{siSwati} &5 &50 &97k &8k &7.2k &11.3k &4.1k &3.8k\\
		{isiNdebele} &5 &50 &110k &9k &9.3k &13.8k &5k &4.8k\\
		{Sepedi} &3 &30 &86k &5.2k &4.9k &10.9k &1.7k &2.3k\\
		{Setswana} &4 &40 &110k &6.4k &5.7k &5.2k &2.2k &1.1k\\
		{Sotho} &3 &30 &100k &5.9k &5.4k &10.2k &2.4k &2.5k\\
		{Afrikaans} &5 &50 &98k &6.7k &6.2k &7.5k &3.1k &2.8k\\
		{English} &5 &50 &116k &5k &5k &6.7k &2.2k & 2.5k\\
		{Xitsonga} &3 &30 &116k &7.6k &8.4k &5.8k &2.1k & 2.1k\\
		{Tshivenda} &3 &30 &117k &8k &7.1k &7.4k &2.2k &3.3k\\    
		\bottomrule		
	\end{tabularx}
	\label{tbl:related_data_splits}
\end{table}
We recognise that languages from the same language family, especially the ones related at the lowest level (Figure~\ref{fig:tree}), might share words among their vocabulary.
As an initial investigation of language similarity, we therefore examine the amount of overlap between the training sets of all languages.
Figure~\ref{fig:overlap} shows the percentage of unique words for each language that also appear in the training set of every other language.
As expected, we observe the most overlap between languages that belong to the same language family, illustrated by the darker shade blocks around the diagonal.
For most of the language, with the exception of English and Afrikaans, limited text resources are available on the internet.
Therefore, a large part of the text was obtained from South Africa's government website, containing content in all the official languages.
Given the nature of the prompts, many English words are present in those recordings and are noticed by the consistent overlap between English words in the training sets of all the other languages.



Of the eleven languages, we treat six of them as well-resourced in our experiments that follow: Xhosa (Xho$^*$), siSwati (Ssw$^*$), isiNdebele (Nbl$^*$), Sepedi (Nso$^\dagger$), Setswana (Tsn$^\dagger$) and English (Eng$^\ddagger$).\footnote{We use superscripts to indicate the different language families: $^*$Nguni, $^\dagger$Sotho-Tswana, $^\ddagger$Germanic.}
We use labelled data from the well-resourced languages to train a single supervised multilingual model and then apply the model to the target zero-resource languages.
More specifically, we extract 100k true positive word pairs using forced alignments for each training language.
We select Zulu (Zul$^*$), Sotho (Sot$^\dagger$), Afrikaans (Afr$^\ddagger$), and Xitsonga (Tso) as our zero-resource languages and use another language, Tshivenda (Ven), for validation of each model.
Training and evaluation languages are carefully selected such that for each evaluation language at least one language from the same family is part of the training languages, except for Tso which is in a group of its own.

\subsection[Models]{Models}
\label{ssec:related_models}

We use the \system{ContrastiveRNN} presented in Chapter~\ref{chap:contrastive} as the embedding function for all experiments in this chapter.
The \system{ContrastiveRNN} configuration and hyperparameter selection is the same as the  multilingual \system{ContrastiveRNN} models trained in Chapter~\ref{chap:contrastive}.
The encoder unit consists of three unidirectional RNNs with 400-dimensional hidden vectors, with an embedding size of $M = 130$ dimensions.
All models are optimised using Adam optimisation~\cite{kingma_adam_2017} with a learning rate of $0.001$. 
The temperature parameter $\tau$ in Equation~\ref{eqn:contrastive_loss} is set to $0.1$.

\subsection[Evaluation]{Evaluation}
\label{ssec:related_eval}

We again use the same-different word discrimination task (Section~\ref{sec:background_evaluation}) to measure the quality of the AWEs. 
Additionally, we perform QbE analysis to measure the performance of the AWEs in a downstream zero-resource task (Section~\ref{sec:related_qbe}).

\subsubsection[Word discrimination]{Word discrimination}
\label{sssec:related_eval_samediff}
As a recap from Section~\ref{sec:background_evaluation}, to evaluate a particular AWE model, a set of isolated test word segments is embedded.
The number of word pairs in the test set for each language is shown in Table~\ref{tbl:related_data_splits}.
For every word pair in this set, the cosine distance between their embeddings is calculated.
Two words can then be classified as being of the same or different type based on some distance threshold, and a precision-recall curve is obtained by varying the threshold.
The area under this curve is used as the final evaluation metric, referred to as the average precision (AP).

\subsubsection[Query-by-example]{Query-by-example}
\label{sssec:related_eval_qbe}

We evaluate a particular AWE in the QbE speech system described in Section~\ref{sec:related_qbe}.
For each evaluation language, we use approximately two hours of test utterances as the search collection.
Sub-segments for the utterances in the speech collection are obtained by embedding
windows stretching from 20 to 60 frames with a 3-frame overlap. 
For each evaluation language, we randomly draw instances of 15 spoken query word types from a disjoint speech set (the development set---which we never use for any validation experiments) where we only consider query words with at least 5 characters
for Afr and Zul and 3 for Sot.
There are between 6 and 51 occurrences of each query word and appears 110 to 49 times in the search collection.
For each QbE test, we ensure that the relevant multilingual AWE model has not seen any of the search or query data during training or validation.

To evaluate the QbE retrieval accuracy, we calculate the precision at ten ($P@10$), which is the fraction of the ten top-scoring retrieved utterances from the search collection that contains the given query. For example, assume the utterances retrieved by a QbE system that contains the query are at rank 1,2,4,7,8 among the ten top-ranked utterances.
For that single query, the $P@10$ will be $\frac{5}{10}$, given that only five utterances contain the occurrence of the query, while the other five were false retrievals.  
The reported $P@10$ values are the averages calculated over all query types,

\begin{equation}
	\text{average} \quad P@10 = \frac{1}{K} \sum_{k=1}^{K} (P@10)_{k},
	\label{eqn:pat}
\end{equation}
where $K$ is the number of query types and $(P@10)_k$ indicates the average precision at ten for all query items from type $k$.

{As a baseline, we also give the results where DTW (Section~\ref{ssec:background_dtw}) is used directly on the MFCCs of the segmented speech collection. } 

\mysection{Experiments}{Experiments}
\label{sec:results}


\begin{figure}[!b]
	\centering
	\includegraphics[width=0.9\linewidth]{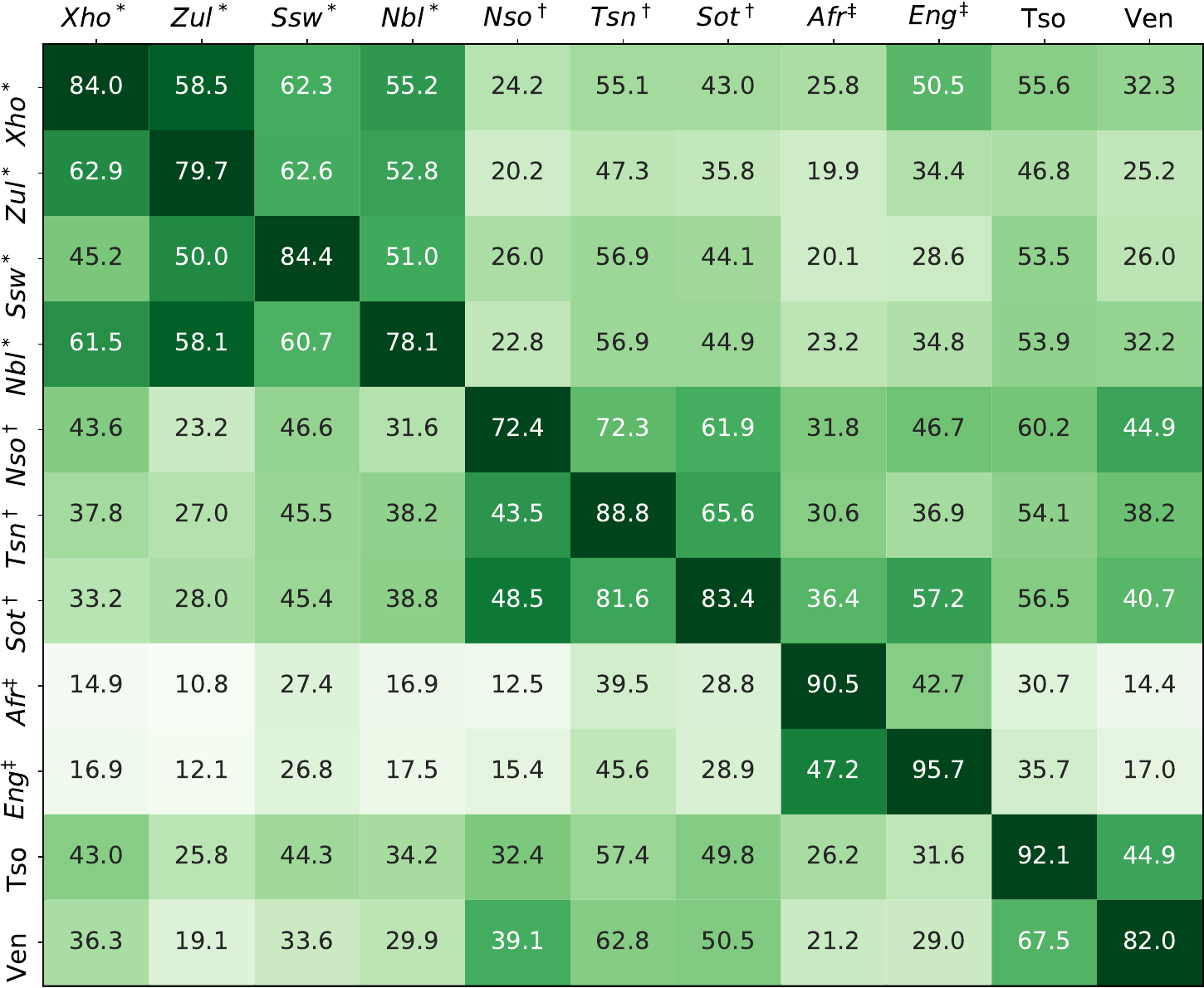}

	\caption{AP (\%) 
		when training a monolingual supervised \tablesystem{ContrastiveRNN} on each language (rows) and then evaluating it on each of the other languages (columns). Heatmap colours are normalised for each evaluation language (i.e. per column).}
	\label{fig:heatmap}
\end{figure}

\subsection{Cross-lingual evaluation}

Before looking at multilingual modelling, we first consider a cross-lingual evaluation where we treat each language as a training language, train a supervised monolingual AWE model, and then apply it to every other language.
This allows us to see the effect of training on related languages in a pairwise fashion.
The results are shown in Figure~\ref{fig:heatmap}.
The result for training and evaluating a model on the same language is given along the diagonal, showing the top-line performance. 
For each evaluation language excluding Tso and Ven, which are in family groups of their own,  
the best results are achieved from models trained on a language from the same family.
For example, on Zul, Xho is the best training language giving an AP of 58.5\%.
Eng is the only exception where the model trained on Sot performs better than using Afr, the other Germanic language.
Although Ven is in its own group at the lowest layer of the family tree in Figure~\ref{fig:tree}, some of the best results when evaluating on Ven are obtained using models trained on Sotho-Tswana languages (Sot, Tsn, Nso), which are in the same family at a higher level.
We also see that for all nine Bantu evaluation languages, the worst performance is obtained from the two Germanic models (Afr, Eng).

\begin{figure}[!t]
	\centering
	\includegraphics[width=0.9\linewidth]{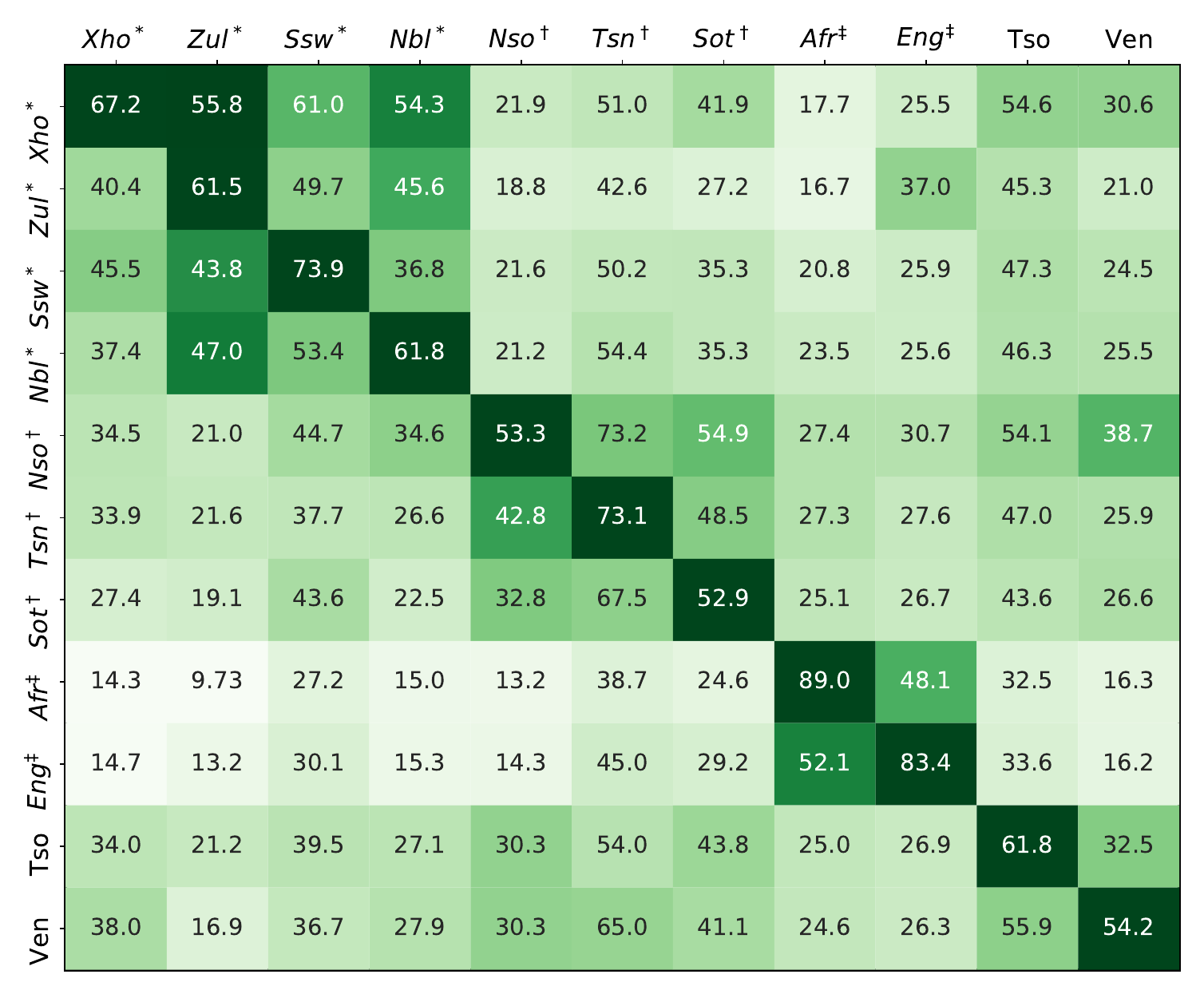}

	\caption{AP (\%) 
		when training a monolingual supervised \tablesystem{ContrastiveRNN} on each language (rows) and then evaluating it on each of the other languages (columns) with overlapping words removed from training and evaluation data. Heatmap colours are normalised for each evaluation language (i.e. per column).}
	\label{fig:heatmap_no_overlap}
\end{figure}

Comparing Figure~\ref{fig:heatmap} to Figure~\ref{fig:overlap}, we notice a high correlation between the AP and the percentage overlap between the training data of two languages.
For example, monolingual models trained on Zul and Nbl with $14.14\%$ and $13.24\%$ overlap gives an AP of $62.9\%$ and $61.5\%$. respectively. 
On the other hand, a model trained on Ssw and evaluated on Xho with only $6.84\%$ overlap gives an AP of $45.2\%$.
Although all three languages belong to the same language family as Xho, the two with a higher percentage overlap outperform the one with less overlap.
Moreover, we suspect that the model trained on Sot outperforms the model trained on Afr (from the same language family as Eng) when evaluating on Eng because of the high number of Eng words present in the training set of Sot, showing a  $9.69\%$ overlap.


This begs the question of whether the increased performance of training on a language related to the target language (Figure~\ref{fig:heatmap}) is related to the phonetic properties shared among them, or because a model is exposed to training examples from the target language.
We investigate this uncertainty by reproducing the experiment yielding the results in Figure~\ref{fig:heatmap}, this time removing the overlapping word segments from the training and validation sets.
The results without overlapping data are displayed in Figure~\ref{fig:heatmap_no_overlap}.
A monolingual model is trained on data on each language where all words that overlap with any other language is removed, then applied to every other language where words from the training language are removed from the evaluation set.

First, we observe a performance decrease when training and evaluating on the same language (shown along the diagonal), especially for the Bantu languages.
Secondly, for most languages, the AP considerably decreased when training a model on one language and applying it to another from the same language family.
Importantly, although the performance decreased, the best results are still achieved by training on related languages, except for the one model where training on Ven outperforms a model trained on Nbl by $0.6\%$ when evaluated on Xho.
Finally, for most languages, the AP decreased on Eng after the Eng words were removed from the training data of all the other languages.
For example, the AP of Sot dropped from $57.2\%$ to $26.7\%$, with Afr (same language family as Eng) now achieving the best AP ($11.1\%$ higher than second best).
This experiment indicates the potential of using related languages in multilingual AWE modelling which we explore in the following section.





\subsection{Multilingual evaluation}

The cross-lingual experiment above was in large part our inspiration for the subsequent analysis of multilingual models.
Concretely, we hypothesise that even better performance can be achieved by training on multiple languages from the same family as the target zero-resource language and that this would be superior to multilingual models trained on unrelated languages.
Focusing on two evaluation languages, Zul and Sot from distinct language families, we
investigate this hypothesis
by training multilingual models with different language combinations, as shown in Table~\ref{tbl:multilingual}.
Here we use the original data without disregarding any particular words since it better simulates a real-world scenario.   
Firstly, we see the best result on a 
language when all the training languages come from the same family as the target.
Secondly, we see how the performance 
gradually decrease as the number of training languages related to the target drop.
Furthermore, notice the performance boost from 
including even just one training language related to the target compared to not including any.
For instance, on Sot we see an increase of more than 12\% absolute when adding just one related language (from 52.5\% and 51.9\% to 64.8\%). 

To further demonstrate the benefit of using training languages from the same family, 
we train a multilingual model for each evaluation 
language on all its related languages using a 10\% subset of the original data.
For both Zul and Sot, 
the subset models outperform the models where no related languages are used. 
E.g.\ on Zul, the Xho+Ssw+Nbl subset model outperforms the full Tsn+Nso+Eng model (no related languages) by more than 20\% in AP.
Moreover, this subset model (58.6\% in AP) even outperforms the Xho+Nso+Eng model (55.7\%) where all the training data from one related language are included and almost matches the AP when using two related languages (Xho+Ssw+Eng, 60.9\%).

These comparisons do more than just show the benefit of training on related languages: they also show that it is beneficial to train on a diverse set of related languages.

\begin{table}[!t]
	\mytable
	\caption{AP (\%) on test data for multilingual models trained on different combinations of well-resourced languages. Models are applied to two zero-resource languages from different language families, Nguni and Sotho-Tswana. For each training language, 100k word pairs were extracted.}
	
	\begin{tabularx}{0.65\linewidth}{Lcc}
		\toprule
		Multilingual model & Zul$^*$ & Sot$^\dagger$ \\
		\midrule
		\underline{\textit{Nguni:}} & &\\[2pt]        
		Xho$^*$ + Ssw$^*$ + Nbl$^*$ &\textbf{68.6} &---  \\
		Xho$^*$ + Ssw$^*$ + Eng$^\ddagger$ &60.9 &---  \\
		Xho$^*$ + Nso$^\dagger$ + Eng$^\ddagger$ &55.7 &---  \\
		Tsn$^*$ + Nso$^\dagger$ + Eng$^\ddagger$ &37.5 &---  \\
		Xho$^*$ + Ssw$^*$ + Nbl$^*$ {(\scriptsize{$10$\% subset})} & 58.6 &--- \\ [2pt]
		\underline{\textit{Sotho-Tswana:}} & & \\ [2pt]
		Nso$^\dagger$ + Tsn$^\dagger $&--- &\textbf{76.7} \\
		Nso$^\dagger$ + Eng$^\ddagger$ &--- &64.8 \\		
		Xho$^*$ + Ssw$^*$ &--- &51.9 \\
		Xho$^*$ + Eng$^\ddagger$ &--- &52.5 \\		 
		Nso$^\dagger$ + Tsn$^\dagger$ {(\scriptsize{$10$\% subset})} &--- & 58.4 \\       
		\bottomrule		
	\end{tabularx}
	\label{tbl:multilingual}
\end{table}

\subsection{Adding more languages} 

In the above experiments, we controlled for the amount of data per language and saw that training on languages from the same family improves multilingual transfer.
But this raises a question:
will adding additional unrelated languages harm performance? 
This question is of practical relevance in cases where related data is either unavailable or the system developer is unsure for which language AWEs will be extracted. 
To answer this, we systematically train two sequences of multilingual models on all six well-resourced languages, 
evaluating each target language as a new training language is added.

\begin{figure}[!t]	
	\centering
	\centerline{\includegraphics[width=.95\linewidth]{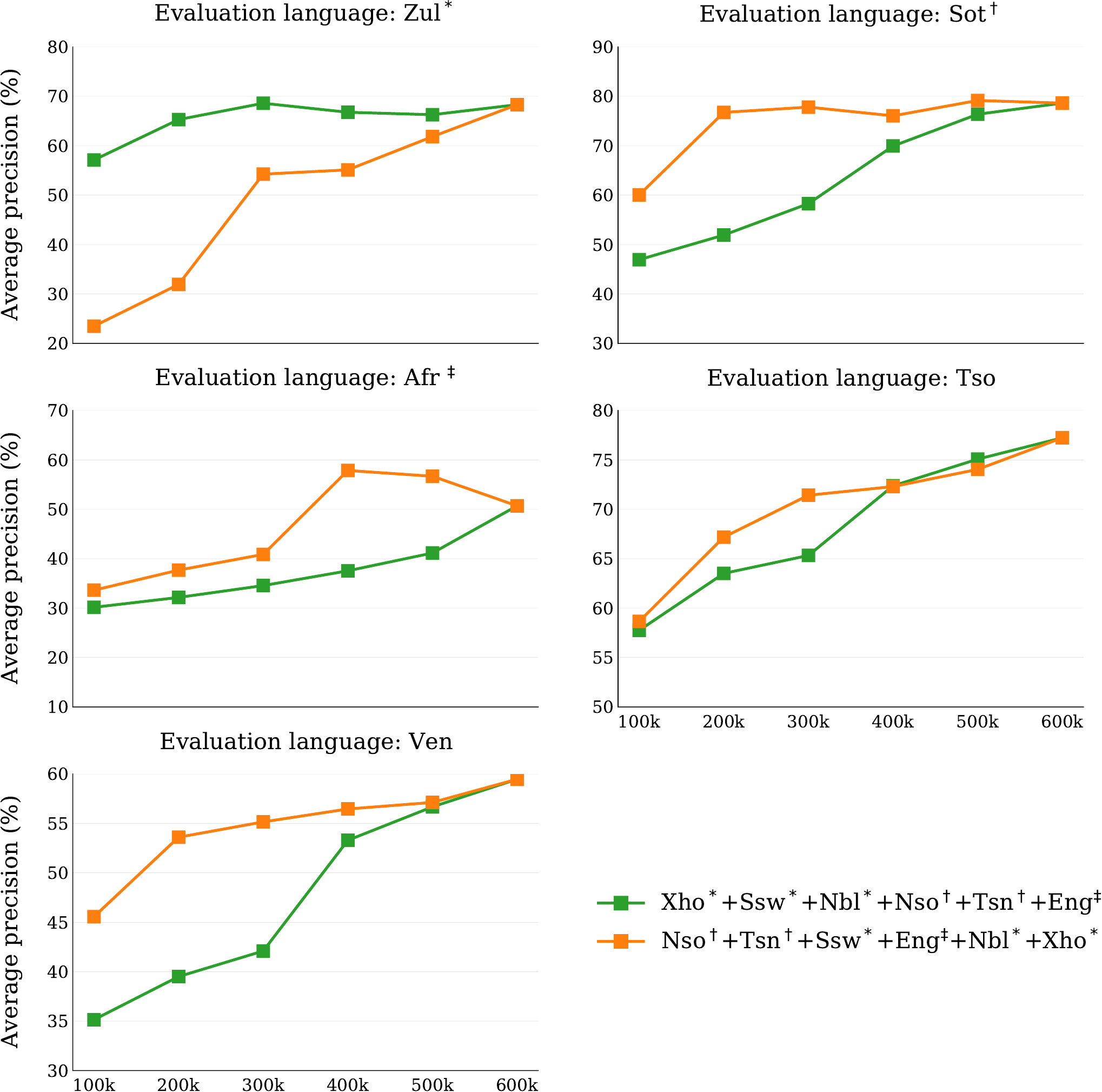}}
	\vspace{2mm}
	\caption{Same-different results from two sequences of multilingual models, trained by adding one language at a time. For each training language, 100k positive word pairs are used, which is indicated on the $x$-axis.}	\label{fig:multi_increment}
\end{figure}

\begin{figure}[!t]	
	\centering
	\centerline{\includegraphics[width=.95\linewidth]{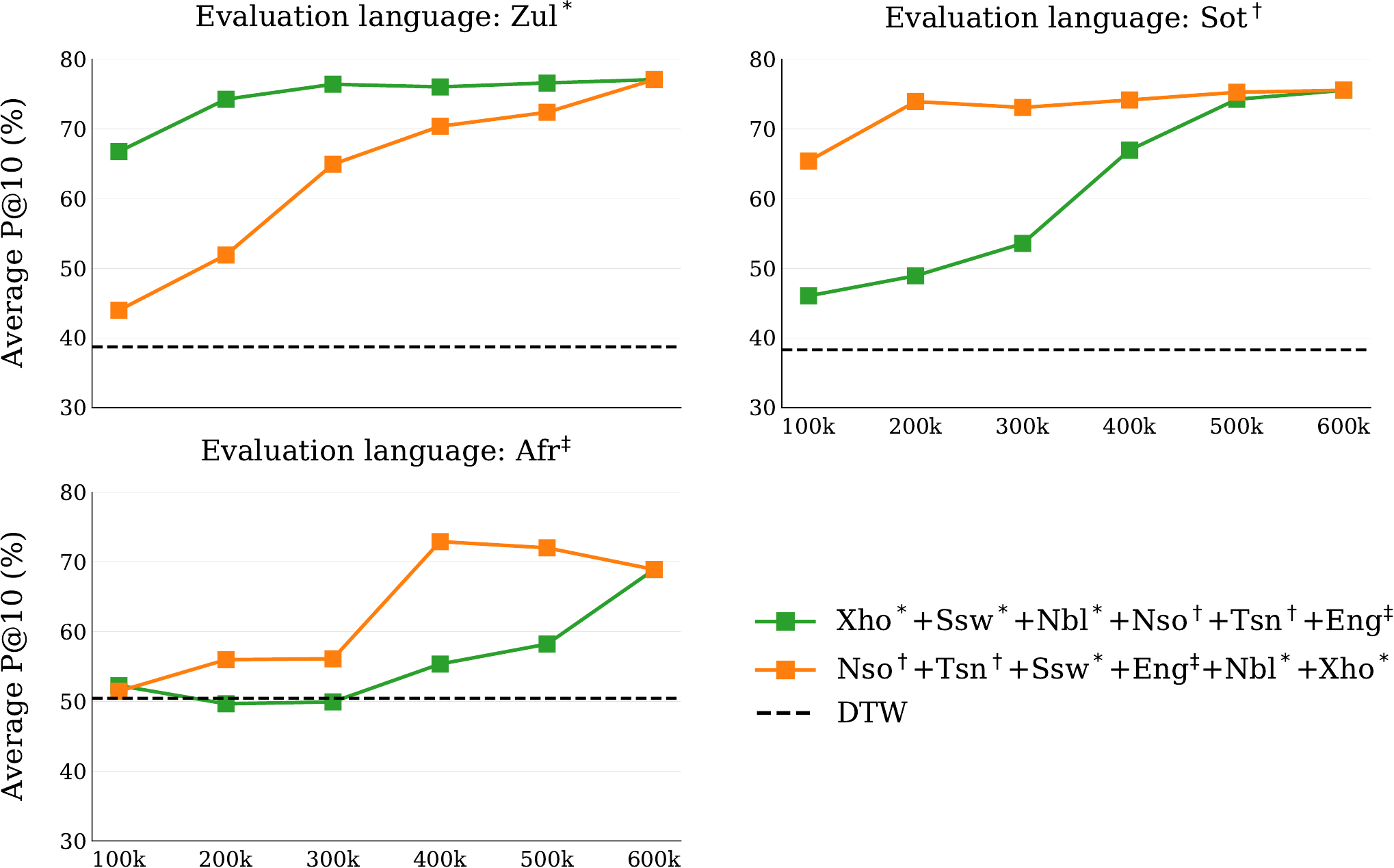}}
	
	\vspace{2mm}
	\caption{QbE results on Zul, Sot and Afr, using the same sequences of multilingual models as in Figure~\ref{fig:multi_increment}. DTW result are shown for baseline performance.}
	\label{fig:qbe_increment}
\end{figure}

Same-different results for all five evaluation languages are shown in Figure~\ref{fig:multi_increment} and QbE results for Zul, Sot and Afr are shown in Figure~\ref{fig:qbe_increment}. 
Notice how the trends in the same-different and QbE results track each other closely for corresponding evaluation languages.
In the first sequence of multilingual models (green), we start by adding the three Nguni languages (Xho, Ssw, Nbl), followed by the two Sotho-Tswana languages (Nso, Tsn), and lastly the Germanic training language (Eng).
The second sequence (orange) does not follow a systematic procedure.



On Zul, the green sequence, which starts with a related language (Xho), initially achieves a higher score compared to the orange sequence in both Figures~\ref{fig:multi_increment} and~\ref{fig:qbe_increment}.
Then, the score gradually increase by adding more related languages (Ssw, Nbl). 
Thereafter, adding additional unrelated languages (Nso, Tsn, Eng) show no performance increase.
In fact, AP decreases slightly after adding the two Sotho-Tswana languages (Nso, Tsn), but not significantly.
The orange sequence starts low on Zul until the first related language (Ssw) is added, causing a sudden increase.
Adding the Germanic language (Eng) has little effect.
Adding the last two related languages (Nbl, Xho) again causes the score to increase.
A similar trend follows for Sot and Afr, where adding related languages causes a noticeable performance increase, especially when adding the first related language; after this, performance seems to plateau 
when adding more unrelated languages.
(Afr is the one exception, with a drop when adding the last language in the orange sequence).
On Tso, which does not have any languages from the same family in the training set, AP gradually increases in both sequences without any sudden jumps.
Although Ven is not in the same family as Nso and Tsn at the lowest level of the tree in Figure~\ref{fig:tree}, it belongs to the same family {(Sotho-Tswana)} at a higher level. 
This explains why it
closely tracks the Sot results.


Summarising these results, we see that adding unrelated languages generally does not decrease scores, but also does not provide a big benefit (except if it is one of the earlier languages in the training sequence, where data is still limited). In contrast, it seems that training on languages from the same family is again beneficial; this is especially the case for the first related language, irrespective of where it is added in the sequence.

\mysection{Chapter summary}{Chapter summary}
\label{chap:related_summary}


We investigated the effect of training language choice when applying a multilingual AWE model to a zero-resource language.
Using word discrimination and QbE search tasks on languages spoken in South Africa, we showed that training a multilingual model on languages related to the target is beneficial.
We observed gains in absolute scores, but also in data efficiency: you can achieve similar performance with much less data when training on multiple languages from the same family as the target. 
We showed that even including just one related language already gives a large gain.
From a practical perspective, these results indicate that one should prioritise collecting data from related languages (even in modest quantities) rather than collecting more extensive datasets from diverse unrelated families, when building multilingual acoustic word embedding models for a zero-resource language.
In the following chapter, we present a framework for producing semantically organised AWEs using multilingual models developed in this chapter.
\mychapter{Hate speech detection using multilingual acoustic word embeddings}{Hate Speech Detection Using Multilingual Acoustic Word Embeddings}
\label{chap:hatespeech}

Leading up to this chapter, we mainly aimed at improving the quality of AWEs for low-resource languages: we developed a new AWE model (Chapter~\ref{chap:contrastive}), explored multilingual adaptation techniques (Chapter~\ref{chap:adaptation}), and considered language choice in multilingual AWE modelling (Chapter~\ref{chap:related}).
In Chapter~\ref{chap:related} we briefly measured AWE performance in a downstream query-by-example (QbE) speech search task.
However, this was performed with limited development data in a controlled environment.
In this chapter, we evaluate the practicality of using multilingual AWEs for real-world applications, beyond the experimental environments considered up to now.

The content of this chapter originates from a collaborative project with VoxCroft, a company specialising in developing speech applications in low-resource settings.
We address a specific case study involving the development of keyword spotting (KWS) systems to monitor radio broadcast audio for low-resource languages.
Here, we confront the challenge of lacking transcribed data in the target language and dealing with radio recordings that differs significantly from experimental audio data.
In this challenging environment, we develop an ASR-free approach, utilising a pre-trained multilingual AWE model in a KWS system.
Our main objective is to compare the performance of this AWE-based KWS system to the conventional ASR-based KWS systems, which mostly rely on transcribed data from the target language.
We are particularly concerned about the trade-off between accuracy and feasibility for rapid deployment.
To measure performance in this in-the-wild environment, we perform our final experiments on real Swahili radio broadcast audio, scraped from local radio stations in Kenya.
Moreover, our KWS systems are specifically developed to detect instances of hate speech in radio broadcasts.
Therefore, to search the audio, we use a set of keywords carefully selected by domain experts that are potential markers of hate speech.



\begin{tcolorbox}[width=\linewidth, colback=white!95!black, boxrule=0.5pt]
	\small
	\textit{The content of this chapter was presented at Interspeech:} \\
	C. Jacobs, N. C. Rakotonirina, E. A. Chimoto, B. A. Bassett, and H. Kamper, ``Towards hate speech detection in low-resource languages: Comparing ASR to acoustic word embeddings on Wolof and Swahili,'' in \textit{Proceedings of Interspeech}, 2023. \\

	\textit{Contributions per author:} \\
	C. Jacobs developed the AWE-based KWS systems (Section~\ref{ssec:hatespeech_awe}). \\
	N. C. Rakotonirina performed the ASR experiments (Section~\ref{ssec:hatespeech_asr}). \\
	E. A. Chimoto evaluated radio recordings flagged by our KWS systems (Section~\ref{sssec:hatespeech_wild_eval}). \\
	B. A. Bassett and H. Kamper supervised the project and helped with editing the paper.
\end{tcolorbox}

\mysection{Background}{Background}
\label{sec:hatespeech_background}

The rise of the internet and communication platforms are increasing the spread of content, some of which include the use of derogatory or discriminatory language targeting individuals or groups based on their identity or characteristics.
This form of expression is commonly referred to as \textit{hate speech}. 
Given the overwhelming volume of content being generated on online platforms, several automatic monitoring systems have been explored to curb the spread of hate speech (usually in the form of written text), especially for resource-rich languages.

However, in many developing countries where there is a lack of technological advancement, radio remains the primary medium for communicating news and information to the public~\cite{chalk_hate_1999}.
These radio stations play a vital role as platforms for public discourse, facilitating freedom of expression~\cite{somerville_violence_2011}.
Therefore, it is essential to establish appropriate mechanisms for monitoring these radio stations to ensure responsible broadcasting practices.

A striking example of the consequences of hate speech over radio is the 1994 genocide in Rwanda where propaganda was disseminated through local radio stations targeting individuals from an ethnic minority.
The dehumanising language used on air actively fuelled violence, resulting in devastating consequences.
Similarly, in the 2007-2008 post-election violence in Kenya, vernacular radio stations played a significant role in exacerbating the conflict~\cite{odera_radio_2015}.

In response to this ongoing issue, the United Nations has taken proactive measures to address hate speech on radio~\cite{noauthor_united_nodate}.
By actively monitoring hate speech claims, authorities can promptly respond and mitigate instances, preventing tension escalation and the dissemination of harmful ideologies.
The efforts of this chapter are in line with the objective of combating hate speech on radio broadcasts, specifically for low-resource languages.

Hate speech detection can be supported by using a \textit{keyword spotting} (KWS) system that can effectively search an audio corpus for a predetermined set of keywords or phrases that are indicative of hate speech.
Such a KWS system has the potential ability to quickly scan through large amounts of audio content and flag parts of the audio that have been identified as potential markers of hate speech. 
It is important to note that a KWS system solely relying on keyword or phrase identification  may not be sufficient for comprehensive hate speech detection. 
For instance, given the keyword ``blood", a KWS system may retrieve utterances ``we must spill their blood and eradicate them from our society" and ``she donated blood to help save lives during the blood drive."
The former contains hate speech whereas the latter highlights a positive action.
Therefore, the context in which these potential hate speech markers occur should be carefully considered and analysed by domain experts.

\mysection{Related work}{Related work}
\label{sec:hatespeech_related}
 
The conventional KWS approach involves transcribing the target audio with an automatic speech recognition (ASR) system and then searching for keywords in the output text~\cite{larson_spoken_2012, mandal_recent_nodate}.
For low-resource languages, the absence of large quantities of transcribed speech in the target language can make it difficult to train high-quality ASR models.
Several studies have looked into this problem, proposing methods to train ASR models on limited amounts of labelled data~\cite{chen_small-footprint_2014, audhkhasi_end--end_2017, saeb_very_2017, menon_fast_2018, mekonnen_end--end_2022}.

\citet{menon_radio-browsing_2017} were the first to evaluate and report results of a KWS system monitoring radio broadcasts in low-resource languages spoken in Uganda.
Their system was tailored to identify specific keywords, contributing to the progress of development and relief initiatives in rural areas.
Their system utilises an ASR model trained on around nine hours of transcribed data from the target language. 
However, relying on this amount of data poses a challenge as it will require a significant amount of time to collect and transcribe the audio.


\citet{saeb_very_2017} extended the work of \citet{menon_radio-browsing_2017} and implemented an ASR-based KWS system that combines multilingual acoustic modelling and semi-supervised training relying on much less transcribed data from the target language.
Here, they only use twelve minutes of annotated data from the target language to train an ASR model. 
However, they use several hours of untranscribed audio from the target language, coupled with a self-supervised learning strategy to improve their system.

Rapidly deploying ASR KWS systems for low-resource languages remains challenging due to the time-consuming process of collecting even a few minutes of transcribed data, which can take several weeks.
Moreover, multiple hours of untranscribed speech are not necessarily available for all target languages for self-supervised training.

When we want to rapidly develop and deploy a KWS system for hate speech detection in a new setting, it might be better to opt for an ASR-free approach.
In a truly zero-resource setting, ASR-free KWS systems are favourable, where KWS systems are developed without the need for transcribed data.
The most popular methodology extends QbE (Section~\ref{ssec:background_qbe}), which uses a \textit{spoken} instance of a keyword to search through an audio corpus, without the need for transcribing the audio into text.
To use QbE for KWS, we need a small number of spoken templates to serve as queries for the keywords of interest; this is the approach followed in~\cite{menon_fast_2018, van_der_westhuizen_feature_2022} and the methodology we also use for our AWE-based KWS system in this chapter.

DTW (Section~\ref{ssec:background_dtw}) is typically used to match the speech features of a spoken query to a search utterance~\cite{park_unsupervised_2008, hazen_query-by-example_2009, zhang_unsupervised_2009, jansen_indexing_2012}.
However, DTW has drawbacks in terms of its high time complexity and limitations, such as poor speaker independence, often requiring a large template set.

AWEs have been proposed as an alternative approach to finding matches in QbE where search segments and query segments are jointly mapped to the same fixed vector space~\cite{anguera_query_2014, levin_segmental_2015, settle_query-by-example_2017, yuan_query-by-example_2019}. 
This allows for faster comparisons between query and search segments in linear time complexity.
Various neural networks have been considered to obtain these AWEs as described in Section~\ref{sec:background_models}.

As demonstrated in Chapter~\ref{chap:contrastive} and Chapter~\ref{chap:related}, one approach to quickly obtain robust AWE for a new language is to train a multilingual AWE model on labelled data from multiple well-resourced languages, and then apply it to the target low-resource language.
Although these multilingual AWE models have proven successful in controlled experiments, there has been limited work investigating the effectiveness of these systems beyond the experimental environment.

In this chapter, for the first time, we develop and evaluate a KWS system utilising multilingual AWEs in a real-world scenario.
For this experiment, we apply the system to real radio broadcast audio, specifically targeting the detection of instances of hate speech.
The radio data was collected from stations in Kenya, a Sub-Saharan country in Africa, where the predominant language is Swahili.
Additionally, we conducted developmental experiments where we included Wolof, another low-resource language spoken mainly in Senegal, Sub-Saharan Africa.

We compare the AWE-based KWS system to ASR-based KWS systems.
We train our ASR models, leveraging advancements in self-supervised speech representation learning (SSRL) (Section~\ref{ssec:background_features}). 
These models enable fine-tuning of ASR models using limited transcribed data from the target language.
Our main research question is to compare the AWE-based KWS system with ASR-based KWS systems, focusing on the trade-off between accuracy and rapid deployment.


Before we describe the different KWS systems, we first briefly give the overall methodology that we follow towards our end goal of hate speech detection on Swahili radio broadcasts.

\mysection{Overview of methodology}{Overview of methodology}
\label{sec:hatespeech_overview}


We start by developing Wolof and Swahili KWS systems in a controlled environment where we use parallel audio-text data from experimental datasets to train and test our models.
These datasets~\cite{gauthier_collecting_2016, ardila_common_2020} contain high-quality, noise-free read speech.
We then test our models on held-out test data from these corpora.
We refer to these experiments as \textit{in-domain} because the test data comes from the same domain as the training data.
In this controlled setup, we perform KWS using a set of keywords drawn from development data.
These are not necessarily hate speech keywords.
The benefit of this controlled setting is that we have transcriptions for the test data, and can therefore quantify performance exactly.

We then apply the models to the \textit{out-of-domain} Swahili radio broadcasts.
These recordings were obtained from three radio stations in Kenya.
They consist of a mix of read and spontaneous speech, and also include non-speech audio like music and advertisements.
Additionally, there may be instances of different accents and languages present within the recordings.
This is the type of real-world audio that we would like to monitor for hate speech using KWS.
As our keywords in this context, we utilise actual hate speech keywords.
These were labelled as inflammatory words by native Swahili experts familiar with the media environment.
We should emphasise that, when we detect occurrences of these words, they might not necessarily correspond to hate speech.
For example, the English sentence ``do not kill your neighbour'' is not hate speech despite containing the word ``kill''.
In a deployed system the utterances flagged by our KWS approach would be passed on to human moderators for further review and hate speech assessment.
In this work we also use a human moderator, but only to mark whether a detected keyword actually occurred in an utterance. We do this for our in-the-wild test since we do not have transcriptions for the radio broadcasts. 

In the following section, we provide a detailed overview of the KWS systems that we developed.
This will be followed by the implementation details of the KWS systems (Section~\ref{sec:hatespeech_setup}).
\mysection{Low-resource keyword spotting}{Low-resource keyword spotting}
\label{sec:hatespeech_asr_awe}

The ASR-based KWS systems of this chapter were developed by a colleague.
Therefore, in Section~\ref{ssec:hatespeech_asr}, we provide a high-level overview of this approach.
The AWE-based KWS system we developed as an ASR-free alternative, is discussed in detail in the follow-up section.

\subsection[Keyword spotting with ASR]{Keyword spotting with ASR}
\label{ssec:hatespeech_asr}

\begin{figure}[!t]
	\centering
	\includegraphics[width=0.99\linewidth]{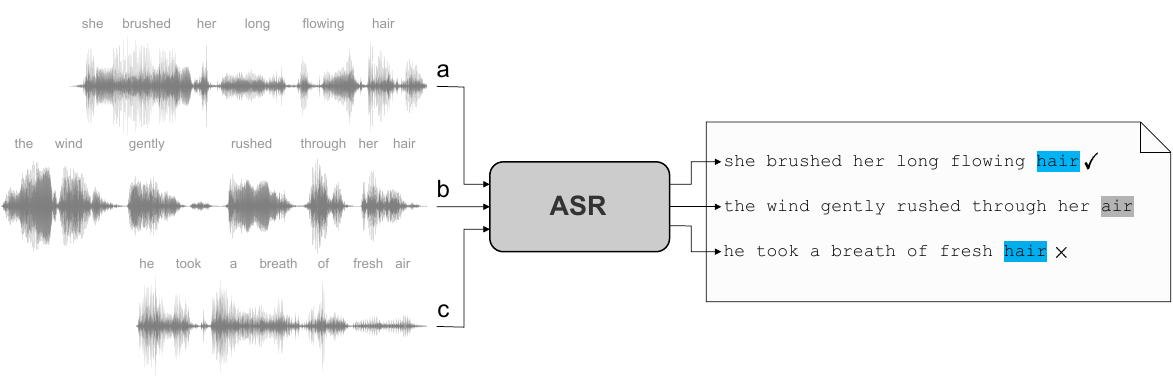}
	\caption{An illustration of ASR-based keyword spotting.}
	\label{fig:hatespeech_asr}
\end{figure}

In an ASR-based KWS system, the target corpus is first transcribed to text.
Occurrences of a keyword in the corpus can be efficiently identified by simply matching the keyword tokens with the output transcriptions.
An example of this ASR-based KWS is illustrated in Figure~\ref{fig:hatespeech_asr}.
In this example, when using the keyword ``hair'' to search the output transcriptions, the KWS system will correctly retrieve the first utterance (a).
In the third utterance (c), ``air'' was incorrectly transcribed as ``hair'' and will therefore be an incorrect retrieval (although the keyword matches the transcription). 
In the second utterance (b), the word ``hair'' in the audio was incorrectly transcribed as ``air'' and will not be retrieved by the KWS system.
Also, note that in this example the ASR model correctly transcribed all the other words and would therefore have a low word error rate (WER), but WER does not always correspond to KWS performance.
In Section~\ref{ssec:hatespeech_eval} we give a more detailed explanation of how we evaluate KWS performance.

Typically, to train an ASR model we need audio with transcriptions from the target language.
In a low-resource setting, a lack of large quantities of transcribed speech data from the target languages prohibits the rapid development of ASR-based KWS systems.

For our ASR systems, we build on recent advancements in ASR models leveraging SSRL (Section~\ref{ssec:background_features}).
These SSRL models are trained in a self-supervised fashion using large amounts of unlabelled audio recordings to produce high-quality latent frame-level speech representations.
It has been shown that these pre-trained SSRL models can be fine-tuned in an end-to-end fashion for ASR by mapping the latent frame representations to target text tokens.
Fine-tuning SSRL models for ASR requires using much less training data than training an ASR model from scratch.  

In our experiments, we use the pre-trained Wav2Vec2.0 XLSR model~\cite{baevski_wav2vec_2020, babu_xls-r_2022}.
This SSRL model is a large-scale cross-lingual speech model pre-trained on half a million hours of unlabelled speech data in 128 languages (including Swahili but not Wolof).
The speech representation learned during cross-lingual pre-training improves performance for low-resource languages by leveraging data from the high-resource languages on which it is pre-trained.
Using the pre-trained model as a starting point, the model is fine-tuned on labelled data from the language of interest.

Previous work has shown that this fine-tuning approach leads to competitive ASR performance even with limited amounts of labelled data~\cite{babu_xls-r_2022}.
In our experiments (Table~\ref{tbl:hatespeech_asr} specifically) we show that the sample efficiency of the XLSR model also extends to KWS when training on just five minutes of labelled data.

\subsection[Keyword spotting with multilingual acoustic word embeddings]{Keyword spotting with multilingual acoustic word embeddings}
\label{ssec:hatespeech_awe}

\begin{figure}[!t]
	\centering
	\includegraphics[width=0.99\linewidth]{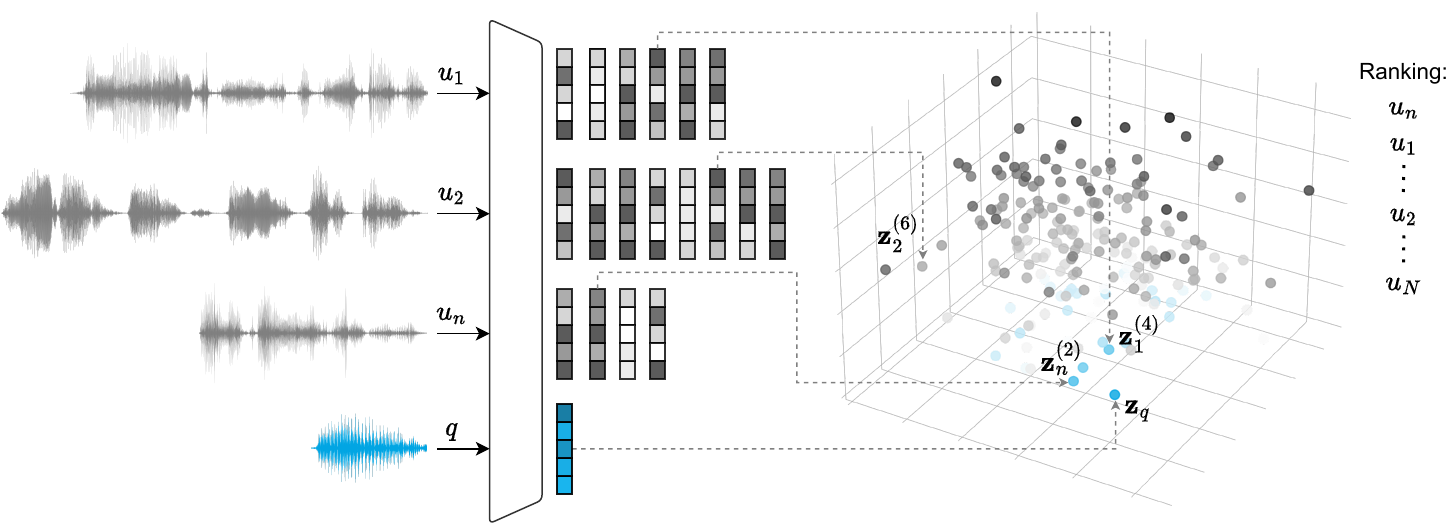}
	\caption{Search utterances ($u_1, u_2, \ldots, u_N$) are segmented and mapped to a fixed-dimensional space, along with a query segment $q$, to produce embeddings $\bigl\{\mathbf{z}_n^{(i)}\bigr\}$ and $\mathbf{z}_q$, respectively.
		The subscript $n$ represents a unique utterance within the search collection, and superscript $i$ represents a segment extracted from utterance $\mathbf{z}_n$.    
		Subsequently, search utterances are ranked based on the shortest distance between a search segment $\mathbf{z}_n^{(i)}$ and the query vector $\mathbf{z}_q$.
	}
	\label{fig:kws_awe}
\end{figure}

For a severely under-resourced language, there might not even be limited amounts of labelled data available for ASR fine-tuning.
In this scenario, collecting only a few spoken instances of a keyword of interest would allow for KWS through QbE search (Section~\ref{ssec:background_qbe}).
QbE is the task of retrieving utterances from a search corpus using a spoken instance of a keyword, instead of a written keyword.

In the previous chapter, our results (on development data) showed improved performance in QbE search by using multilingual AWEs instead of DTW.
We use a similar setup to the QbE experiments described in Section~\ref{sec:related_qbe} for experiments in this chapter.
The workings of a QbE system utilising AWEs are displayed in Figure~\ref{fig:kws_awe}.
First, the search corpus is segmented into word-like segments.
Search utterances are segmented by splitting into overlapping segments of some minimum to some maximum duration (see Figure~\ref{fig:related_qbe}).
Concretely, utterances from the search corpus ($u_1, u_2, \ldots, u_N$) are segmented to produce a set of variable-length speech segments $S = \bigl\{s_n^{(i)}\bigr\}_{n=1}^{N}$, where $i$ is the $i$th search segment extracted from utterance $n$.

A multilingual AWE model (trained on labelled data from multiple well-resourced languages but not the target language) is then applied to the segmented search corpus $S$ and query segment $q$, mapping them to the same fixed-dimensional vector space.
In this experiment, we use the \system{CAE-RNN} network (Section~\ref{ssec:background_cae}) and train a multilingual \system{CAE-RNN} AWE model.
This model is trained on word pairs using labelled data from multiple languages (but does not include the target language).
This multilingual AWE model can be transferred and applied to any unseen language to embed word segments of arbitrary length.
Utilising such a universal, language-independent model would enable fast deployment of practical QbE-based KWS systems.

Applying the AWE model to search segments $S$ and query segment $q$ produces search embeddings $\big\{\mathbf{z}_n^{(i)}\bigr\}$ and query embedding $\mathbf{z}_q$, respectively.
The utterances in the search corpus are ranked based on comparing the query-segment distances $d(\mathbf{z}_q, \mathbf{z}_n^{(i)})$.
For each query, the lowest distance for each utterance is used as the final utterance score.
We can then predict that all query-utterance pairs with a distance below some threshold are positive matches (this requires a threshold to be tuned on development data).
With these predictions, we can calculate common KWS metrics and compare them to those achieved by the ASR-based KWS systems.


\mysection{Experimental setup}{Experimental setup}
\label{sec:hatespeech_setup}

In this section, we present the implementation details of both the ASR-based (Section~\ref{ssec:hatespeech_asr}) and AWE-based (Section~\ref{ssec:hatespeech_awe}) KWS systems. 
It is important to note that the overview provided in Section~\ref{sec:hatespeech_overview} serves as a reminder for the reader.
By the end of this section, the difference between the developmental experiments conducted in a controlled environment and our in-the-wild KWS testing, particularly concerning data and evaluation methods, should be clear.

\subsection{Data}
\label{ssec:hatespeech_data}

Here we emphasise the difference between the training and evaluation data we use during development and the data we use for our in-the-wild testing.

\subsubsection{Development}
\label{sssec:hatespeech_data_controlled}
For our controlled experiments, we train and evaluate our models on Swahili and Wolof data, collected from the Common Voice~\cite{ardila_common_2020} and ALFFA~\cite{gauthier_collecting_2016} datasets, respectively.
Common Voice is an open-source data collection initiative by Mozilla Foundation where volunteers can contribute by recording their voices while reading prompts collected from a variety of sources. 
These recordings are then verified for quality and accuracy.
For our experiments we use the Swahili data from the Common Voice Corpus 8.0 dataset which gives train, test, and developing sets totalling 150 hours.
In order to match the sampling rate used during the training of the XLSR model, the original recordings, which are sampled at 48 kHz, are downsampled to 16 kHz.

The current data repository of Common Voice does not include Wolof. 
We therefore use a separate dataset from the ALFFA project, which contains Wolof recordings and their corresponding transcriptions.
Wolof is primarily spoken in Senegal, another country in Sub-Saharan Africa. 
However, the official language in Senegal is French, which serves as the lingua franca in all governmental institutions.
Due to the prevalence of French as the official language, there are limited written documents available for Wolof to construct parallel training data. 
For the Wolof language, we have designated training, testing, and development sets, consisting only of 16 hours, 2 hours, and 2 hours, respectively.

For both datasets, we use the default train, development, and test splits.
In our controlled experiments, we evaluate in-domain KWS performance on the Swahili and Wolof test splits using available transcriptions. 
The test split of both languages serves as the search collection and is not used during model development.
The Swahili search collection contains 8~941 utterances and has a duration of approximately 14 hours, while the Wolof set contains 2~000 utterances and has a duration of approximately 2 hours. 

\subsubsection{In-the-wild}
We perform our in-the-wild test on recorded radio broadcasts (collected by VoxCroft) from three different radio stations in Kenya: Ghetto Radio, Radio Salaam, and Radio Taifa.
We do not have any transcriptions, speaker information or content information on these audio clips.
The nature of these radio recordings has distinct characteristics that are different from the prompted read speech we use during development.
These broadcasts can contain a variety of elements, including music in the background, fast-paced speech by radio hosts, advertisements, background music, and dialogues involving multiple speakers.

Our systems are only applied to these out-of-domain recordings and are nowhere used during the development of our KWS systems.
Before applying our systems to this data we first apply a preprocessing step where we segment the recordings using an unsupervised diarisation system that also attempts to remove music.
Apart from this preprocessing step, we simply apply our systems to this out-of-domain data without further changes or calibration.
After this preprocessing step, 19\,716 utterances are obtained, which is our out-of-domain search collection.
The duration of all utterances is between three and 30 seconds.
Given that we do not have transcriptions of these recordings, we cannot evaluate system performance in the same way we do in the controlled experiments (where we have audio with corresponding transcriptions).
As we explain below (Section~\ref{sssec:hatespeech_wild_eval}), we use a domain expert to evaluate our system's performance on this set.

\subsection{Models}
\label{ssec:hatespeech_models}

We now provide the implementation details of the ASR model and the multilingual AWE model for ASR-based (Section~\ref{ssec:hatespeech_asr}) and AWE-based KWS (Section~\ref{ssec:hatespeech_awe}), respectively.

\subsubsection{ASR model}
\label{sssec:hatespeech_asr_model}

To train our ASR models, we fine-tune the pre-trained XLSR model by adding a projection head to the final transformer layer. 
Subsequently, the model is trained using transcribed data from the target language by mapping the frame-level representations to character tokens. 
The model is optimised using the connectionist temporal classification (CTC) loss~\cite{graves_connectionist_2006}. 
The CTC loss is effective for aligning input sequences and producing output sequences that may not be temporally aligned, as is the case with speech and text.
After the fine-tuning step, the model produces a sequence of characters given an arbitrary audio recording, that is then decoded to give the predicted transcription.


For the ASR-based KWS systems, we control the amount of training data and train three ASR models for Wolof and Swahili.
Specifically, we fine-tune the XLSR model~\cite{babu_xls-r_2022} for each language on 30 hours, one hour, and five minutes of labelled training data.
These models are used for our development experiments before applying them to the radio broadcast recordings.

\subsubsection[Multilingual acoustic word embedding model]{Multilingual acoustic word embedding model}
\label{sssec:hatespeech_awe_model}

For the AWE-based KWS system (Section~\ref{ssec:hatespeech_awe}), we train a multilingual AWE model on five different Common Voice languages: Abkhazian (ab), Czech (cs), Basque (eu), Swedish (sv), Tamil (ta).
Word boundaries are obtained using the Montreal forced-aligner~\cite{mcauliffe_montreal_2017}. 
We pool the data from all languages and extract 300k positive word pairs which we use to train a multilingual \system{CAE-RNN} AWE model similar to the CAR-RNN model of Chapter~\ref{chap:contrastive}.
We use another well-resourced language, Bulgarian, for validation during training using the same-different evaluation (Section~\ref{sec:background_evaluation}). 
The encoder and decoder each consist of three unidirectional RNNs with 400-dimensional hidden vectors and an embedding size of 100 dimensions.  
As inputs to the encoder, we use the XLSR model to extract speech features (Section~\ref{ssec:background_features}) instead of MFCCs used for AWE modelling in previous chapters.
We use the output of the $12$th transformer layer, producing input features with 1024 dimensions.
In the work of ~\cite{sanabria_analyzing_2023}, they show that the lower layers of the XLSR transformer model capture more discriminative phonetic properties than the final layers when simply using the model to extract frame-level speech representations.
Preliminary experiments showed using these features outperforms MFCCs in AWE modelling.
We provide a short analysis in Section~\ref{ssec:hatespeech_xlsr_vs_mfcc} comparing multilingual AWEs trained on MFCC and XLSR.
All models are optimised using Adam~\cite{kingma_adam_2017} with a learning rate of $0.001$.


\subsection{Evaluation}
\label{ssec:hatespeech_eval}

We outline our KWS evaluation methods for the ASR-based and AWE-based KWS systems for the two scenarios: controlled in-domain experiments with available transcriptions and out-of-domain in-the-wild tests where transcriptions are unavailable for evaluation.

\subsubsection[Controlled experiments]{Controlled experiments}
\label{sssec:hatespeech_controlled_eval}
To measure in-domain KWS performance, we apply the ASR-based and AWE-based KWS systems to the same search corpora.
Each utterance in the search corpus is assigned a label of 1 if the keyword is detected by the respective KWS system, or 0 if the keyword is not present.
We evaluate our KWS systems by using the predicted labels and compare them against the true labels obtained from available transcriptions.
Following this approach, we can calculate precision and recall scores.
 
Precision measures the accuracy of a KWS system in predicting whether an utterance contains a keyword.
It is calculated as the number of utterances correctly predicted as keyword-containing utterances (\# hits) divided by the total number of utterances predicted to contain a keyword (\# keywords detected):
\begin{equation}
	P = \frac{\text{\# hits}}{\text{\# keywords detected}}
\label{eqn:hatespeech_precision}
\end{equation}
Recall quantifies the KWS system's ability to retrieve keyword-containing utterances from the search corpus.
It is calculated as the number of correctly predicted keyword-containing utterances (\# hits) divided by the total number of keyword-containing utterances in the search corpus (\# keywords):
\begin{equation}
	R=\frac{\text{\# hits}}{\text{\# keywords}}
\end{equation}
In addition to precision and recall, we report the $F_1$-score to quantify the trade-off between these two metrics.
The $F_1$-score represents the balanced combination of precision and recall, calculated as the harmonic mean of the two metrics:
\begin{equation}
	F_1 = 2 \times \frac{P \times R}{P + R}
\end{equation}

To evaluate our ASR KWS systems, we simply check whether a keyword occurs in the predicted transcript after applying it to the search corpus and label each utterance accordingly (1 or 0) to calculate the metrics above.

The AWE-based KWS system does not output transcribed text but a query-utterance score for each query segment (bottom of Section~\ref{ssec:hatespeech_awe}).
First, sub-segments for utterances in the search collection are obtained by extracting windows ranging from 20 to 35 frames with a 5-frame overlap (we experimented with different window and overlap sizes on development data where this configuration worked best).
Then, an AWE model is applied to all the sub-segments and query segment.
Finally, the similarity between a query and search segment is calculated using cosine distance.

We use a set of ten templates of 36 unique keywords with a minimum character length of five for Swahili, and 15 unique keywords with a minimum character length of four, for Wolof.
These keywords were randomly sampled from a larger set of words that appear at least ten times in both the development and test sets.
These keywords do not represent hate speech keywords.
Query templates are drawn from development data, embedded, and averaged to obtain a single AWE embedding representing a keyword.
Preliminary experiments showed that averaging all templates from the same keyword improves KWS performance compared to using a single template for searching.
The threshold value in the AWE-based KWS is tuned for the highest $F_1$-score across all keywords on the controlled test data (more in Section~\ref{ssec:hatespeech_th}).

\begin{table}[!t]	
	\mytable
	\caption{Swahili hate speech keywords used for in-the-wild KWS on Swahili radio broadcasts, with their English translations.}
	\addtolength{\tabcolsep}{10pt}
	\begin{tabularx}{.8\linewidth}{Ll|Ll}
		\toprule				
		Swahili & English & Swahili & English \\ [0.5mm]
		\midrule
		vita & war & wezi & thieves \\
		damu & blood & majimbo & states \\
		hama  & move & wakora & conmen  \\
		kabila & tribe & panga & machete \\
		utapeli & fraud & takataka & garbage \\
		mende & cockroaches &  mjinga & stupid \\
		kitendawili & riddle & fala & stupid\\
		\bottomrule		
	\end{tabularx}
	\label{tbl:keywords}
\end{table}

\subsubsection{In-the-wild}
\label{sssec:hatespeech_wild_eval}
For the in-the-wild KWS, we apply the systems directly to the out-of-domain Swahili radio broadcasts.
We use a set of keywords labelled as inflammatory by expert analysts from VoxCroft who are familiar with the media environment for the purpose of hate speech detection.
These keywords with their English translations are given in Table~\ref{tbl:keywords}.
For the AWE-based KWS system, ten query templates per keyword are extracted from the in-domain Swahili data.
We evaluate KWS performance by asking each approach to give 100 utterances out of the search collection that are most likely to contain any of the hate speech keywords.
In the absence of per-word confidence scores for the ASR models, here we simply take 100 random utterances that were predicted to contain a keyword 
(for most models, the total number of utterances was around 150).

For the AWE-based KWS, we use the 100 highest-ranked utterances. 
Because we do not have transcriptions for this data, we provide a native Swahili speaker with untranscribed recordings of the utterances that were predicted to contain a keyword. 
They then mark whether the keyword was indeed present.
Note that due to the absence of transcriptions for the radio recordings, we are unable to calculate recall or the $F_1$-score.
Therefore, we can only measure performance by calculating precision, based on 100 retrieved utterances per system (Table~\ref{tbl:hatespeech_wild}).

\mysection{Experiments}{Experiments}
\label{sec:hatespeech_results}

We first look at the results from our controlled test on in-domain Swahili and Wolof data.
We then look at the results on the out-of-domain data, which is of particular interest to us as it would indicate the effectiveness of both these systems for real-world application.

\subsection[KWS results in a controlled environment]{KWS results in a controlled environment}
\label{ssec:results_in}

\begin{table}[!t]	
	\mytable
	\caption{
		ASR and AWE KWS results (\%) on in-domain test data.
		For the ASR systems, the XLSR model is fine-tuned on each target language, controlling the amount of training data.
		For the AWE KWS system, a supervised multilingual AWE is trained on multiple well-resourced languages and applied to the two low-resource target languages.
	}
	\begin{tabularx}{\linewidth}{L@{\extracolsep{10pt}}cccccc}
		\toprule				
		\multirow{4}{1cm}{Model} & \multicolumn{3}{c}{Swahili} & \multicolumn{3}{c}{Wolof} \\
		\cmidrule(r){2-4} 
		\cmidrule(l){5-7}
		& Prec. &  Rec. & $F_1$ & Prec. &  Rec. & $F_1$ \\ 
		\midrule
		\underline{\textit{ASR}:} & & & & & & \\ [1mm]
		XLS-R (30-h) & 97.6 & 98.6 &  {98.0} &  93.1 & 87.1 & 90.0 \\
		XLS-R (1-h) &96.4 & 76.1 & {85.1}  & 93.1 & 74.3 & 82.7 \\
		XLS-R (5-min)
		& 95.4 & 42.6 &  58.9  & 88.0 & 45.6 & 60.0 \\
		\addlinespace
		\underline{\textit{Multilingual AWE}:} & & & & & & \\ [1mm]
		\tablesystem{CAE-RNN} (ab+cs+eu+sv+ta)  & 57.1 &  56.1 & 56.6 & 44.0 & 58.4 & 50.2 \\
		\bottomrule		
	\end{tabularx}
	\label{tbl:hatespeech_asr}
\end{table}


The first three lines in Table~\ref{tbl:hatespeech_asr} report the ASR KWS results on in-domain Swahili and Wolof data.
The results show that fine-tuning XLSR on only five minutes of labelled data achieves decent KWS performance, with $F_1$-scores of {59\%}
and 60\% for Swahili and Wolof, respectively.
The multilingual AWE KWS system using roughly one minute of template data, performs worse on most metrics.
The only metric on which the AWE system	is better is recall, where it achieves higher scores than the 5-minute ASR system on both languages.
In practice, the AWE system could therefore be a better option if recall is important.

We briefly consider how ASR performance is affected by the amount of training data.
For reference, WERs for the 30-hour, 1-hour, and 5-minute models are respectively 9\%, 36\%, and 62\% on Swahili.
For Wolof, the WERs are 27\%, 44\%, and 68\%.
As expected, more training data gives better WER and KWS scores.
But it is noteworthy that with just five minutes of training data, we can already spot keywords with high precision (95\% and 88\% for Swahili and Wolof, respectively). 
This is especially useful for our use case, where we want to rapidly develop KWS applications in severely low-resourced settings.
The other noteworthy finding from the ASR results is that scores for Swahili are not notability higher than Wolof, although the former is one of XLSR's pretraining languages.

\begin{table}[!t]
	\mytable	
	\caption{
		Supervised AWE KWS results (\%) on in-domain Swahili test data.
		The supervised monolingual \system{CAE-RNN} AWE model is trained using labelled Swahili data.
		It is applied with two segmentation configurations: using ground-truth word boundaries
		(true segm.), and using a variable-length window which is swept across the search collection (random segm.). 
		Here we also report the standard QbE metrics $P@10$ and $P@N$ (Section~\ref{sssec:related_eval_qbe}).
	}	
	\begin{tabularx}{1\linewidth}{L@{\extracolsep{3pt}}ccccc}
		\toprule				
		Model & Prec. & Rec. & $F_1$ & $P@10$& $P@N$ \\ 
		\midrule
		\underline{\textit{Supervised monolingual}:} & & & & &  \\ [1mm]
		\tablesystem{CAE-RNN} (true segm.) & 95.5 & 90.8 &{93.1} & 98.6 & 94.3 \\
		\tablesystem{CAE-RNN}  (random segm.)  & 79.2 &  76.1 & {77.6} & 92.2 & 90.9 \\
		\addlinespace
		\underline{\textit{Supervised multilingual}:} & & & & & \\ [1mm]
		\tablesystem{CAE-RNN}  ab+cs+eu+sv+ta (random segm.)  & 57.1 &  56.1 & {56.6} & 87.8 & 64.7 \\
		\bottomrule		
	\end{tabularx}
	\label{tbl:hatespeech_awe}

\end{table}

We now turn to the AWE-based approach.
Specifically, we ask what the upper bound on performance would be if we had more training data or a more idealised search setting.
Therefore, for a moment, we assume we have labelled data available and perform top-line experiments on the in-domain Swahili data, shown in Table~\ref{tbl:hatespeech_awe}.   
A supervised Swahili AWE model is trained {on 30 hours of labelled data} and applied to the search collection that is segmented using true word boundaries.
Results are shown in the first row of Table~\ref{tbl:hatespeech_awe}, serving as the top-line performance for AWE KWS, with a high $F_1$-score of 93\%.
Compared to the 30-hour ASR system (Table~\ref{tbl:hatespeech_asr}), the idealised AWE system comes closer to the $F_1$ of 98\%.
The second row in Table~\ref{tbl:hatespeech_awe} shows KWS performance using a supervised Swahili AWE model but applied to search segments extracted by sliding a variable-length window across the search collection---the way we apply the AWE approach in practice and in Table~\ref{tbl:hatespeech_asr}.
Segmentation without true word boundaries incurs a significant penalty, resulting in an $F_1$-score drop to 77\%.
It is clear that the sliding window approach has a large effect on downstream performance, so future work should consider more sophisticated unsupervised word segmentation techniques.

\subsection[KWS results in the wild]{KWS results in the wild}
\label{ssec:results_out}

We now turn to our main research question: comparing ASR to multilingual AWE-based KWS on real-life, out-of-domain audio in a low-resource setting. 
As mentioned in Section~\ref{sssec:hatespeech_wild_eval}, systems are applied to out-of-domain Swahili radio broadcasts, after which, for each system, the top 100 utterances predicted to contain a hate speech keyword (Table~\ref{tbl:keywords}) are manually reviewed.
The results are given in Table~\ref{tbl:hatespeech_wild}.

The table reports precision: the proportion of retrieved top-100 utterances that correctly contain a hate speech keyword.
Surprisingly, we see that the multilingual AWE KWS system achieves a precision of 45\%---better than the 5-minute and even the 1-hour ASR system in this in-the-wild test.
This is in contrast to the in-domain KWS results, where the 1-hour ASR model outperformed the AWE KWS system (Table~\ref{tbl:hatespeech_asr}).
Further investigation is required to understand exactly why the relative performance of the ASR and AWE systems are affected differently when applied to out-of-domain data.
However, it is worth noting that several studies have shown that ASR system performance can drop dramatically when it is applied to data outside of its training domain~\cite{seltzer_investigation_2013, likhomanenko_rethinking_2021, hsu_robust_2021}.
One example where this can be seen in this case is how many of the ASR retrievals contain music (which we asked the human annotator to mark).

The in-the-wild search collection has been diarised automatically, which included a step to remove music segments. 
But this preprocessing step is not perfect: the search collection still ends up with some music (which neither the ASR nor AWE systems have seen in training).
Table~\ref{tbl:hatespeech_wild} shows that, out of the top-100 utterances for the 30-hour ASR system, 30\% contained music. 
This decreases for the 1-hour (29\%) and 5-minute (17\%) ASR systems. 
But the AWE KWS system only retrieves two utterances containing music.
And of these two, one utterance actually did contain a hate speech keyword in the music lyrics (there are also examples of such correct matches in the ASR music retrievals).
Nevertheless, this shows that the AWE approach seems to be more robust to domain mismatch compared to training an ASR system on one hour or five minutes of labelled data.
We therefore conclude that implementing an AWE-based KWS system as an ASR-free alternative for rapid deployment in an unseen language proves to be fruitful.

\begin{table}[!t]	
	\mytable
	\caption{
		In-the-wild KWS results (\%) on Swahili radio broadcasts. 
		For each system, we report the percentage of utterances retrieved containing a keyword (precision) and the percentage of utterances retrieved containing music.
	}
	\begin{tabularx}{\columnwidth}{L@{\extracolsep{0pt}}cc}
		\toprule		
		Model & Precision & Music \\
		\midrule
		\underline{\textit{ASR}:} & & \\ [1mm]
		XLS-R (30-h) & 52 & 30 \\
		XLS-R (1-h) & 42 &  29 \\
		XLS-R (5-min) & 36 & 17 \\
		\addlinespace
		\underline{\textit{Multilingual AWE}:} & & \\ [1mm]
		\tablesystem{CAE-RNN}  ab+cs+eu+sv+ta & 45 & 2 \\		
		\bottomrule		
	\end{tabularx}
	\label{tbl:hatespeech_wild}
\end{table}

\mysection{Further analysis}{Further analysis}

We briefly examine the transition from MFCCs to XLSR features as input to our multilingual AWE model.
We also investigate the potential of selecting a more optimal threshold value for the AWE-based KWS system (Section~\ref{ssec:hatespeech_awe}).

\begin{figure}[!t]
	\centering
	\begin{subfigure}[b]{0.49\textwidth}
		\centering
		\includegraphics[width=\textwidth]{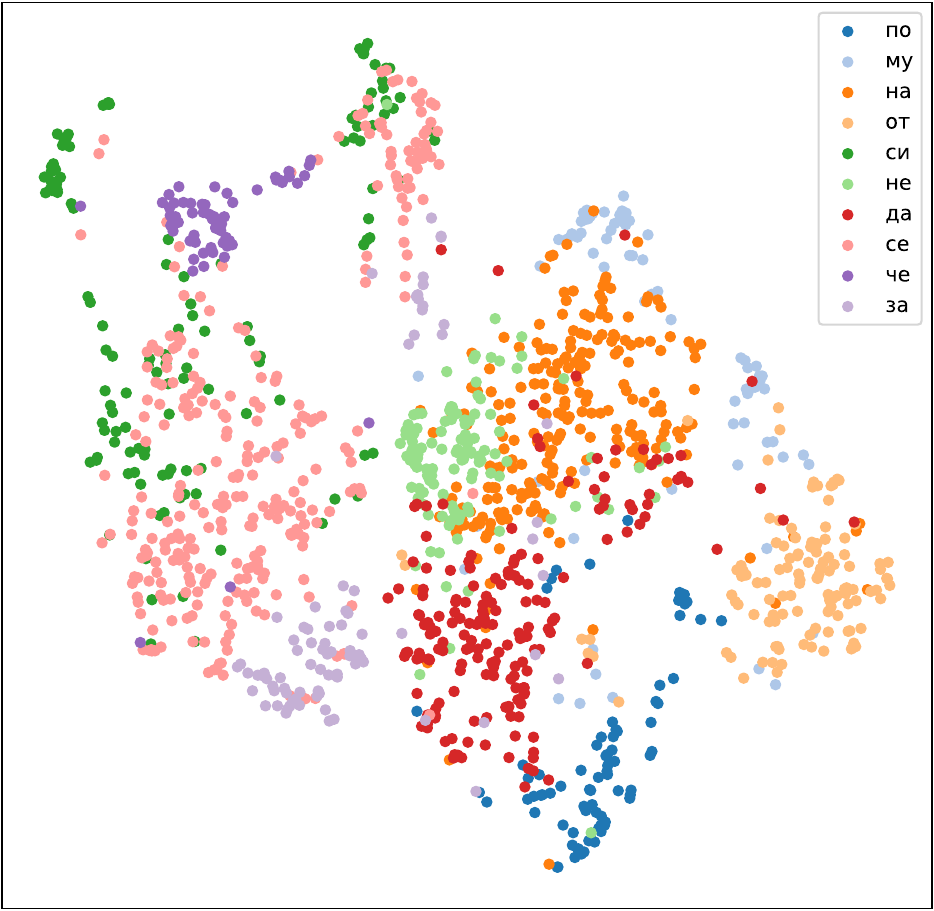}
		\caption{MFCC}
		\label{fig:mfcc_short}
	\end{subfigure}
	\hfill
	\begin{subfigure}[b]{0.49\textwidth}
		\centering
		\includegraphics[width=\textwidth]{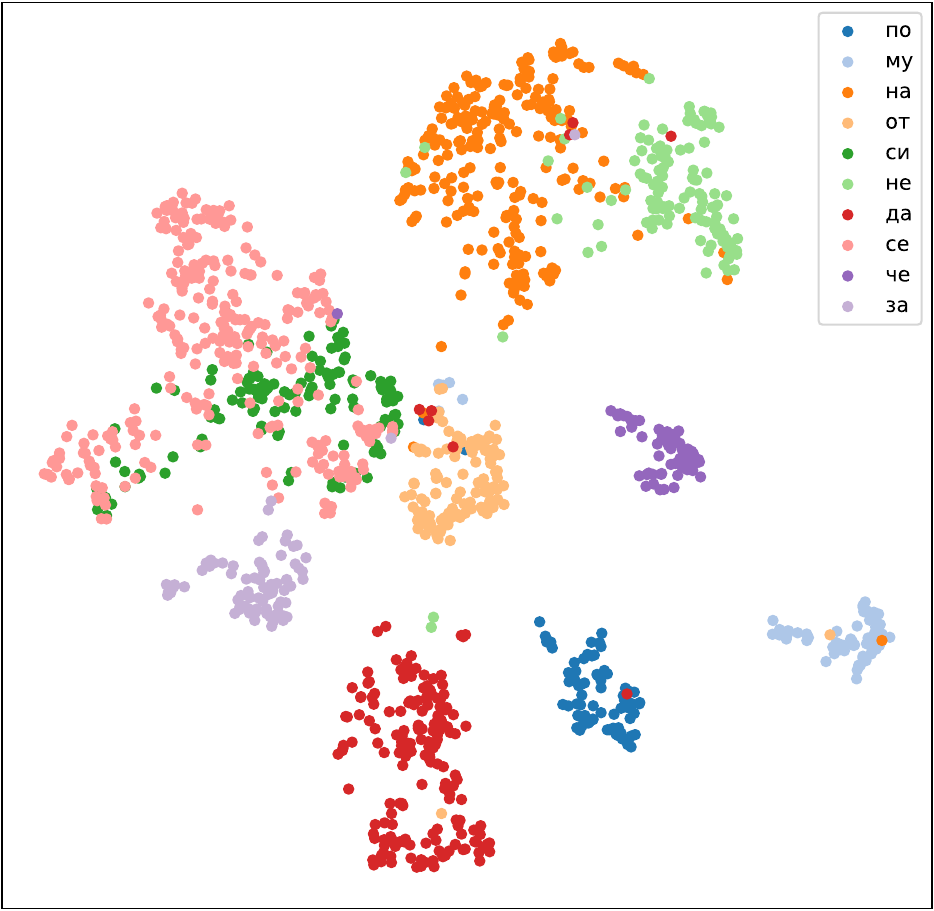}
		\caption{XLSR}
		\label{fig:xlsr_short}
\end{subfigure}
	\caption{
	UMAP~\cite{mcinnes_umap_2018} visualisation of the ten most frequent two-character word segments in Bulgarian development data. 
	These plots show the difference between AWEs produced by a multilingual AWE model trained on (a) MFCC features and (b) XLSR features.
	}
\label{fig:hatespeech_analysis}
\end{figure}

\subsection{XLSR vs MFCCs}
\label{ssec:hatespeech_xlsr_vs_mfcc}
For our experiments in this chapter we substituted MFCCs for XLSR speech features as input to our AWE models.
Here we briefly look at the difference in embedding quality produced between a multilingual AWE model trained on XLSR features compared to MFCCs.
We use the same training setup in Section~\ref{sssec:hatespeech_awe_model} and train a multilingual AWE using MFCCs.
For evaluation, we use the same developmental language, Bulgarian, and perform the same-different word task.
On the same set of evaluation pairs, we obtain an AP of 29\% and 58\% for the MFCC and XLSR multilingual AWE models, respectively.
With further investigation, we found that a model trained on XLSR features is much better at learning higher-quality embeddings for shorter word segments.
Figure~\ref{fig:hatespeech_analysis} shows the difference between the quality of embeddings produced for two character words when using (a) MFCCs and (b) XLSR as input features.
Overall, we see word segments from the same class to be much better clustered together and separated from word segments from different classes.
Using XLSR features as input clearly improves the quality of AWEs, leading to improved performance in the AWE-based KWS approach.

%

\subsection{Threshold}
\label{ssec:hatespeech_th}

\begin{figure}[!t]
	\centering
	\includegraphics[width=0.99\linewidth]{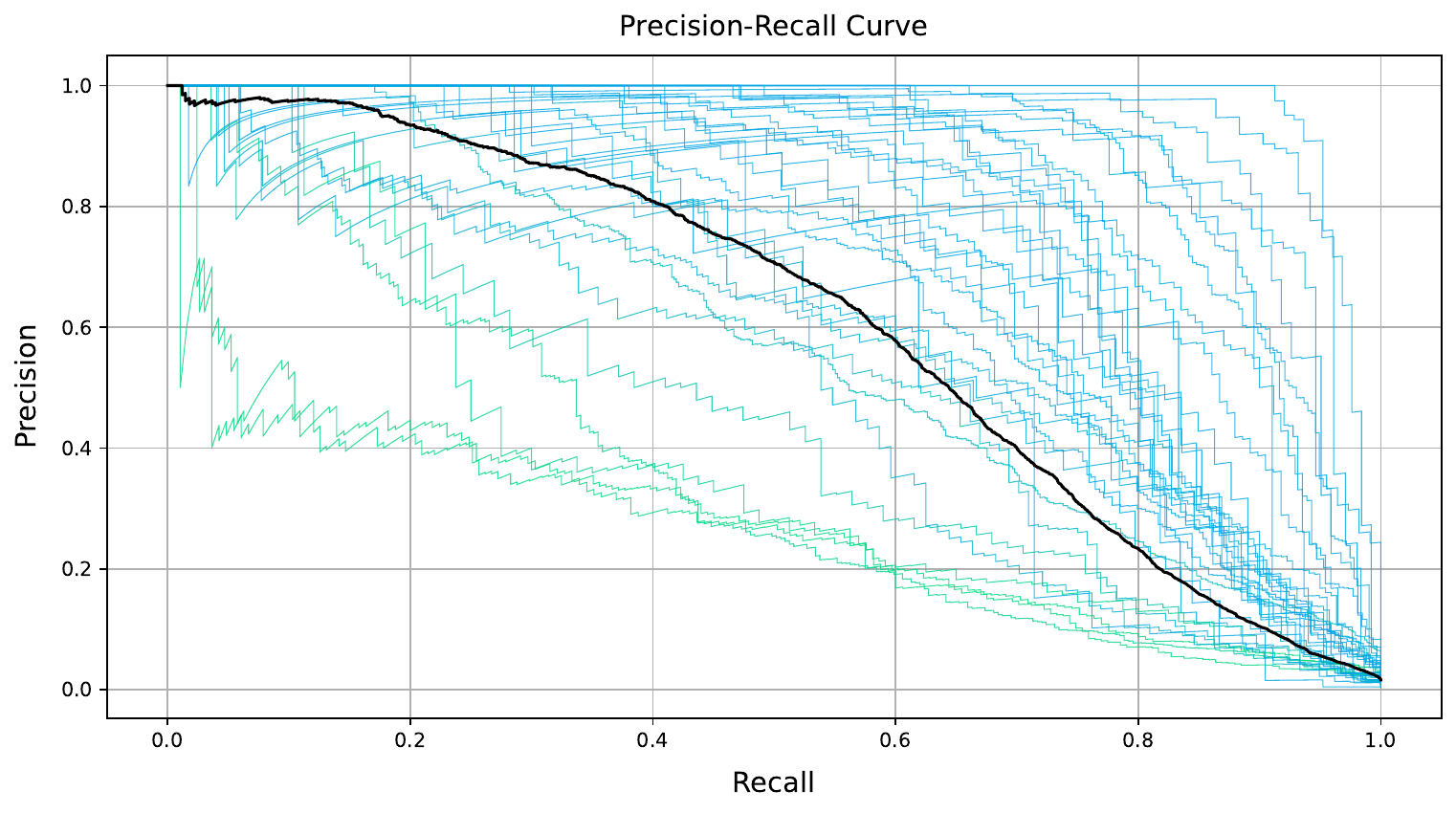}
	\caption{Precision-recall curves from in-domain AWE-based KWS on in-domain Swahili test data using the same 36 keywords as in Section~\ref{ssec:results_in}.
	}
	\label{fig:hatespeech_curve}
\end{figure}

As mentioned in Section~\ref{ssec:hatespeech_awe}, the threshold value selected for the AWE-based KWS system was tuned on the $F_1$-score by varying the threshold value over all keywords.
For interest, we tune the threshold value per keyword and plot a precision-recall curve for each of the 36 keywords, shown in Figure~\ref{fig:hatespeech_curve}.
We also plot the precision-recall curve (in black) using the same threshold value for all keywords.
From the figure we see that most per-keyword scores are higher than the average (blue lines above black line).
When we average the $F_1$-scores using the optimal threshold value per keyword, we obtain an $F_1$ score of 69\% compared to 57\% (bottom of Table~\ref{tbl:hatespeech_asr}) when using the same threshold for all keywords.
This raises the question of the selected threshold value and whether simply treating keywords as equal might be a suboptimal approach.
Practically, however, a per-keyword threshold can not be calculated given that for a new arbitrary keyword in a new unseen language we do not have development data to tune this.

But there is a possibility that a threshold can be selected based on other properties of a new keyword.
These factors include the length of the speech segment, the number of phones present, and the relative distances between other unknown word segments in the acoustic space.
For example, we may find that, in general, the threshold value for shorter word segments is lower than for longer speech segments.
Future work should investigate the correlation between the threshold value of keywords and the potential factors mentioned above.

\mysection{Chapter summary}{Chapter summary}
\label{sec:hatespeech_summary}

This chapter addressed the problem of keyword spotting (KWS) for the purpose of hate speech detection in low-resource languages.
We compared two KWS systems for data from Swahili and Wolof: a fine-tuned automatic speech recognition (ASR) model using different amounts of training data, and an ASR-free KWS system that utilises multilingual AWEs. 
The results show that fine-tuning a pre-trained multilingual ASR model using even a small amount of labelled data can outperform an AWE-based KWS system in controlled environments where training and test data come from the same domain.  
However, the AWE-based KWS system is more robust on out-of-domain radio broadcast data and achieves comparable results to an ASR model fine-tuned on 30 hours of labelled data. 
In the end, the merits of ASR vs AWE KWS will come down to the practical setting: it will depend on whether labelled training data can be collected from the target domain or not, and whether precision or recall is more important.

\mychapter{Leveraging multilingual transfer for unsupervised semantic acoustic word embeddings}{Leveraging Multilingual Transfer for Unsupervised Semantic Acoustic Word Embeddings}
\label{chap:semantic}

In the previous chapters, we focussed on producing vector representations for spoken word segments where different acoustic realisations from the same word type should have similar embeddings.
This can clearly be seen in the way we evaluated the embeddings: we use a word discrimination task (Section~\ref{sec:background_evaluation}), a QbE speech search task (Section~\ref{sec:related_qbe}), and a keyword spotting task (Section~\ref{ssec:hatespeech_awe}), all of which require exact matching of spoken word segments. 
The performance of both these tasks relies on the discriminative ability between embeddings from different word types.
Specifically, we aimed at producing these embeddings in low-resource settings, where annotated data in the target language is not available; here a multilingual transfer approach demonstrated superior performance.

In this chapter, we shift our focus to a different type of representation learning: \textit{semantic acoustic word embeddings}.
In a semantic-acoustic space, embeddings should not only be similar for different realisations of the same word type.
Additionally, embeddings from distinct word types should also be close to each other if the respective words share meaning.
These embeddings, mapping to spoken words, should reflect the same semantic properties as word embeddings trained on text data (Figure~\ref{fig:intro_twe}).
In other words, proximity in this new acoustic-semantic space should reflect word-type similarity as well as semantic relatedness between embeddings from different word types.

Again, we consider a low-resource scenario where word labels in the target language are unavailable.
In this setting, there have been limited efforts to address the challenging problem of producing semantic AWEs.
As a contribution, we propose a novel semantic AWE modelling approach: leveraging a pre-trained multilingual AWE model to assist semantic AWE modelling.
This marks  the first instance of using a multilingual transfer strategy in this context.
Furthermore, existing research on semantic AWEs lacks comprehensive evaluation.
We therefore perform our experiments in a constrained setup and evaluate our embeddings in an intrinsic word similarity task (measuring semantic relatedness between words), and an extrinsic semantic QbE search.
In the latter, a system should retrieve utterances containing different realisations of the query as well as utterances containing word segments that are semantically related to the query.

\begin{tcolorbox}[width=\linewidth, colback=white!95!black, boxrule=0.5pt]
	\small
	\textit{Parts of this chapter are presented in the journal paper:} \\
	C. Jacobs and H. Kamper, ``Leveraging multilingual transfer for unsupervised semantic acoustic word embeddings,'' \textit{IEEE Signal Processing Letters}, 2023.
\end{tcolorbox}

\mysection{Background}{Background}
\label{sec:semantic_background}

\begin{figure}[!t]
	\begin{subfigure}[c]{.4\linewidth}
		\centering
		\includegraphics[width=0.99\linewidth]{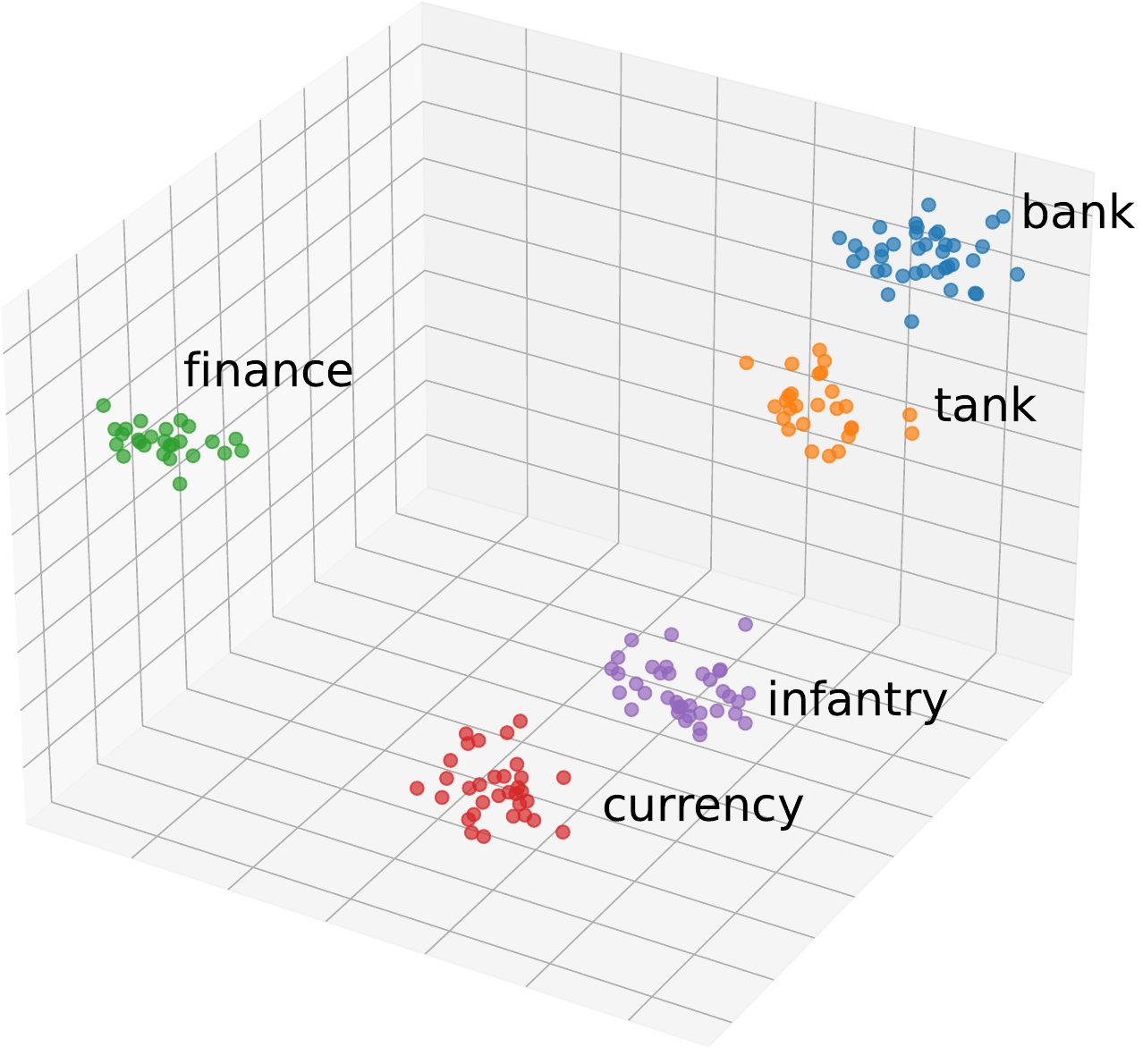}
		\caption{Acoustic word embeddings}
		\label{fig:semantic_p}
	\end{subfigure}
	\hfill
	\begin{subfigure}[c]{0.4\linewidth}
		\centering
		\includegraphics[width=0.99\linewidth]{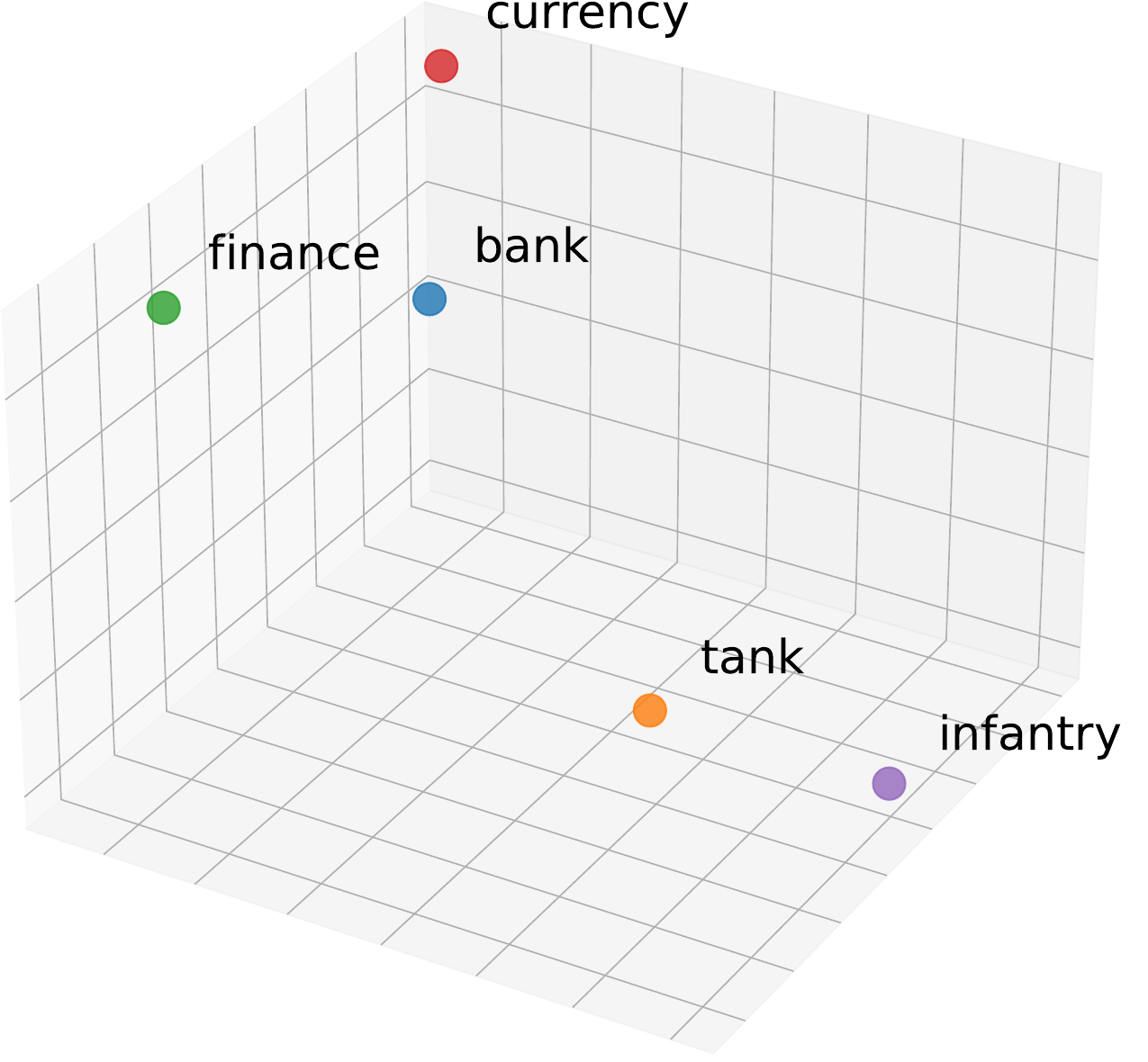}
		\caption{Textual word embeddings}
		\label{fig:semantic_s}
	\end{subfigure}
	
	\caption{
		A visualisation of the fundamental difference between (a) AWEs and (b) TWEs.
		(a) Preserves word-type similarity among different acoustic realisations of five different word types.
		(b) Preserves semantic relatedness among the same word types mapping to written words.
	}
	\label{fig:semantic_goal}
\end{figure}

We remind the reader of the properties and differences between AWEs and textual word embeddings described in Section~\ref{sec:intro_awe}.
In the zero-resource setting, unsupervised representation learning of whole word speech segments primarily focuses on producing embeddings that allow discrimination between word segments of different word types.
In this context, we train our AWE models using isolated word pairs.
For example, the CAE-RNN (Section~\ref{ssec:contrastive_cae_rnn}) is trained to reconstruct a different instance of the same word type as the input, and the \system{ContrastiveRNN} (Section~\ref{sec:contrastive_rnn}) is explicitly trained to minimise the distance between instances of the same word type and maximise the distance between all instances of different word types.
Consequently, different instances of the same word type should have similar embeddings and will be different from instances from different word types.
These representations are learned without considering the context in which they appear within a spoken corpus.
{From this point onwards, we refer to these representations, which only encapsulate the acoustic properties linked to word type, as phonetic AWEs.
}
An example vector space containing these phonetic AWEs is shown in Figure~\ref{fig:semantic_p}.
Here, different spoken instances of the same word type, for example ``bank'', end up close to each other.
Different spoken instances of the word ``tank'' will also end up close to each other.
Since ``bank'' and ``tank'' are phonetically similar, instances of both word types might also end up close to each other in vector space.
These are the type of embeddings generated in all our AWE experiments from previous chapters.

In contrast, in the text domain, word embedding models map written words to fixed vectors, capturing the contextual meaning of words within a text corpus.
These models learn from the co-occurrence information in large unlabelled text corpora.
For example, the Skip-gram model of~\citet{mikolov_efficient_2013} employs a neural network model that, given a target word, aims to predict the neighbouring words in that sentence (more in Section~\ref{sec:semantic_skipgram}).
As a result, words that are related in meaning end up having similar embeddings.
Moreover, these textual word embedding (TWE) models assign a unique embedding to each word type within a fixed vocabulary.
Figure~\ref{fig:semantic_s} illustrates an example semantic space, where the written words ``finance'', ``bank'', and ``currency'' are mapped nearby, and ``tank'' and ``infantry'', reflecting their shared contextual usage.

Instead of written words, our goal in this chapter is now to map spoken word segments to vector representations that preserve this contextual relationship among words described by~\citet{mikolov_efficient_2013} (Figure~\ref{fig:semantic_s}).
A unique aspect of our problem lies in the continuous nature of speech, where each spoken word segment will be different from another.
Consequently, each spoken segment will have a unique vector representation, which is different to TWEs where a  single vector is assigned to each word type.

\begin{figure}[!b]
		\centering
		\includegraphics[width=0.4\linewidth]{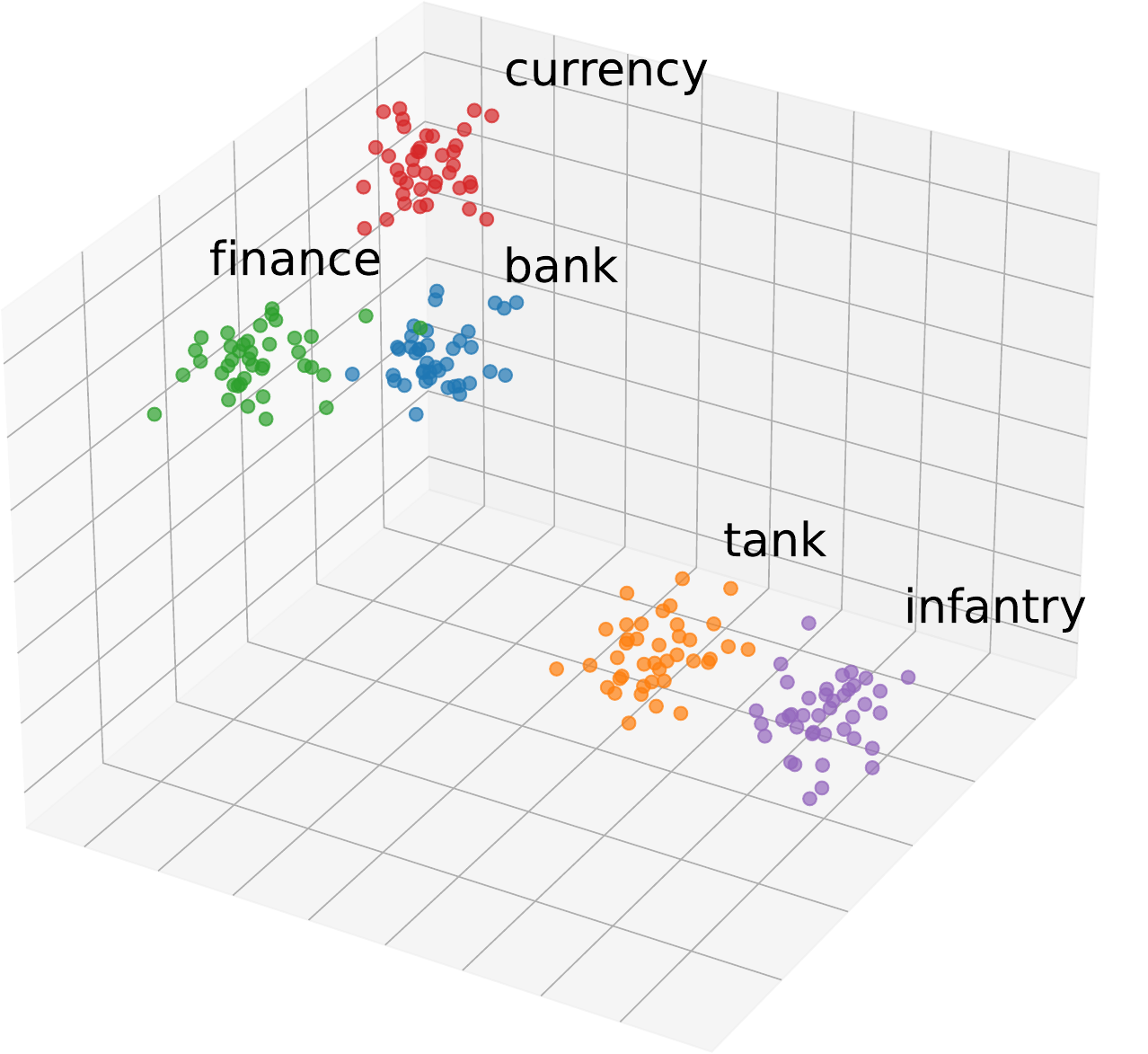}
		\caption{
		A visualisation of a semantic acoustic word embedding space.
		}
		\label{fig:semantic_ps}
\end{figure}

For instance, we would now like to see spoken instances of ``finance'', ``currency'', and ``bank'' mapped close to each other and, ``tank'' and ``infantry'', to be in close proximity in the resulting acoustic-semantic space, as illustrated in Figure~\ref{fig:semantic_ps}.




As part of our approach to create semantic AWEs, we will extend the textual word embedding approach called skip-gram.
This approach has also been used as the basis for other existing semantic AWE approaches.
We therefore present the skip-gram approach in detail below, before going into further  details on previous work on semantic AWEs.

\mysection{Skip-gram}{Skip-gram}
\label{sec:semantic_skipgram}




The Skip-gram model of ~\citet{mikolov_efficient_2013} is trained on large amounts of text corpora.
The objective of this model is to learn word representations that are useful for predicting neighbouring words in a sentence.
Formally, given a sequence of training words, $w_1, w_2, w_3, \ldots, w_T$, the model aims to maximise the average log probability~\cite{mikolov_distributed_2013},
\begin{equation}
	\frac{1}{T} \sum_{t=1}^{T} \sum_{-c \le j \le c, j \ne 0} \text{log} \, p(w_{t+j}|w_t),
	\label{eqn:word2vec_objective}
\end{equation}
where $c$ is the number of words before and after the target centre word $w_{t}$.
Consider an example of multiple pairs of target and context words as training samples for a single sentence in a text corpus, shown in Table~\ref{tbl:semantic_ex_pairs}.
\begin{table}[!t]
	\caption{Training samples extracted from a sentence using a context size of two. Example sentence: ``two boys are kicking a ball to each other in the park".}	
	\begin{tabularx}{1\linewidth}{@{\extracolsep{60pt}}Lll}
		\toprule
		Context size = 2 & Target word  & Context words \\
		\midrule
		\text{[\textbf{two} boys are]} & two & boys, are \\
		\text{[two \textbf{boys} are kicking]} & boys & two, are, kicking \\
		\text{[two boys \textbf{are} kicking a]} & are & two, boys, kicking, a\\
		\text{[boys are \textbf{kicking} a ball]} & kicking & boys, are, a, ball \\
		\ldots & \ldots & \ldots\\
		\text{[each other \textbf{in} the park]} & in & each, other, the, park \\	
		\text{[other in \textbf{the} park]} & the & other, in, park \\
		\text{[in the \textbf{park}]} & park & in, the \\		 
		\bottomrule		
	\end{tabularx}
	\label{tbl:semantic_ex_pairs}
\end{table}
For the training sample with input (the target word) ``kicking" the model will learn to predict the words ``boys", ``are", ``a", ``ball". 

\begin{figure}[!b]	
	\centering
	\centerline{\includegraphics[width=.95\linewidth]{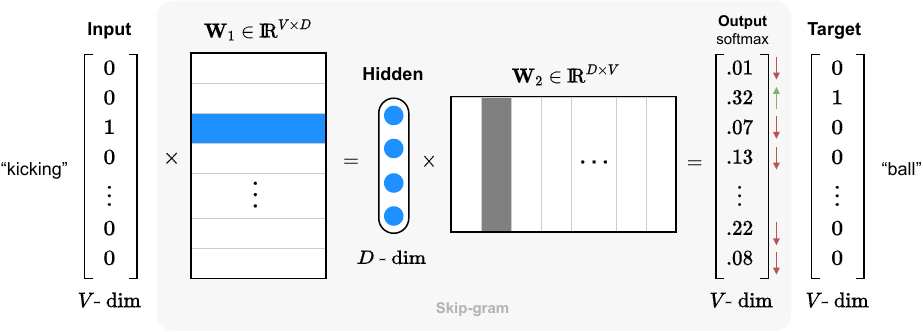}}
	\caption{The Skip-gram model architecture.}
	\label{fig:semantic_skipgram}
\end{figure}

Figure~\ref{fig:semantic_skipgram} shows the Skip-gram model architecture.
The model consists of two fully connected feedforward layers with weight parameters $\mathbf{W}_1 \in \mathbb{R}^{(V \times D)}$ between input and hidden layer, and $\mathbf{W}_2 \in \mathbb{R}^{(D \times V)}$ between hidden layer and output, where $V$ is the vocabulary size (number of unique words in text corpus) and $D$ is the dimension of the hidden layer.
For each training sample, the target and context words are encoded into one-hot vectors where the non-zero index corresponds to a word's index in the vocabulary set.
Given a one-hot vector as input, the model produces a latent embedding of size $D$.
The embedding is then given as input to the second layer to produce an output vector of size $V$.
A softmax function is applied to the output vector to normalise the output to a probability distribution over $V$ classes, where the value at each index shows the prediction of the model given the input word.
The model then updates weight parameters $\mathbf{W}_1$ and $\mathbf{W}_2$ to maximise the objective function in Equation~\ref{eqn:word2vec_objective}.
Note that the one-hot input vector selects the $i$-th row from weight parameter $\mathbf{W}_1$ to produce the latent embedding.
The one-hot vector essentially acts as a lookup table for each word in the vocabulary.
Therefore, after training, $\mathbf{W}_1$ represents the learned word embeddings where the $i$th row of size $D$ represents the $i$th word in the vocabulary set.

Keep in mind that the meaning of the words itself is not important, but rather the context in which they appear with other words, such that the words can be replaced with any unique character sequence (i.e. label) as long as the sequence in which they appear are retained.
This allows a model to be trained on a text corpus with unknown word labels.
In one of our speech models, we make use of this property by inferring unknown word labels for spoken word segments.


The training strategy of the Skip-gram model, which involves predicting surrounding words from a target centre word, served as an inspiration to the speech community as a means to obtain semantic embeddings from unlabelled spoken word segments. 
In the following section, we explore the existing body of work addressing this problem.

\mysection{Related work}{Related work}
\label{sec:semantic_related}

Learning semantic AWEs from speech is extremely difficult due to channel variability, noise, and speaker-specific information that are not present in written text.
Moreover, the presence of inherent phonetic interference poses a significant obstacle to semantic AWE modelling.
For example, distinguishing between spoken instances of ``steal'' and ``steel'' is challenging, even with the context provided, since from a model's perspective they would appear identical.
Some studies, therefore, consider the use of another modality as a grounding signal, for example, images~\cite{harwath_deep_2015, kamper_visually_2017, kamper_semantic_2019} or text labels~\cite{abdullah_integrating_2022} as a weak form of supervision.

Only a handful of studies have looked at learning semantic AWEs from unlabelled speech alone.
\citet{chung_speech2vec_2018}  were the first to present and gain widespread recognition for producing semantic AWEs without word labels.
They introduced a model named \system{Speech2Vec}, borrowing the training methodology of the Skip-gram model, by training on pairs of context words.
\system{Speech2Vec} uses an RNN decoder-encoder structure to accommodate variable-length speech segments as input and target output (similar to the \system{CAE-RNN}), while the Skip-gram model uses a two-layer fully connected feedforward network with one-hot encoded vectors as input and output.

\system{Speech2Vec} is evaluated as follows: After training, all word segments from the training data are embedded using the encoder RNN. 
Each word type is then represented by a single embedding by averaging all instances of that type.
The averaged embeddings are then used in multiple intrinsic word similarity tasks measuring semantic relatedness.
With this setup, \system{Speech2Vec} achieved surprisingly good results, even outperforming textual word embeddings (trained on the corresponding transcripts) in some tasks.
This certainly raised our suspicion, as well as others~\cite{chen_reality_2023}, concerning the validity of the embeddings produced by this model.
In Section~\ref{ssec:semantic_scratch_s2v} we describe the model in more detail and address some of the concerns with \system{Speech2Vec}.
Nevertheless, \system{Speech2Vec} has not been evaluated for single sample performance nor word segments that were not included in the model's training data. 
The quality of embeddings produced for unseen segments is essential to determine the usefulness of embeddings for downstream applications, such as semantic QbE.



\citet{chen_phonetic-and-semantic_2019} proposed a two-step framework towards semantic AWE modelling.
Speech segments are first given to a speaker disentanglement network (trained using the unlabelled word segments), which outputs two embeddings, separating speaker and phonetic information.
The idea is that the phonetic embedding should only contain information related to word type (the same property of our multilingual AWEs).
To produce semantic embeddings, they construct context training pairs using the phonetic embeddings  and train a network optimising a contrastive loss similar to the \system{ContrastiveRNN} (Equation~\ref{eqn:contrastive_loss}).
Their results are not compared to \system{Speech2Vec} and also use averaged embeddings from the training data for evaluation.

\system{Speech2Vec}~\cite{chung_speech2vec_2018} and the work of \citet{chen_phonetic-and-semantic_2019} only consider the unlabelled speech from the target language during model development.
In this chapter, we propose strategies for semantic AWE modelling, utilising a pre-trained multilingual AWE model.
We evaluate our models on an intrinsic word similarity task, measuring the quality of embeddings for single word segments not used during training.
To demonstrate the practical applicability of our embeddings, we also apply them to a downstream semantic QbE task.

In both cases~\cite{chung_speech2vec_2018, chen_phonetic-and-semantic_2019}, the problem is simplified by assuming that we know where words start and end in the target language (but the word classes are still unknown).
However, in a true zero-resource setting, word boundaries will be unknown.
In the experiments presented in this chapter, we also assume word boundaries are known.
This problem needs to be addressed in the future to consider complete unsupervised semantic AWE modelling (Section~\ref{sec:conclusion_future}).

\mysection{Semantic AWEs trained from scratch}{Semantic AWEs trained from scratch}
\label{sec:semantic_scratch}

Until now, our AWE models mapped speech segments from the same word type nearby in vector space (Figure~\ref{fig:semantic_p}).
In our notation, a speech segment $X = \left( \mathbf{x}_1, \mathbf{x}_2, \ldots, \mathbf{x}_T \right)$ is projected to a fixed vector $\mathbf{z}$.
From this point onwards we add the subscript \text{p} to the AWEs from previous chapters.
In other words, $\mathbf{z}_{\text{p}}$ indicates that the embedding preserves phonetic information related to word type only.

The goal of this chapter is to produce semantic embeddings $\mathbf{z}_{\text{s}}$.
These embeddings should not only reflect phonetic similarity but also capture word meaning (Figure~\ref{fig:semantic_ps}).
\system{Speech2Vec}~\cite{chung_speech2vec_2018} and the work of~\citet{chen_phonetic-and-semantic_2019} achieve this by only training on context pairs extracted from unlabelled speech data (from the target language).
We use the training strategies presented in these two methods as a baseline for our multilingual enhanced semantic AWE models (Section~\ref{sec:semantic_multilingual}).

\subsection{Speech2Vec}
\label{ssec:semantic_scratch_s2v}
\begin{figure}[!t]	
	\centering
	\centerline{\includegraphics[width=.95\linewidth]{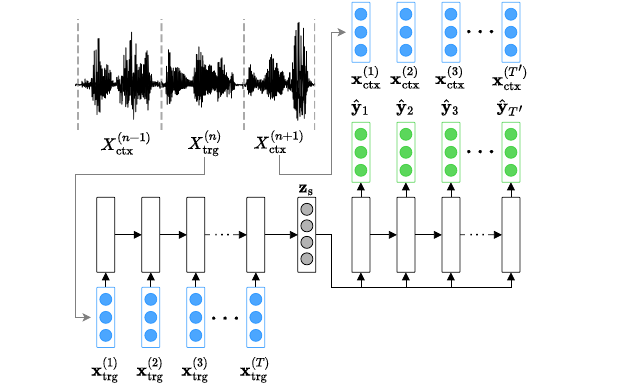}}
	\caption{The \system{Speech2Vec} model architecture. 
		This is similar to the \system{CAE-RNN} model architecture (Figure~\ref{fig:cae_rnn}) but adapted for context word pairs $(X_{\text{trg}}, X_{\text{ctx}})$ instead of positive pairs $(X, X^{\prime})$. Dashed lines in the audio indicate true word boundaries.}
	\label{fig:semantic_s2v}
\end{figure}

\system{Speech2Vec} is a variant of the \system{CAE-RNN} AWE model (Section~\ref{ssec:contrastive_cae_rnn}).
The \system{CAE-RNN} is an encoder-decoder model that takes a speech segment as input and aims to reconstruct a different instance of the same word type.
\system{Speech2Vec} uses the same encoder-decoder framework, but instead of training on pairs of instances from the same word type (as for phonetic AWE modelling), it uses context word pairs $(X_{\text{trg}}, X_{\text{ctx}})$.
$X_{\text{trg}}$ is a target centre word segment while $X_{\text{ctx}}$ is a context word appearing somewhere in a window around the centre.
These context pairs are constructed similarly to those used for training the Skip-gram model on text data (see Table~\ref{tbl:semantic_ex_pairs}). However, in this case, $X_{\text{trg}}$ and $X_{\text{ctx}}$ represent spoken words.
Importantly, these context pairs are constructed without word labels by only considering the relative position of words within an utterance (requiring word boundaries).
Figure~\ref{fig:semantic_s2v} shows the \system{Speech2Vec} model for a single training pair.
\system{Speech2Vec} optimises the reconstruction loss of Equation~\ref{eqn:contrastive_cae_loss}, where $X^\prime$ in that equation is replaced with $X_{\text{ctx}}$.
Ideally the resulting embeddings $\mathbf{z}_{\text{s}}$ should therefore be similar for words that co-occur in the speech data.
Although this approach seems theoretically sound, concerns have been raised regarding the performance of this model. We address this now.

\textbf{Criticism.}
The authors of \system{Speech2Vec}~\cite{chung_speech2vec_2018} did not make their code available, despite multiple requests from the public.
They did however release a set of embeddings produced by the model.\footnote{\href{https://github.com/iamyuanchung/speech2vec-pretrained-vectors}{https://github.com/iamyuanchung/speech2vec-pretrained-vectors}}
With numerous attempts to replicate the model's embeddings exactly as described in the paper, and reaching out to the authors without receiving a response, we have not been able to achieve results close to theirs.
We are also not aware of any credible publications that successfully replicated \system{Speech2Vec} to obtain similar quality semantic embeddings on the same evaluation tasks. 
Other failed attempts to replicate the results of \system{Speech2Vec} include the works of  Jang\footnote{\href{https://github.com/yjang43/Speech2Vec}{https://github.com/yjang43/Speech2Vec}}, Wang\footnote{\href{https://github.com/ZhanpengWang96/pytorch-speech2vec}{https://github.com/ZhanpengWang96/pytorch-speech2vec}} and EarthSpecies\footnote{\href{https://github.com/earthspecies/audio-embeddings/issues/6}{https://github.com/earthspecies/audio-embeddings/issues/6}}.

\citet{chen_reality_2023} recently performed an in-depth analysis questioning the authenticity of the original \system{Speech2Vec} model.
On a homophone inspection analysis, they show that \system{Speech2Vec} is immune to phonetic interference.
For example, the model will be able to perfectly discriminate between spoken instances of ``steal'' and ``steel'' despite being phonetically identical, which should inevitably affect model performance.
By inspecting the vocabulary of the released embeddings, they found it exactly matches the vocabulary of a word embedding model trained on the corresponding transcripts. 
They show this should not be true given that some information (words) are lost when applying a forced-aligner to segment the audio.
Based on these findings, there is compelling evidence that the embeddings released by \system{Speech2Vec} have been influenced by word labels in some way.

Moreover, the \system{CAE-RNN} AWE was originally designed to learn embeddings that are invariant to speaker identity, channel characteristics, etc. (Section~\ref{ssec:contrastive_cae_rnn}).
In the \system{Speech2Vec} setup, training pairs are from the same speaker and utterance (the way context pairs are constructed), so that the model is not explicitly able to learn embeddings invariant to those properties.
Despite the scepticism of \system{Speech2Vec} we include the model as a baseline in our experiments.
We use our own version of the model and perform experiments in a different setup (but still refer to it as \system{Speech2Vec}).


\subsection{Semantic ContrastiveRNN}
\label{ssec:semantic_scratch_crnn}

Similar to how the \system{CAE-RNN} AWE model is adapted for semantic AWE modelling (\system{Speech2Vec}), we modify the \system{ContrastiveRNN} AWE model (Section~\ref{sec:contrastive_rnn}) to learn semantic AWEs.
We substitute $X_a$, $X_p$ and $X_{n_{1}}, \ldots, X_{n_{K}}$ in Equation~\ref{eqn:contrastive_loss}, with $X_{\text{trg}}$ (target centre), $X_{\text{ctx}}$ (context), and out-of-context words.
Now, by minimising the contrastive loss of Equation~\ref{eqn:contrastive_loss}, the model should optimise relative distances between embeddings such that it reflect semantic similarity.
This is similar to the contrastive optimisation for semantic AWEs by~\citet{chen_phonetic-and-semantic_2019}, without speaker disentanglement (requiring speaker labels).

For both \system{Speech2Vec} and semantic \system{ContrastiveRNN}, a semantic AWE model is trained from scratch, therefore requiring the models to learn phonetic and semantic similarity simultaneously.
We now look at new semantic AWE modelling strategies that we propose, utilising a pre-trained multilingual AWE model.



\mysection{Our approach: Semantic AWEs using multilingual transfer}{Our approach: Semantic AWEs using multilingual transfer}
\label{sec:semantic_multilingual}
This section introduces a novel semantic AWE modelling approach.
We propose incorporating phonetic word knowledge through a pre-trained multilingual AWE model during semantic modelling.
Incorporating phonetic details may assist the model to prioritise learning word semantics by filtering out unrelated characteristics, for example, speaker information.
Specifically, we propose three strategies utilising a pre-trained multilingual AWE model in various ways.



\subsection[ContrastiveRNN with multilingual initialisation]{ContrastiveRNN with multilingual initialisation}
\label{ssec:semantic_mul_init}

Instead of training semantic models from scratch (Section~\ref{sec:semantic_scratch}) we warm-start them using the learned weights of a pre-trained multilingual AWE model.
In our experiments, we use the learned weights of a multilingual AWE model's encoder to initialise the encoder RNN of the \system{ContrastiveRNN} and subsequently train a semantic \system{ContrastiveRNN} (Section~\ref{ssec:semantic_scratch_crnn}).
This learned parameter initialisation method is similar to the multilingual adaptation strategy explored in Chapter~\ref{chap:adaptation}: a pre-trained multilingual AWE model is adapted to a target low-resource language using discovered word pairs.
Here, we adapt the multilingual weights to produce semantic AWEs by training on context pairs from the target language.



\subsection[Projecting multilingual AWEs]{Projecting multilingual AWEs} 
\label{ssec:semantic_mul_proj}

\begin{figure}[!t]
	\centering
	\includegraphics[width=0.8\linewidth]{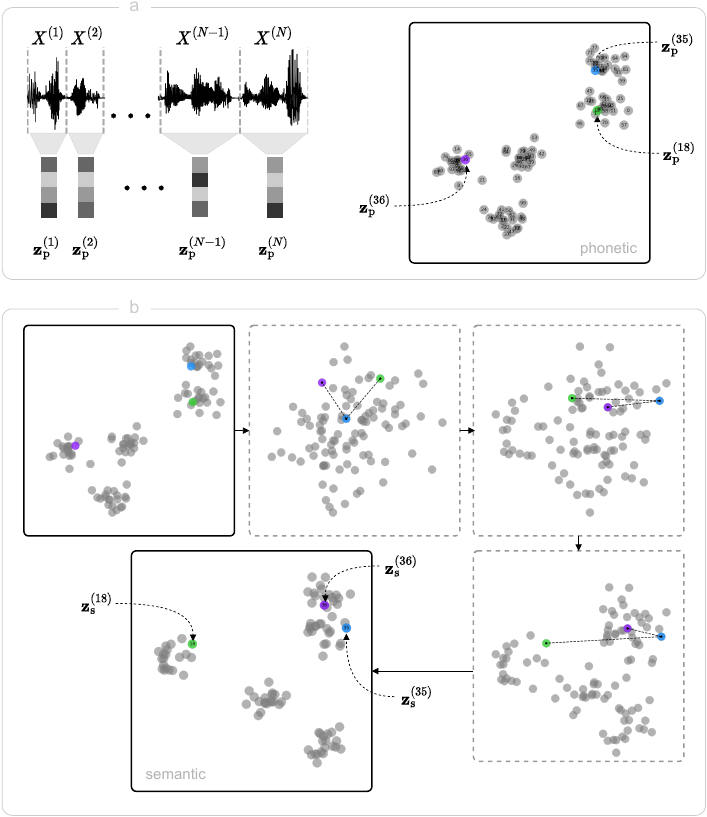}
	\caption{
		An illustration of the \system{Project AWE} semantic strategy, as detailed in Section~\ref{ssec:semantic_mul_proj}.	
		}
	\label{fig:semantic_projection}
\end{figure}

An alternative approach would be to first apply the multilingual AWE model to the segmented speech segments $\{X^{(n)}\}^{N}_{n=1}$ where $N$ is the number of word tokens in the training corpus.
Then, train a projection network, mapping the resulting phonetic AWEs to a new semantic AWE space.
Figure~\ref{fig:semantic_projection} illustrates this mapping process.
Applying the multilingual model to the unlabelled speech segments $\{X^{(n)}\}$ produces a set of phonetic AWEs $\{\mathbf{z}^{(n)}_{\text{p}}\}$ (upper part in figure).
In the phonetic space, we note that word segments are mapped to unknown word classes (five in this simplified example).
This implies $\mathbf{z}_{\text{p}}^{(35)}$ (blue) and $\mathbf{z}_{\text{p}}^{(71)}$ most likely belong to the same unknown word class, as well as, $\mathbf{z}_{\text{p}}^{(18)}$ (green) and $\mathbf{z}_{\text{p}}^{(87)}$.
Groups containing  $\mathbf{z}_{\text{p}}^{(35)}$ and $\mathbf{z}_{\text{p}}^{(18)}$ are in close proximity, indicating they might also share phonetic similarities.

In the semantic space, we would like to see embeddings of segment $X^{(35)}$ and $X^{(36)}$ to be positioned close to each other (assuming the unknown classes associated with these segments frequently co-occur). 
We do this by training a projection network, mapping the phonetic AWEs to a new set of semantic embeddings $\{\mathbf{z}^{(n)}_{\text{s}}\}$.
We train this network using context pairs and the contrastive loss of Equation~\ref{eqn:contrastive_loss}, optimising the distances between the output embeddings $\mathbf{z}_{\text{s}}$.
The lower part of Figure~\ref{fig:semantic_projection} illustrates the transformation of the phonetic space into a new semantic space, through optimising the contrastive loss.
We now observe neighbouring word segments $X^{(35)}$ and $X^{(36)}$ end up having similar semantic representations $\mathbf{z}^{(35)}_{\text{s}}$ and $\mathbf{z}^{(36)}_{\text{s}}$.



This approach is similar to the semantic AWE method presented by~\citet{chen_phonetic-and-semantic_2019}.
Here they used data from the target language to train a network separating speaker and phonetic information (which requires speaker labels).
Instead, we use a pre-trained multilingual AWE model designed for generating phonetic representations for unlabelled word segments.



%


\subsection{Cluster+Skip-gram} 
\label{ssec:semantic_mul_cluster}

\begin{figure}[!b]
	\centering
	\includegraphics[width=0.75\linewidth]{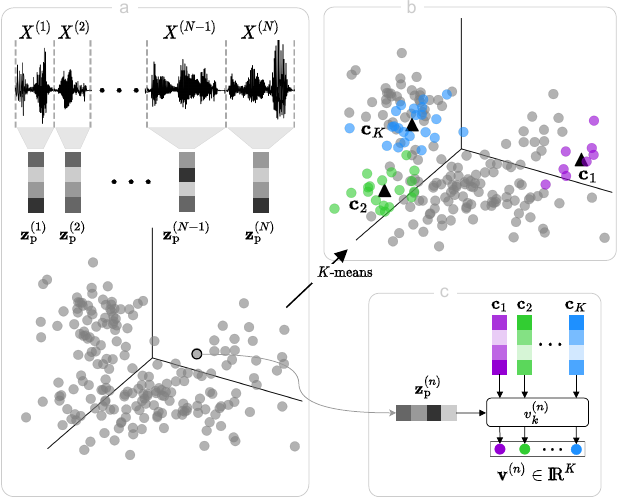}
	\caption{ Our \system{Cluster+Skip-gram} semantic AWE approach.
		Speech segments $X^{(n)}$ are represented by soft pseudo-word label vectors $\mathbf{v}^{(n)}$ 
		which are then used to train a Skip-gram model.}
	\label{fig:cluster+skipgram}
\end{figure}

This new approach is based on the Skip-gram model
(Section~\ref{sec:semantic_skipgram}).
The Skip-gram model uses a fixed dictionary of discrete word class labels to create input and output vectors for training on text data.
In the low-resource setting we do not have word labels for speech segments.

Instead, we use the phonetic similarities in the original AWE space to derive a soft pseudo-word label for each speech segment. 
This process is illustrated in Figure~\ref{fig:cluster+skipgram}.
First, a pre-trained multilingual AWE model is applied to the segmented speech corpus $\{X^{(n)}\}$, producing a set of phonetic AWEs $\{\mathbf{z}_\text{p}^{(n)}\}$.
The colourless embedding space represents the unsupervised setting where embeddings have no associated word types or labels.
We then apply $K$-means clustering to the phonetic embedding space, producing a set of centroids $\{\mathbf{c}_{k}\}^{K}_{k=1}$, with $K$ representing the predetermined number of clusters for partitioning the vector space.
\begin{figure}[!b]
	\centering
	\centerline{\includegraphics[width=0.55\linewidth]{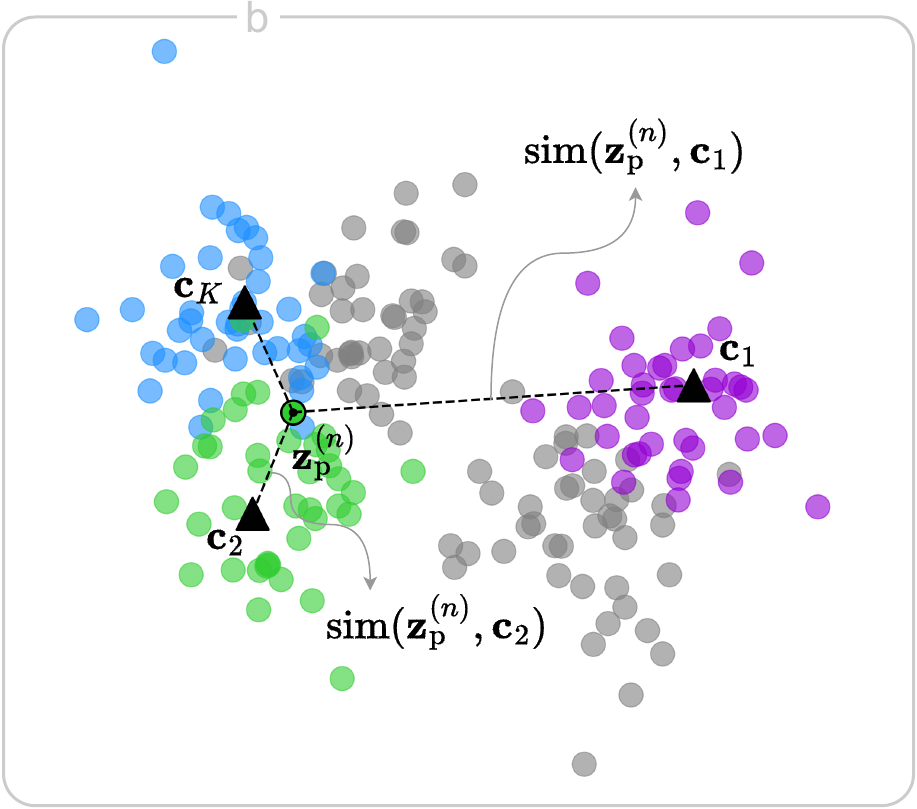}}
	\caption{Given an acoustic embedding $\mathbf{z}_{\text{p}}^{(n)}$ (assigned to cluster 2), instead of using a one-hot encoded input $\mathbf{x} = \{0, 1, 0, \cdots, 0\}$, the value at each index is set using Equation~\ref{eqn:semantic_soft_label}, for example, $\mathbf{v} = \{0.03, 0.45, 0.01, \cdots, 0.43\}$. Note that the values at index $2$ and $K$ are the highest given that $\mathbf{z}_{\text{p}}^{(n)}$ is the closest to cluster means $\mathbf{c}_2$ and $\mathbf{c}_K$. Black triangles indicate cluster means.}
	\label{fig:semantic_cluster_mean_dist}
\end{figure}

Ideally, these clusters should resemble distinct word classes.
For example, all spoken instances of ``apple'' should be assigned to cluster one (purple) and all instances of ``water'' should be assigned to cluster two (green).
Note, with a perfect clustering, where all embeddings from an unknown word label are exclusively assigned to a unique cluster, training a Skip-gram model using the cluster labels would be equivalent to training on the corresponding text labels (see bottom of Section~\ref{sec:semantic_skipgram}).
However, due to clustering inaccuracies, not all embeddings belonging to the same unknown word class will be assigned to the same cluster.
The main challenge arises from embeddings of the same word class being assigned to multiple clusters; a consequence of the uncertainty regarding vocabulary size and word class frequency range.
For example, embeddings assigned to cluster $2$ (green) and cluster $K$ (blue) in Figure~\ref{fig:semantic_cluster_mean_dist} might represent the same word class given that there is no clear separation between them.
We experimented with cluster labels as training labels for Skip-gram modelling, but this gave very poor performance on development data.

Rather than assigning a hard label to each segment based on its associated cluster number, we calculate a soft vector label of an AWE belonging to each cluster:
\begin{equation}
	v_{k}^{(n)} = \frac{\exp\left(-\text{sim}(\textbf{z}_{\text{p}}^{(n)}, \textbf{c}_k)/\sigma^2\right)}{\sum_{j=1}^{K}\exp\left(-\text{sim}(\textbf{z}_{\text{p}}^{(n)}, \textbf{c}_j)/\sigma^2\right)}
	\label{eqn:semantic_soft_label}
\end{equation}
where $\text{sim}(\cdot)$ denotes cosine similarity and $\sigma$ is a hyperparameter controlling the
influence of distant centroids.
This non-linear scaling approach assigns more weight to embeddings closer to the target embedding $\mathbf{z}^{(n)}_{\text{p}}$.
Figure~\ref{fig:semantic_cluster_mean_dist} extends the soft labelling calculation illustration in the bottom right in Figure~\ref{fig:cluster+skipgram}. 
Now, each word segment $X^{(n)}$ is associated with a new unique vector $\mathbf{v}^{(n)} \in \mathbb{R}^{K}$, where the value at index $k$ indicates the likelihood of the segment being contained in cluster $k$.
Even if segments from the same unknown word class are assigned to multiple clusters, their relative position in the embedding space is captured in $\mathbf{v}$.

\begin{figure}[!b]
	\centering
	\centerline{\includegraphics[width=0.95\linewidth]{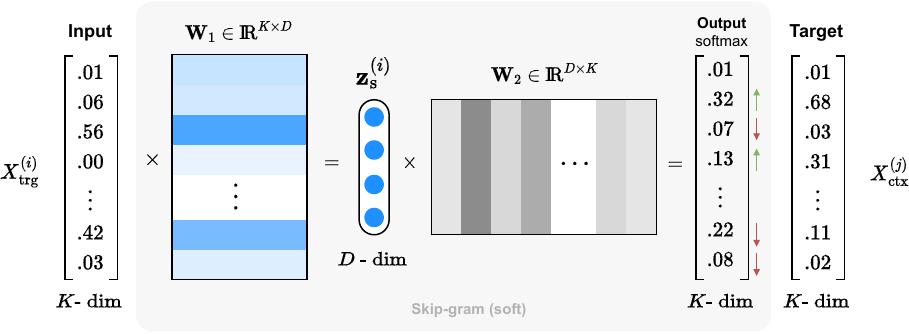}}
	\caption{The skip-gram model using soft inputs and targets.}
	\label{fig:semantic_skipgram_soft}
\end{figure}

Lastly, we use these soft vectors as inputs and targets to train a Skip-gram model, illustrated in Figure~\ref{fig:semantic_skipgram_soft}. 
Compared to the Skip-gram model trained on text data (Figure~\ref{fig:semantic_skipgram}), there is a key difference: using these soft representations as input, compared to a one-hot vector. This means that $\textbf{W}_1$ do not act as a look-up table, instead, each row contributes to the hidden embedding with corresponding weight (visually illustrated with shading in Figure~\ref{fig:semantic_skipgram_soft}).
In other words, the hidden layer contains the weighted sum of all rows in $\textbf{W}_1$, which we take as the semantic embedding $\textbf{z}_{\text{s}}$.
The same applies to $\textbf{W}_2$ when using soft targets.


During inference, a unique embedding $\textbf{z}_{\text{s}}$ is produced for a word segment by first obtaining a phonetic AWE using the multilingual model, then calculating representation $\mathbf{v}$ using the trained cluster means, and finally applying it to the learned weight matrix $\mathbf{W}_1$.

\mysection{Semantic query-by-example}{Semantic query-by-example}
\label{sec:semantic_qbe}

In Chapter~\ref{chap:related} and Chapter~\ref{chap:hatespeech}, we performed QbE experiments using AWEs to calculate the phonetic similarity between speech segments. 
This involved searching an unlabelled audio corpus with a query word segment, aiming to retrieve utterances containing instances of the query. We refer to this as exact QbE, where the query only retrieves utterances if it contains a word sharing the same class label.
In this chapter, apart from intrinsic AWE evaluations, we consider the task of \textit{semantic} QbE as a downstream application for semantic AWEs.
In semantic QbE the goal is to retrieve utterances from an audio collection that not only match segments containing the exact query segment but also utterances which contain word segments that are semantically similar to the query.
For example, using a query segment, ``football'', the system should retrieve utterances containing exact query matches for instance, ``the football player is injured'', as well as utterances containing word segments semantically related to the query for instance, ``two kids play soccer in the field'', and ``the player was offside''. 


We perform our semantic QbE experiments using a similar setup as in~\cite{kamper_semantic_2019-1}.
They performed experiments to produce semantic embeddings on the same data corpus we use for our semantic AWE experiments.
However, during training, they use images as a weak form of supervision to produce semantic embeddings.
To the best of our knowledge, we are the first to perform semantic QbE using semantic AWEs trained on unlabelled speech segments.

\mysection{Setup}{Setup}
\label{sec:semantic_setup}

In this section we outline the data corpus we use for our semantic AWE modelling.
We provide implementation details for the two semantic models, \system{Speech2Vec} and \system{ContrastiveRNN}, trained from scratch (Section~\ref{sec:semantic_scratch}).
Subsequently, we present the specifics of the three semantic AWE models leveraging a pre-trained multilingual AWE model we introduced in Section~\ref{sec:semantic_multilingual}.
Finally, we discuss the evaluation process for our semantic AWEs.

\subsection[Data]{Data}
\label{ssec:semantic_setup_data}

\begin{figure}[!b]
	\centering
	\includegraphics[width=0.99\linewidth]{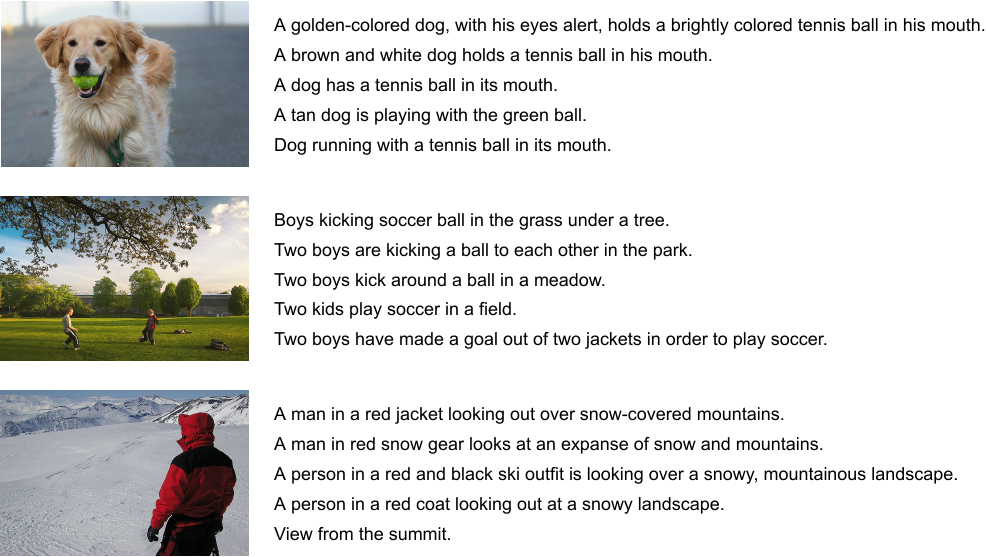}
	\vspace{2mm}
	\caption{Examples of images with corresponding spoken captions from the FACC.}
	\label{fig:flickr_samples}
\end{figure}

For this experiment, we use the Flickr8k Audio Captions Corpus (FACC)~\cite{harwath_deep_2015}.
The corpus contains 40k spoken captions in English describing the content of a Flickr image~\cite{rashtchian_collecting_2010}.
These images depict scenes for which humans have given verbal descriptions.
Example images with corresponding spoken captions are shown in Figure~\ref{fig:flickr_samples}.
Note, the text in Figure~\ref{fig:flickr_samples} is only for the purpose of illustration and not used during training (the corresponding speech are used). 
The corpus is divided into train, development, and test splits containing $30\,000$, $5\,000$, $5\,000$ utterances respectively.

Although the amount of training data in the Flickr8k corpus is relatively low, the content of the data suits our semantic experiments well: images come from a fairly narrow domain, with five different spoken captions describing the content of each image.
Therefore, semantic concepts reoccur in different utterances.
This dataset has also been used for acoustic semantic modelling by others~\cite{harwath_deep_2015, kamper_semantic_2019,kamper_semantic_2019-1}.
We train and evaluate our models using two sets of speech features. 
Firstly, we employ 13-dimensional MFCCs. 
Secondly, as in Chapter~\ref{chap:hatespeech}, we conduct experiments with self-supervised speech features: we use the 12th transformer layer of the multilingual XLSR model~\cite{babu_xls-r_2022} to get 1024-dimensional features.
Utterances are normalised per speaker and segmented using true word boundaries from forced alignments~\cite{mcauliffe_montreal_2017}.
We use these word segments from the training data to construct context word pairs.
For all the semantic models, we use a context window of three words before and after a centre word: this yields roughly two million context pairs in our training data.

\subsection[Semantic models trained from scratch]{Semantic models trained from scratch}

The encoder and decoder of our \system{Speech2Vec} (Section~\ref{ssec:semantic_scratch_s2v}) implementation each consists of three unidirectional RNNs with 400-dimensional hidden vectors and an embedding size of 100.
The semantic \system{ContrastiveRNN} (Section~\ref{ssec:semantic_scratch_crnn}) uses the same encoder structure.
We train both these models using the context pairs extracted from the training set.
For the \system{ContrastiveRNN} we randomly sample word segments outside the context window as negatives; for each positive, we sample 20 negatives.

\subsection[Semantic models using multilingual transfer]{Semantic models using multilingual transfer}
\label{ssec:semantic_setup_mul}

All the multilingual transfer strategies presented in Section~\ref{sec:semantic_multilingual} use a pre-trained multilingual AWE model.
The multilingual model is trained in the same way multilingual models were trained in previous chapters, with embedding dimension set to 100.
We train a multilingual \system{CAE-RNN} AWE model on labelled data from four different well-resourced languages included in Common Voice~\cite{ardila_common_2020}: Italian, Dutch, Russian, and Czech.
Note, the entire speech corpus is in English and therefore not included as a training language.
We pool the data from all four languages and extract 300k training pairs.

\textbf{Multilingual initialisation. (Section~\ref{ssec:semantic_mul_init})}
For the multilingual weight initialisation method, we use the trained parameters of the multilingual \system{CAE-RNN} encoder to initialise the \system{ContrastiveRNN}'s encoder.
Furthermore, we freeze the weights of the first two encoder layers to preserve the phonetic structure of input words during further training. 
This initialisation approach is similar to how we initialised AWE models for adaptation in Chapter~\ref{chap:adaptation}: a multilingual AWE model is adapted using discovered word pairs from the target language.
Here, we adapt for semantic AWEs by training on context pairs from the target corpus.

\textbf{Project AWE. (Section~\ref{ssec:semantic_mul_proj})}
For our projection network, we use a feed-forward network of two linear layers with an inner dimension of 1024 and input and output dimensions of 100.
The input dimension is set to match the embedding dimension produced by the multilingual AWE model.
The inner and output dimensions are chosen based on experimenting on development data. 
Again, we train on context pairs and sample 20 negatives for each positive.
The network optimises by minimising the contrastive loss of Equation~\ref{eqn:contrastive_loss}.

\textbf{Cluster+Skip-gram. (Section~\ref{ssec:semantic_mul_cluster})}
For $K$-means clustering, we use $K = 5\,000$ clusters and set $\sigma = 0.01$ in Equation~\ref{eqn:semantic_soft_label}.
The number of unique word types in the target corpus is $6\,307$ where $2\,150$  appears once.
We experimented with different values for $K$ ranging from $1\,000$ to $10\,000$ which produced similar results.
We initialise our cluster means with a random selection of acoustic embeddings, perform 50 permutations, and keep the best means.
We reimplement the Skip-gram model~\cite{mikolov_efficient_2013} described in Section~\ref{sec:semantic_skipgram}, setting the latent embedding dimension to $D=100$.
The model is trained using the soft-vectors as input and optimises a negative log-likelihood with class probabilities set to the context target word's soft-vector.
The model is optimised using Adam~\cite{kingma_adam_2017}, with a learning rate of $0.001$.

\subsection[Evaluation]{Evaluation}

We evaluate the quality of resulting semantic embeddings in an intrinsic word similarity task.
To measure the applicability for real-life applications, we also apply the semantic AWEs in a downstream semantic QbE task for extrinsic evaluation.

\subsubsection[Intrinsic semantic AWE evaluation: Word similarity]{Intrinsic semantic AWE evaluation: Word similarity} 
\label{sssec:semantic_eval_sim}

\begin{table}[!t]	
	\mytable
	\caption{Word-pair similarity ratings extracted from the WordSim353~\cite{hill_simlex-999_2015} evaluation set.
		Similarity scores indicate word relatedness based on human judgements.}
	\addtolength{\tabcolsep}{10pt}
	\begin{tabularx}{.5\linewidth}{llr}
		\toprule				
		{money} & {bank} & {8.50} \\
		{physics} & {proton} & {8.12} \\ 
		{tennis}  & {racket} & {7.56} \\
		{journey} & {car} & {5.85} \\ 
		{smart} & {student} & {4.62} \\
		{problem} & {airport} &  {2.38} \\
		{professor} & {cucumber} & {0.31} \\
		\bottomrule		
	\end{tabularx}
	\label{tbl:semantic_eval_wordsim}
\end{table}

In the text domain, word embeddings are evaluated for semantics by measuring the similarity between word pairs through vector distance calculations. 
These vector distances are then compared to standard reference sets containing similarity ratings assigned by humans. 
As an illustration, Table~\ref{tbl:semantic_eval_wordsim} shows some reference word-pair similarities from the WordSim353~\cite{hill_simlex-999_2015} evaluation set.
For instance, in a resulting semantic space, the distance between the embedding of ``tennis'' and ``racket'' should be smaller than the distance between ``journey'' and ``car'', but greater than the distance between ``physics'' and ``proton.'' Furthermore, ``money'' and ``bank'' should have the smallest distance than all the shown word-pair distances.
Spearman's $\rho$ is used to quantify the similarity between the set of word-pair distances $(\text{WP}_{\text{embed}})$ and the set of word-pair human judgement ratings $(\text{WP}_{\text{human}})$, calculated as follows:

\begin{equation}
	\rho = \frac{\text{cov}(R(\text{WP}_{\text{embed}}), R(\text{WP}_{\text{human}}))}{\sigma_{R(\text{WP}_{\text{embed}})}\sigma_{R(\text{WP}_{\text{human}})}}
	\label{eqn:semantic_rho}
\end{equation}
with $R$ converting the scores to ranks, and $\text{cov}(\cdot\,, \cdot)$ and $\sigma_{\cdot}$ denotes, covariance and standard deviation, respectively.

\begin{figure}[!t]
	\begin{framed}[.8\linewidth]
		For the sentence below, select all words that could be used to search for the described scene.
		
		\bigskip
		\textbf{Sentence:} a skateboarder in a light green shirt.
		
		\bigskip
		
		\begin{enumerate*}[label=$\square$]
			\item dogs \medspace
			\item beach \medspace
			\item children \medspace
			\item white \medspace
			\item swimming
		\end{enumerate*}
		
		\smallskip
		\begin{enumerate*}[resume, label=$\square$]
			\item wearing \medspace
			\item skateboard \medspace
			\item {None of the above}
		\end{enumerate*}
		
	\end{framed}
	\caption{An example annotation job presented to a human for manual semantic labelling, with keyword options. In this case, ``skateboard'' was selected by all five annotators and ``wearing'' by four. Reproduced from~\cite{kamper_semantic_2019-1}.}
	\label{fig:semantic_eval_job}
\end{figure}

In our controlled experiment, our vocabulary is limited, causing many words from the standard evaluation sets typically used to assess word embeddings trained on large amounts of text to be absent from our training and evaluation data.
To address this, we compile our own list of word pairs that frequently appear in our dataset.
Instead of using human judgement, we assign similarity scores to those pairs based on the distance between textual word embeddings generated by an off-the-shelf Skip-gram model trained on the corresponding transcribed utterances.
Given that our goal is to obtain semantic embeddings similar to Word2Vec, we believe this approach is reasonable.

To obtain a single semantic embedding for each word class, we calculate the average of all embeddings from the same class and report $\rho_{\text{avg}}$. 
Given that we are particularly interested in obtaining semantic embeddings for individual word segments, single-sample performance is also measured by randomly selecting one instance of each spoken word.
This is repeated ten times and averaged to get a single score $\rho_{\text{single}}$.


\subsubsection[Semantic QbE]{Semantic QbE} 

\begin{table}[!b]
	\mytable
	\caption{
		The number of annotators (out of five) that selected the keyword for that utterance is shown.
		The majority vote indicates whether the keyword is deemed semantically relevant to an utterance, according to the labelling.
		Adapted from~\cite{kamper_semantic_2019-1}.}
	\begin{tabularx}{\linewidth}{llcc}
	\toprule				
	Keyword & Example utterance & Count \hphantom{0} &  Majority vote \\ [0.5mm]
	\midrule
	jumps & biker jumps off of ramp & 5 / 5 \hphantom{0} & Y \\ 
	ocean & man falling off a blue surfboard in the ocean & 5 / 5 \hphantom{0} & Y\\
	race & a red and white race car racing on a dirt racetrack & 5 / 5 \hphantom{0} & Y\\
	snowy & a skier catches air over the snow & 5 / 5 \hphantom{0} & Y\\
	bike & a dirt biker rides through some trees & 4 / 5 \hphantom{0} & Y\\
	children & a group of young boys playing soccer & 4 / 5 \hphantom{0} & Y\\
	young & a little girl in a swing laughs & 4 / 5 \hphantom{0} & Y\\
	field & two white dogs running in the grass together & 3 / 5 \hphantom{0} & Y\\
	carrying & small dog running in the grass with a toy in its mouth & 2 / 5  & Y\\ 
	face & a man in a gray shirt climbs a large rock wall & 2 / 5 &  N \\
	large & a group of people on a zig path through the mountains & 1 / 5  & N \\
	sitting & a baby eats and has food on his face & 1 / 5  & N \\
	hair & two women and a man smile for the camera & 0 / 5   & N \\
	\bottomrule		
\end{tabularx}
\label{tbl:semantic_eval_utts}
\end{table}

\label{sssec:semantic_eval_qbe}
We use the same setup as \cite{kamper_semantic_2019} to evaluate downstream semantic QbE performance.
Semantic labels for 1\,000 test utterances from FACC were collected from human annotators, using a set 67 keyword classes~\cite{kamper_semantic_2019-1}.
Specifically, each of the 1\,000 utterances was labelled by five annotators, indicating whether a particular keyword is semantically relevant to that utterance (regardless of whether the word instance appears verbatim in the utterance).
An example annotation job is shown in Figure~\ref{fig:semantic_eval_job}.

We use the majority decision to assign a hard label for whether a keyword is relevant to an utterance as in~\cite{kamper_semantic_2019-1}.
Table~\ref{tbl:semantic_eval_utts} shows annotations for some of the utterances.
For instance, a correct retrieval using the keyword ``jump'' would be ``bike jumps of ramp'', while an incorrect retrieval using the keyword ``face'' would be ``a man in a gray shirt climbs a large rock wall''.
Using these hard labels, we calculate semantic $P@10$, $P@N$, EER (metric details in Section~\ref{sec:related_qbe}).

Given the ambiguity among annotators when determining if a keyword is semantically related to an utterance, we also use the counts as soft labels and calculate Spearman's $\rho$ (Equation~\ref{eqn:semantic_rho}). 
Here, $\rho$ measures the correlation between a system's ranking and the actual number of annotators who deemed a query keyword relevant to an utterance. 

To simplify the QbE task, we still assume that ground truth word boundaries are known: a query AWE is therefore compared to AWEs for the word segments in an unlabelled search utterance.
\mysection{Experimental results}{Experimental results}
\label{sec:semantic_results}


We first compare embedding quality through measuring word similarity between the semantic AWE models trained from scratch (Section~\ref{sec:semantic_scratch}) and the semantic AWE models we proposed leveraging multilingual AWEs (Section~\ref{sec:semantic_multilingual}).
We then apply one of our proposed semantic AWE models in a semantic QbE search for extrinsic evaluation.
This is followed by a qualitative analysis of our semantic AWEs.

\subsection[{Intrinsic evaluation: Word similarity}]{Intrinsic evaluation: Word similarity}
\label{ssec:semantic_resuls_ieval}

\begin{table}[!t]
	\mytable
	\caption{    
		Spearman's $\rho$ (\%) for semantic word similarity of AWEs on development data. 
		Embeddings are either averaged over word segments of the same class, or a single segment is sampled per class. AWE models take either MFCC or XLSR features as input.
	}
	\captionsep
	\begin{tabularx}{\linewidth}{@{\extracolsep{20pt}}Lcccc}
		\toprule				
		& \multicolumn{2}{c}{MFCC} & \multicolumn{2}{c}{XLSR} \\
		\cmidrule(l){2-3} 
		\cmidrule(l){4-5}
		Model & { $\rho_{\text{single}}$ }&   $\rho_{\text{avg}}$ &  $\rho_{\text{single}}$ &   $\rho_{\text{avg}}$ \\ 
		\midrule		
		\multicolumn{2}{@{}l}{\underline{\textit{Trained from scratch:}} (Section~\ref{sec:semantic_scratch})} \\ 
		\system{Speech2Vec} & \hphantom{0}1.0 & \hphantom{0}5.3 & \hphantom{0}7.2 & 21.3 \\
		\system{ContrastiveRNN} & \hphantom{0}0.7 & \hphantom{0}5.1 & \hphantom{0}4.6 & 25.9 \\[1mm]
		
		\multicolumn{2}{@{}l}{\underline{\textit{Using multilingual transfer:}} (Section~\ref{sec:semantic_multilingual})} \\
		\system{ContrastiveRNN} multilingual init. & \hphantom{0}0.6 & \hphantom{0}5.5 & \hphantom{0}6.1  & 24.5 \\
		\system{Project AWE} & \hphantom{0}3.6 & 18.2 & 18.3 & 33.6 \\
		\system{Cluster+Skip-gram} & \textbf{18.0} & \textbf{35.6} & \textbf{35.9} & \textbf{41.7} \\
		\bottomrule		
	\end{tabularx}
	\label{tbl:semantic_results_intrinsic}
\end{table}

Table~\ref{tbl:semantic_results_intrinsic} presents the intrinsic scores of embeddings from the semantic AWE models, trained either from scratch (top section) or using multilingual transfer (bottom).
The benefit of multilingual transfer is evident in the scores of the projection and \system{Cluster+Skip-gram} approaches, with the latter outperforming all other models regardless of the input features used or whether single or averaged embeddings are evaluated.
The single-sample performance $\rho_{\text{single}}$ is particularly significant as it shows that individual representations can be compared accurately---a useful property for downstream applications such as semantic QbE (Section~\ref{sec:semantic_qbe}).

The \system{ContrastiveRNN} is the one exception that does not show a clear gain from initialising with multilingual weights compared to training from scratch.
As a sanity check, we evaluate the phonetic multilingual AWEs before semantic training, in other words, embeddings produced by the foundation multilingual AWE model (top of Section~\ref{ssec:semantic_setup_mul}) used for transfer in the bottom section.
We obtain results for $\rho_{\text{single}} = \text{0.59\%}$ and $\rho_{\text{avg}} = -\text{0.13\%}$. 
As expected, this indicates that phonetic multilingual AWEs do not capture semantic information.

Again, we see the benefit of using self-supervised speech representations as input to AWEs instead of conventional features, as also in previous work~\cite{van_staden_comparison_2021, sanabria_analyzing_2023}.
We therefore use XSLR features for all further experiments.
We apply the semantic AWE model achieving the best results on the word similarity task, \system{Cluster+Skip-gram}, in our semantic QbE search.

\subsection[Extrinsic evaluation: Semantic QbE]{Extrinsic evaluation: Semantic QbE}
\label{ssec:semantic_resuls_eeval}

Table~\ref{tbl:qbe} compares the \system{Cluster+Skip-gram} (semantic) and multilingual AWE (phonetic) models when used in a downstream QbE system.
We evaluate both exact and semantic QbE, where the latter gets awarded for retrieving exact query matches as well as utterances labelled as semantically related to the search query.
To situate results, we use a random baseline model that assigns a random relevance score to each utterance. (The relatively high scores of the random approach are due to the narrow domain of the evaluation data, Section~\ref{ssec:semantic_setup_data}.) 

\begin{table}[!t]
	\mytable
	\caption{
		Exact and semantic QbE results (\%) on test data. 
	}
	\captionsep
	\begin{tabularx}{1\linewidth}{@{}lCCCCCCC@{}}
		
		\toprule				
		& \multicolumn{3}{c}{Exact QbE} & \multicolumn{4}{c}{Semantic QbE} \\
		\cmidrule(l){2-4} 
		\cmidrule(l){5-8}
		Model &  $P@10$ & $P@N$ & EER & $P@10$ & $P@N$ & EER & 
		$\rho$\\ 
		\midrule
		
		\underline{\textit{Baselines}:} \\
		Random & \hphantom{0}5.0 & \hphantom{0}5.0 &  50.0 &  \hphantom{0}9.1 & \hphantom{0}9.1 & 50.0 & \hphantom{0}5.7  \\
		
		Multilingual AWE & \textbf{90.7} & \textbf{82.8} & \textbf{\hphantom{0}4.7}  & \textbf{92.8} & \textbf{59.4} & 24.1 & 17.0  \\ [1mm]
		
		\underline{\textit{Semantic AWE model}:} \\
		\system{Cluster+Skip-gram} & 85.8 & 71.3 & \hphantom{0}9.3  & 88.2 & 52.1 & \textbf{21.6} & \textbf{28.2} \\
		\bottomrule		
	\end{tabularx}
	\label{tbl:qbe}
\end{table}

\begin{table}[!b]	
	\mytable
	\caption{Semantic QbE results (\%) on test data, where any instance of a query is masked from the search utterances.}
	\captionsep
	
	\begin{tabularx}{\linewidth}{@{\extracolsep{5pt}}Lcccc}
		
		\toprule				
		Model & $P@10$ & $P@N$ & EER & Spearman's $\rho$ \\ 
		\midrule
		
		\underline{\textit{Baselines}:} & & & & \\
		Random & \hphantom{0}9.2 & \hphantom{0}9.0 & 50.0 & \hphantom{0}5.4 \\
		Multilingual AWE & 21.5 & 15.6 & 46.1 & \hphantom{0}6.4\\[1mm]
		
		\underline{\textit{Semantic AWE model}:} \\
		\system{Cluster+Skip-gram} & \textbf{29.9} & \textbf{23.1} & \textbf{32.3}  & \textbf{22.9}\\
		
		\bottomrule		
	\end{tabularx}
	\label{tbl:qbe_m}
\end{table}

Looking at the EER and Spearman's $\rho$ for semantic QbE, we see that the \system{Cluster+Skip-gram} model achieves the highest score, outperforming the purely phonetic AWEs from the multilingual AWE model.
The reason why the phonetic multilingual AWE model outperforms the semantic model in $P@10$ and $P@N$ is due to its proficiency in detecting exact matches (which are also correct semantic matches).

In this chapter, we are mainly interested in observing information captured by our embeddings related to word meaning.
However, in the semantic QbE task, a system also gets rewarded for retrieving utterances containing exact instances of the query. 
To get a better sense of the ability of a model to retrieve non-verbatim semantic matches, we mask out all exact occurrences of the query before we segment the search collection.
The system will now only be able to retrieve utterances based on word meaning.
The results are shown in Table~\ref{tbl:qbe_m}.
Now we see a clear benefit in using the \system{Cluster+Skip-gram} model, with the phonetic multilingual AWE model becoming close to a random search given that there are no exact matches.

Our core goal is semantic QbE, but it is worth briefly touching on exact QbE performance of the \system{Cluster+Skip-gram} model in Table~\ref{tbl:qbe}.
In exact QbE, the system should only retrieve utterances containing instances of the query word type.
The \system{Cluster+Skip-gram} model learns to predict frequently co-occurring words, without explicit training for preserving word-type information.
Although only trained for semantics, the \system{Cluster+Skip-gram} still achieves reasonable exact retrieval performance, with only a drop of between 5\% and 10\% in scores compared to the multilingual AWE model. 
It is therefore clear that this semantic model is able to retain phonetic properties for word-type while also capturing semantic information related to context. 


\subsection[Qualitative analysis]{Qualitative analysis}
\label{ssec:semantic_resuls_qeval}

\begin{figure}[!t]
	\centerline{\includegraphics[width=.6\columnwidth]{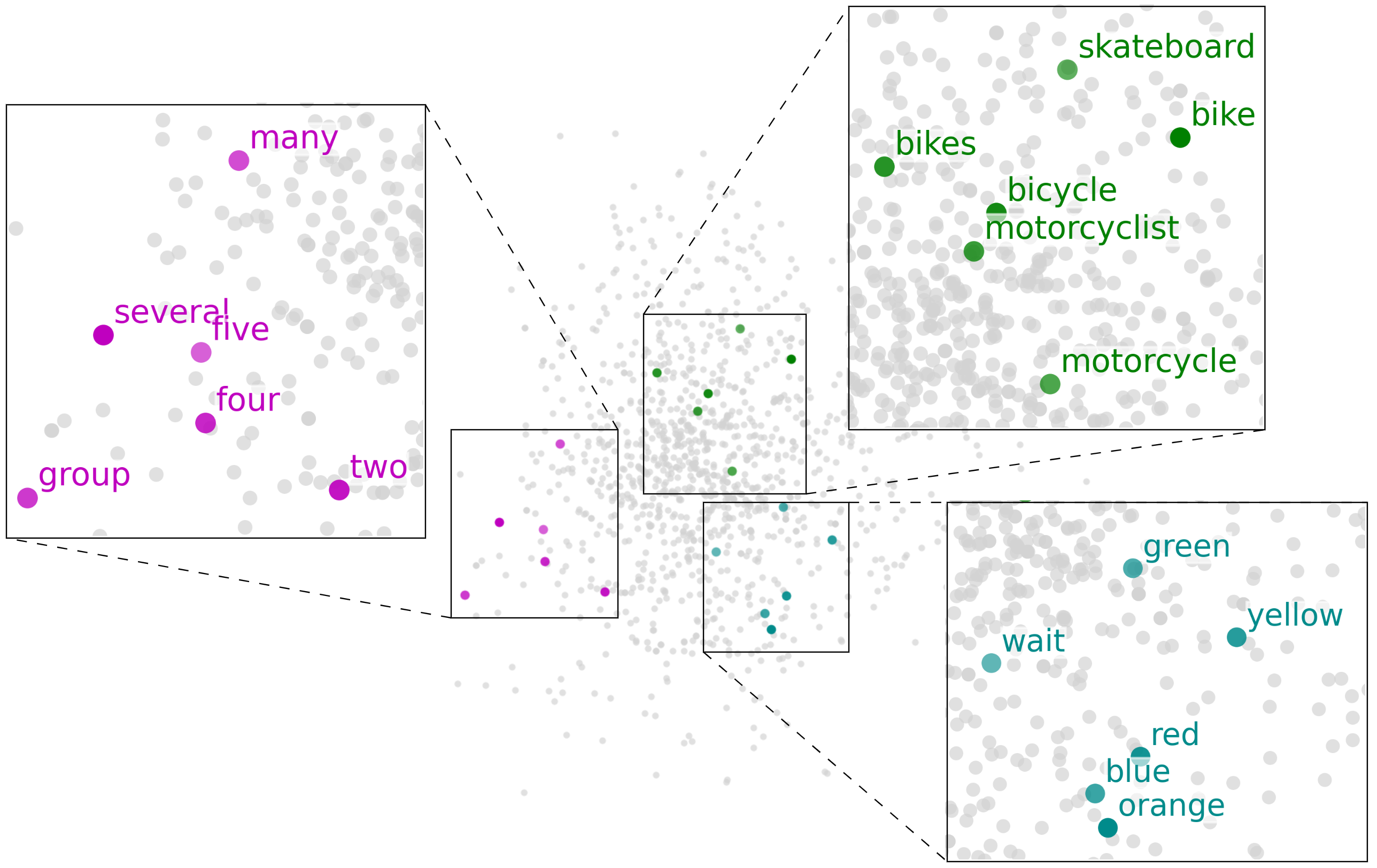}}
	\caption{
		PCA projection of semantic AWEs (averaged), produced by our \system{Cluster+Skip-gram} model on development data.
		We highlight the five nearest neighbours for ``several" (pink), ``bike" (green), and ``orange" (blue). 
	}
	\label{fig:semantic_results_space}
\end{figure}

\begin{table}[!b]
	\mytable
	\caption{
	The top ten closest embeddings for keywords, derived from text (left) and speech segments (right), respectively. Speech embeddings were produced using the \system{Cluster+Skip-gram} model and averaged per word-type before nearest neighbour calculation.
	}
	\captionsep
	\scriptsize
	\begin{tabularx}{\linewidth}{LL@{\qquad}lLL@{\qquad}lLL@{\quad}lLL}
		\toprule				
		\multicolumn{2}{c}{american} && \multicolumn{2}{c}{swimming} && \multicolumn{2}{c}{boys} && \multicolumn{2}{c}{football} \\
		\midrule		
		\textbf{flags} & african && \textbf{pool} & swimmers & &\textbf{children} & \textbf{girls} && opponent & baseball \\ 
		
		spiderman & waving && \textbf{swim} & \textbf{swims} && \textbf{kids} & \textbf{children} && \textbf{soccer} & \textbf{hockey} \\ 
		
		tin & older && water & \textbf{swim} & &\textbf{men} & \textbf{kids} && \textbf{hockey} & \textbf{soccer} \\ 
		
		camouflage & costume & &wet & into && \textbf{girls} & women && \textbf{player} & \textbf{uniform} \\ 
		
		muzzles & \textbf{elderly} && diving & zooms && boy & teens && players & miami \\ 
		participating & \textbf{flags} && headfirst & swings && players & \textbf{men} && volleyball & \textbf{player} \\ 
		\textbf{elderly} & \textbf{asian} && wading & \textbf{pool} && people & are && hitting & approaches \\ 
		tiny & leather && paddling & scratching && dolphins & traveller& & \textbf{uniform} & apportioned \\ 
		instrument & establishment& & \textbf{swims} & scratches && gateway & cans && field & saxophone \\ 
		\textbf{asian} & instruments& & patiently & splashes && girl & adults& & swim &  players\\ 
		\bottomrule		
	\end{tabularx}
	\label{tbl:semantic_eval_words}
\end{table}

Figure~\ref{fig:semantic_results_space} visualises the semantic embedding space produced by the \system{Cluster+Skip-gram} model.
It is clear that the acoustic realisations of semantically related words end up in similar areas in the space.
For example, the model learned that spoken instances of ``orange", ``red", ``blue", ``yellow", and ``green"  should be close to each other.
For additional visualisation, we direct the reader back to Figure~\ref{fig:intro_twe_awe} and Figure~\ref{fig:intro_sem}. Figure~\ref{fig:intro_twe} displays a selected number of TWEs produced by a Skip-gram model trained on the transcribed speech we use for acoustic semantic modelling in this  chapter (Section~\ref{ssec:semantic_setup_data}).
Figure~\ref{fig:intro_awe} displays the AWE produced by the multilingual AWE model we use in our semantic AWE approaches (Section~\ref{sec:semantic_multilingual}).
Finally, Figure~\ref{fig:intro_sem} displays the embeddings (without averaging) produced by our \system{Cluster+Skip-gram} model.
From these visualisations, it is evident that the semantic AWEs satisfactory capture of word-type and semantic information.

The intrinsic results in Table~\ref{tbl:semantic_results_intrinsic} and qualitative analysis in Figure~\ref{fig:intro_sem} show that our semantic embeddings, to some extent, resemble the semantic information captured in the corresponding text embeddings.
To get a better understanding of the difference between the embeddings produced from speech and the corresponding transcripts, we look at the variation in the nearest embeddings of some words.
Table~\ref{tbl:semantic_eval_words} displays chosen keywords, comparing semantic AWEs (on the right) with textual word embeddings (on the left).
Firstly, we notice that across all keywords some words are shared (highlighted in bold).
Secondly, we observe a strong phonetic influence on the semantic embeddings trained on speech. 
They are more likely to have words higher ranked that are phonetically similar to the keyword.
For example, ``american'' and ``football'', have ``african'' and ``baseball'', as their closest neighbour, respectively.
In this instance, we could say they are semantically similar from our own judgement, but they do not appear among the nearest words for the corresponding text embeddings.
Looking at the keyword ``swimming'' we can see where this phonetic disturbance has a negative impact.
For``swimming'', the word ``swims'' appear in the top neighbours for both speech and text. 
Because ``swims'' are phonetically similar to ``zooms'' and ``swings'', the model fails to separate them in vector space.
As a result, ``swimming'' also ends up close to ``zooms'' and ``swings''.
In this instance, we would not regard them as semantically relevant to each other.




\mysection{Chapter summary}{Chapter summary}
\label{sec:semantic_summary}

In this chapter, we addressed a new form of acoustic representation learning, where word embeddings should capture the meaning of words, rather than phonetic information. 
To achieve this, we presented several semantic AWE modelling strategies. 
We specifically promoted transferring knowledge from a pre-trained multilingual AWE model trained for word-class discrimination.
Through intrinsic and extrinsic evaluations, we demonstrated the effectiveness of our strategies in learning semantic representations from unlabelled speech data.
Our best semantic AWE approach involves a soft clustering on the original multilingual AWEs, serving as input to a Skip-gram model.

The main shortcoming of our work (as also in others~\cite{chung_speech2vec_2018, chen_phonetic-and-semantic_2019}), is that the word segmentation is assumed to be known. 
This was reasonable given our goal of comparing different semantic AWE approaches on a sensible benchmark, but future work should look into incorporating unsupervised word segmentation methods~\cite{fuchs_unsupervised_2022, cuervo_contrastive_2022, kamper_word_2023} in order to do fully unsupervised semantic AWE modelling.




\mychapter{Summary and conclusions}{Summary and Conclusions}
\label{chap:conclusion}

Developing robust speech systems for zero-resource languages---where no transcribed speech resources are available for model training---remains a challenge.
Most of these systems, at their core, require speech segments of different duration to be compared.
Previous studies showed the benefit of AWEs over alignment methods for this task.
In a zero-resource setting, the two existing training strategies are unsupervised monolingual learning (Section~\ref{ssec:monolingual}) and supervised multilingual transfer (Section~\ref{ssec:multilingual}).
This thesis mainly contributes to the development of AWEs following the multilingual transfer strategy. 
We presented multiple approaches toward improving the intrinsic quality of AWEs as well as the implementation of AWEs in a downstream speech retrieval task.
Furthermore, we extended the multilingual transfer strategy to semantic representation learning, demonstrating the versatility of multilingual AWEs in upstream tasks.
We first summarise our main findings.
We then provide recommendations on how the work can be extended and applied in future research.


\mysection{Main findings}{Main findings}


In Chapter~\ref{chap:contrastive}, we presented the \system{ContrastiveRNN} AWE model.
This model optimises a distance metric function not considered for AWE modelling in previous work. 
We compared the \system{ContrastiveRNN} to two existing AWE embedding models, \system{CAE-RNN}~\cite{kamper_truly_2019} and \system{SiameseRNN}~\cite{settle_discriminative_2016}.
We re-implemented both existing models and compared our model in the same experimental setup as \citet{kamper_improved_2021}.
On a word-discrimination task measuring the intrinsic quality of the AWEs, the \system{ContrastiveRNN} significantly outperformed the \system{CAE-RNN} and \system{SiameseRNN} following the unsupervised monolingual training strategy, reaching up to 17.8\% absolute increase in average precision on one of the evaluation languages. 
In the supervised multilingual setting, we observed marginal performance increases from the \system{ContrastiveRNN} compared to the \system{CAE-RNN} and \system{SiameseRNN}.




In Chapter~\ref{chap:adaptation}, we proposed a new training strategy that extends the multilingual transfer strategy.
After training a multilingual model, instead of only applying it to an unseen target language, we first fine-tune the multilingual model using discovered word-like pairs obtained from applying an unsupervised term discovery system (UTD) to unlabelled speech data in the target language.
For each of the three multilingual models, we adapt a model to every zero-resource language before applying it.
Here we observe that the \system{ContrastiveRNN} generally adapts better to a target language compared to the \system{CAE-RNN} and \system{SiameseRNN}.

In Chapter~\ref{chap:related}, we investigated the impact a particular choice of training languages have when training an AWE model for a specific target language.
We specifically want to see the impact of including training languages that belong to the same language family as the target language.
We performed multiple experiments {on data from languages in South Africa} where we train multilingual AWEs with different language combinations while controlling for the amount of data.
We show that including related languages over unrelated languages are beneficial, evaluated on an intrinsic word discrimination task.
We also evaluated the multilingual AWEs in a downstream zero-resource speech task, query-by-example, supporting the results of the intrinsic evaluation task.
To summarise the findings in this chapter, including training data from related languages, even in small quantities, are beneficial over using unrelated languages. 
For instance, a multilingual model trained on approximately 6 hours of related languages, outperform a model trained on approximately 60 hours of unrelated languages.
The findings in this chapter would be of great interest to practitioners who want to develop speech applications in a zero-resource setting.

{
In Chapter~\ref{chap:hatespeech}, we developed keyword spotting systems to monitor hate speech in radio broadcast audio in a collaborative effort with a company, VoxCroft.
Our goal was to compare existing ASR-based KWS systems (requiring labelled training data from the target language) to a new multilingual AWE-based KWS system.
We first developed our systems in a controlled environment where we used parallel audio-text data from experimental datasets to train and test our models.
In this in-domain environment, an ASR model trained on only five minutes of transcribed audio outperforms our AWE-based KWS system.
We then applied the models, without further calibrations, to the real radio broadcast audio scraped from radio stations in Kenya.
In this in-the-wild test, our AWE-based KWS system proves to be more robust against domain mismatch and performs similar to an ASR-KWS system using 30 hours of labelled target audio. 
Our findings in this chapter suggest multilingual AWEs can be used as an alternative to existing ASR systems for rapidly deploying KWS systems for languages with limited transcribed audio.
}

{In Chapter~\ref{chap:semantic}, we shifted our attention to a new form of representation learning of spoken word segment: AWEs that reflect word-type information as well as word meaning. 
We specifically proposed using a pre-trained multilingual AWE model to assist in this task compared to existing approaches that only use unlabelled data from the target language~\cite{chung_speech2vec_2018,chen_phonetic-and-semantic_2019}.
One of our models, \system{Cluster+Skip-gram}, show great improvement over previous models in an intrinsic word similarity task, measuring semantic relatedness.
For the first time, we applied these semantic AWEs in a downstream semantic speech retrieval task.
Our results show that semantic information is indeed being captured in the embeddings produced by our new models, indicating promise for future downstream applications.
}


\mysection{Future work}{Future work}
\label{sec:conclusion_future}

Throughout this thesis, we identified multiple opportunities to expand on in future research.

\begin{itemize}
	\item  All our AWE models up to Chapter~\ref{chap:related} took hand-engineered MFCCs as input features.
	In Chapter~\ref{chap:hatespeech} and Chapter~\ref{chap:semantic} we incorporated self-supervised speech features as an alternative to MFCCs for input to our AWE models.
	These learned speech features showed improved performance in a variety of speech tasks~\cite{baevski_wav2vec_2020, conneau_unsupervised_2021, hsu_hubert_2021, chen_wavlm_2022}.
	We also observed the benefits in AWE modelling using these features: improved word discrimination (Section~\ref{ssec:hatespeech_xlsr_vs_mfcc}) and improved semantic representations (Table~\ref{tbl:semantic_results_intrinsic}).
	However, we did not systematically investigate this but merely replaced features without considering alterations to our models or training conditions.
	We hypothesise further improvement by optimising the training setup to accommodate these speech features.
	
	 
	
	\item We notice a large performance gap between the unsupervised monolingual models trained on discovered word pairs (top of Table~\ref{tbl:multi}) and the supervised monolingual models trained on true word segments (top of Table~\ref{tbl:top_line}).
	A more accurate unsupervised segmentation of unlabelled speech data, delivering higher quality discovered pairs, could reduce the gap between the unsupervised and supervised training strategies.
	\citet{kamper_unsupervised_2016} presented a full-coverage segmentation and clustering system---unlabelled speech data are segmented and clustered into unknown word types.
	Their system employs primitive acoustic word embeddings.
	By replacing the AWEs with higher quality AWEs, for instance, embeddings produced by a  multilingual AWE model, better segmentation can be expected.
	Applying this segmentation system to unlabelled speech data in a zero-resource language could potentially yield higher quality discovered word pairs compared to the UTD system of \citet{jansen_efficient_2011} used in this thesis. 
	
	\item Up until now, the choice of training languages in a multilingual transfer setting have not been considered, specifically for AWEs.
	We showed that a more careful selection of training languages can improve performance when developing a multilingual AWE model for a specific target language by including training languages that belong to the same language family as the target.
	We showed that the performance increase is not simply due to the overlap of words in related languages. However, a more in-depth analysis of the phonetic and syntactic properties common to languages from the same language family could give more insight assisting the development of multilingual transfer learning in general.

	\item 
	In Chapter~\ref{chap:hatespeech}, we identified the need for a more optimal threshold value to improve KWS performance.	
	We hypothesise that a per-keyword threshold could be determined based on factors including the length of the speech segment, the number of phones present, and the relative distances between other unknown word segments in the acoustic space.
	In the work of~\citet{matusevych_analyzing_2020}, they analyse an AWE space and discover that the embeddings encode absolute duration and speaker information. Additionally, they observe that the distance between word embeddings increases with phonetic dissimilarity.	
	With such knowledge we could explore potential correlations between these factors stored in AWEs and the distance between instances in the acoustic space.

%

	\item 
	In Chapter~\ref{chap:semantic}, we introduced a novel approach for learning speech representations that reflect semantic meaning instead of acoustic similarity.
	Although our approach exhibits improved performance compared to existing methods, certain limitations remain:
	Similar to others~\cite{chung_speech2vec_2018, chen_phonetic-and-semantic_2019}, we used true word boundaries for segmenting unlabelled audio.
	Future work should consider unsupervised segmentation methods~\cite{fuchs_unsupervised_2022, cuervo_contrastive_2022, kamper_word_2023}, or explore an end-to-end semantic AWE approach that eliminates the need for explicit word boundaries.
	
	Additionally, we performed our experiments within a constrained setup involving a limited vocabulary and a smaller number of words when compared to the millions or even billions used for training textual word embeddings.
	Future work should therefore investigate the scalability of the semantic approaches we have introduced.

\end{itemize}

\chapter*{References}\markboth{}{\scshape References}
\addcontentsline{toc}{chapter}{References}
\renewcommand{\bibsection}{}
\bibliography{mybib, upgradebib}

\end{document}